\pdfoutput=1
\documentclass[11pt,twoside,a4paper,cmspaper,final,collab]{cms-tdr}
\def\svnVersion{e27775b}\def\svnDate{2025/01/30}\def\cmsCernNoTag{CERN-EP-2023-257}\def\cmsCernDate{\today}\def\cmsMessage{Published in the European Physical Journal C as \href{http://dx.doi.org/10.1140/epjc/s10052-024-13606-8}{\doi{10.1140/epjc/s10052-024-13606-8}.}}
\begin{document}\cmsNoteHeader{SMP-21-008}

\newlength\cmsFigWidthOne
\ifthenelse{\boolean{cms@external}}{\setlength\cmsFigWidthOne{0.49\textwidth}}{\setlength\cmsFigWidthOne{0.65\textwidth}}
\newlength\cmsFigWidthTwo
\ifthenelse{\boolean{cms@external}}{\setlength\cmsFigWidthTwo{0.49\textwidth}}{\setlength\cmsFigWidthTwo{0.49\textwidth}}
\ifthenelse{\boolean{cms@external}}{\newcommand{\cmsLeft}{upper\xspace}}{\newcommand{\cmsLeft}{left\xspace}}
\ifthenelse{\boolean{cms@external}}{\newcommand{\cmsRight}{lower\xspace}}{\newcommand{\cmsRight}{right\xspace}}
\ifthenelse{\boolean{cms@external}}{\newcommand{\preprintClearPage}{\relax}}{\newcommand{\preprintClearPage}{\clearpage}}
\ifthenelse{\boolean{cms@external}}{\newcommand{\journalClearPage}{\clearpage}}{\newcommand{\journalClearPage}{\relax}}

\newcommand{\fbinvns}{\ensuremath{\text{fb}^{-1}}\xspace}
\newcommand{\pbinvns}{\ensuremath{\text{pb}^{-1}}\xspace}
\newcommand{\asmz}{\ensuremath{\alpS(m_{\PZ})}\xspace}

\newcommand{\ystar}{\ensuremath{y^{*}}\xspace}
\newcommand{\yboost}{\ensuremath{y_{\text{b}}}\xspace}
\newcommand{\ymax}{\ensuremath{\abs{y}_{\text{max}}}\xspace}
\newcommand{\ymaxNoAbs}{\ensuremath{y_{\text{max}}}\xspace}
\newcommand{\yabs}{\ensuremath{\abs{y}}\xspace}
\newcommand{\ybys}{{(\ystar\!,~\yboost)}}
\newcommand{\pti}[1]{\ensuremath{p_\text{T,#1}}\xspace}
\newcommand{\ptave}{\ensuremath{\langle \pt \rangle_{1,2}}\xspace}
\newcommand{\ptaveNoIndices}{\ensuremath{\langle \pt \rangle}\xspace}
\newcommand{\mjj}{\ensuremath{m_{1,2}}\xspace}

\newcommand{\mjjGen}{\ensuremath{\mjj^\text{gen}}\xspace}
\newcommand{\ymaxGen}{\ensuremath{\ymax^\text{gen}}\xspace}

\newcommand{\mjjRec}{\ensuremath{\mjj^\text{rec}}\xspace}
\newcommand{\ymaxRec}{\ensuremath{\ymax^\text{rec}}\xspace}

\newcommand{\UBar}{\ensuremath{\overline{\mathrm{U}}}\xspace}
\newcommand{\DBar}{\ensuremath{\overline{\mathrm{D}}}\xspace}
\newcommand{\uBar}{\ensuremath{\overline{\mathrm{u}}}\xspace}
\newcommand{\dBar}{\ensuremath{\overline{\mathrm{d}}}\xspace}
\newcommand{\sBar}{\ensuremath{\overline{\mathrm{s}}}\xspace}

\newcommand{\fs}{\ensuremath{f_\mathrm{s}}\xspace}

\newcommand{\intlumi}{\ensuremath{\mathcal{L}_\text{int}}\xspace}

\newcommand{\mur}{\ensuremath{\mu_{\mathrm{R}}}\xspace}
\newcommand{\muf}{\ensuremath{\mu_{\mathrm{F}}}\xspace}
\newcommand{\mufzero}{\ensuremath{\mu_{\mathrm{F},\,0}}\xspace}

\newcommand{\cNP}{\ensuremath{c_{\mathrm{NP}}}\xspace}

\newcommand{\parlevel}{par\-ti\-cle-lev\-el\xspace}
\newcommand{\minbias}{min\-i\-mum-bias\xspace}
\newcommand{\collsafe}{col\-li\-near-safe\xspace}

\providecommand{\NNLOJET}{\textsc{nnlojet}\xspace}
\providecommand{\FASTNLO}{\textsc{fastNLO}\xspace}
\providecommand{\QCDNUM}{\textsc{qcdnum}\xspace}
\providecommand{\LHAPDF}{\textsc{lhapdf}\xspace}

\newcommand{\asmzResult}[7]{%
\ifthenelse{\boolean{cms@external}}{%
  #1 &= #2 &&\pm #3\,(\text{fit})\\
     &     &&\pm #4\,(\text{scale})\\
     &     &&\pm #5\,(\text{model})\\
     &     &&\pm #6\,(\text{param.})\\
     &= #2 &&\pm #7\,(\text{total})}{
  #1 &= #2   \pm #3\,(\text{fit})
             \pm #4\,(\text{scale})
             \pm #5\,(\text{model})
             \pm #6\,(\text{param.})\\
     &= #2   \pm #7\,(\text{total})%
}}

\newcommand{\asmzResultNoTotal}[6]{%
\ifthenelse{\boolean{cms@external}}{%
  #1 &= #2 &&\pm #3\,(\text{fit})\\
     &     &&\pm #4\,(\text{scale})\\
     &     &&\pm #5\,(\text{model})\\
     &     &&\pm #6\,(\text{param.})}{
  #1 &= #2   \pm #3\,(\text{fit})
             \pm #4\,(\text{scale})
             \pm #5\,(\text{model})
             \pm #6\,(\text{param.})%
}}

\cmsNoteHeader{SMP-21-008}

\title{Measurement of multidifferential cross sections for dijet production in proton-proton collisions at \texorpdfstring{$\sqrt{s} = 13\TeV$}{sqrt(s) = 13 TeV}}

\date{\today}

\abstract{
A measurement of the dijet production cross section is reported based on proton-proton collision data collected in 2016 at $\sqrt{s}=13\TeV$ by the CMS experiment at the CERN LHC, corresponding to an integrated luminosity of up to 36.3\fbinv. Jets are reconstructed with the anti-$\kt$ algorithm for distance parameters of $R=0.4$ and 0.8. Cross sections are measured double-differentially (2D) as a function of the largest absolute rapidity $\ymax$ of the two jets with the highest transverse momenta \pt and their invariant mass $\mjj$, and triple-differentially (3D) as a function of the rapidity separation $\ystar$, the total boost $\yboost$, and either $\mjj$ or the average \pt of the two jets. The cross sections are unfolded to correct for detector effects and are compared with fixed-order calculations derived at next-to-next-to-leading order in perturbative quantum chromodynamics. The impact of the measurements on the parton distribution functions and the strong coupling constant at the mass of the $\PZ$ boson is investigated, yielding a value of $\asmz=0.1179\pm0.0019$.
}

\hypersetup{
pdfauthor={CMS Collaboration},
pdftitle={Measurement of multidifferential cross sections for dijet production in proton-proton collisions at {sqrt(s) = 13 TeV}},
pdfsubject={CMS},
pdfkeywords={CMS, QCD, proton, collision, jet, dijet, PDF, NNLO}}

\ifthenelse{\boolean{cms@external}}{
\titlerunning{Multi-differential measurement of the dijet cross section}
}{
}

\maketitle

\section{Introduction\label{sec:intro}}

{\tolerance=1200
The production of jets in high-energy proton-proton ($\Pp\Pp$) collisions
provides an important experimental input for the determination of the
proton structure in terms of parton distribution functions (PDFs),
and for the study of the strong force described by quantum chromodynamics (QCD).
In conjunction with deep-inelastic $\Pe^{\pm}\Pp$ scattering (DIS) measurements~\cite{H1:2009pze,H1:2015ubc},
which strongly constrain the quark PDFs, jet data from $\Pp\Pp$
collisions at the LHC provide sensitivity to the gluon content and
allow the running of the strong coupling constant \alpS to be probed over a
wide range of momentum scales.
Recent progress made in calculating predictions for these processes at
next-to-next-to-leading order (NNLO)
accuracy~\cite{Currie:2017eqf,Gehrmann-DeRidder:2019ibf} in perturbative QCD (pQCD) underscores
the need for precise experimental data up to the highest accessible energies.
\par}

Dijet observables are particularly well-suited for this purpose owing to the
abundant production of jets in hadron-induced processes across a large
phase space, which makes it possible to perform high-precision
multi-differential measurements.
Such measurements performed at the LHC include a triple-differential (3D)
dijet measurement at a center-of-mass energy $\sqrt{s}=8\TeV$~\cite{Sirunyan:2017skj} using
jets reconstructed with the anti-\kt{} clustering
algorithm~\cite{Cacciari:2008gp,Cacciari:2011ma} with a distance parameter
$R = 0.7$, and several double-differential (2D) measurements at
7~and~13\TeV~\cite{Aad:2010ad,Chatrchyan:2011qta,Chatrchyan:2012bja,ATLAS:2013jmu,Aaboud:2017wsi}
for anti-\kt{} jets with $R = 0.4$, 0.6, or 0.7.

In this article, measurements of the dijet production cross section in
$\Pp\Pp$ collisions at $\sqrt{s} = 13\TeV$ from the CMS Collaboration are presented,
using anti-\kt{} jets for two values of the distance parameter,
$R = 0.4$ and 0.8.
Both 2D and 3D measurements are performed as a function of the kinematic
properties of the two jets with the highest transverse momenta (\pt) in the event.

In the 2D case, the cross section is measured as a function
of the largest absolute rapidity $\ymax$ of the two jets and the
invariant mass $\mjj$ of the dijet system, as done for the CMS measurements at
$\sqrt{s} = 7\TeV$ \cite{Chatrchyan:2011qta,Chatrchyan:2012bja}.
For the 3D measurements, the same two angular observables are considered as
for the previous CMS measurement at $\sqrt{s} = 8\TeV$ \cite{Sirunyan:2017skj}:
the dijet rapidity separation $\ystar=\abs{y_1-y_2}/2$
and the total boost of the dijet system $\yboost=\abs{y_1+y_2}/2$,
where $y_1$ and $y_2$ indicate the rapidities of the jets.
The measurements are performed as a function of $\ystar$, $\yboost$, and $\mjj$,
and alternatively as a function of $\ystar$, $\yboost$, and the average \pt of the
two jets, $\ptave$.

The 2D and 3D measurements cover a largely overlapping phase space.
However, each of the two presents different experimental advantages stemming
from the difference in the information content of the respective observables.
The 2D measurement features a more inclusive rapidity binning, leading to an
increased statistical precision and a larger accessible range in $\mjj$. The use of
two angular observables for the 3D measurement provides additional information on the
dijet topology, at the expense of a reduced reach in $\mjj$.
Moreover, the variables $\ystar$ and $\yboost$ encode the dependence on the
partonic scattering angle in the laboratory frame and the imbalance in the
initial-state parton momenta, respectively.
This is advantageous for comparisons to fixed-order pQCD
predictions, which are obtained by convolving the partonic scattering cross
sections and the PDFs.
Specifically, the relation between the dijet invariant mass and the proton momentum
fractions $x_{\pm}$ carried by the incoming partons is given at leading order (LO)
by $x_{\pm} = \mjj~\exp(\pm \yboost) / \sqrt{s}$.
Using the average dijet \pt instead renders the $\ystar$ dependence explicit and
gives $x_{\pm} = 2\ptave~\text{cosh}(\ystar)\exp(\pm \yboost) / \sqrt{s}$.

This article is organized as follows.
A brief description of the CMS detector is given in Section~\ref{sec:detector}.
Section~\ref{sec:samples} presents the samples of recorded and simulated events
used for the measurement.
In Section~\ref{sec:selection}, the reconstruction of the event content is described,
and the selection criteria applied to events entering this analysis are
given.
Sections~\ref{sec:measurement} and~\ref{sec:unfolding} detail the measurement of
the 2D and 3D dijet cross sections using the reconstructed jets, and the unfolding
of the resulting spectra to correct for detector effects, respectively.
The different sources of experimental uncertainty in the measurement are outlined
in Section~\ref{sec:uncertainties}.
The measurements are compared to fixed-order predictions obtained at NNLO
accuracy in pQCD, which are discussed in Section~\ref{sec:theory}.
A comparison of the measurements to the predictions obtained for several global
PDF sets is presented in Section~\ref{sec:results}.
In Section~\ref{sec:fits}, the impact of including the present measurements in
determinations of PDFs and the strong coupling constant at the scale of the
\PZ boson mass, $\asmz$, is investigated.
Finally, a summary of the main findings is given in Section~\ref{sec:summary}.

Tabulated results are provided in the HEPData record for this analysis~\cite{hepdata}.

\section{The CMS detector\label{sec:detector}}

The central feature of the CMS apparatus is a superconducting solenoid of
6\unit{m} internal diameter, providing a magnetic field of 3.8\unit{T}.
Within the solenoid volume are a silicon pixel and strip tracker, a lead
tungstate crystal electromagnetic calorimeter (ECAL), and a brass and
scintillator hadron calorimeter (HCAL), each composed of a barrel and two
endcap sections.
Forward calorimeters extend the pseudorapidity ($\eta$) coverage provided by the barrel
and endcap detectors.
Muons are detected in gas-ionization chambers embedded in the steel flux-return
yoke outside the solenoid.

Events of interest are selected using a two-tiered trigger
system~\cite{Khachatryan:2016bia}.
The first level, composed of custom hardware processors, uses information
from the calorimeters and muon detectors to select events at a rate of around
100\unit{kHz} within a fixed latency of about 4\mus~\cite{Sirunyan:2020zal}.
The second level, known as the high-level trigger (HLT), consists of a processor farm
running a compact version of the full event reconstruction software,
optimized for fast processing, and reduces the event rate to around 1\unit{kHz}
before data storage.

A more detailed description of the CMS detector, together with a definition of
the coordinate system used and the relevant kinematic variables, can be found
in Ref.~\cite{Chatrchyan:2008zzk}.

\section{Data and simulated samples\label{sec:samples}}

The measurements presented in this article are based on $\Pp\Pp$ collision data
recorded by the CMS detector in 2016 at $\sqrt{s} = 13\TeV$,
corresponding to an integrated luminosity of up to 36.3\fbinv.
Collision events containing jets are identified during data taking by dedicated
trigger algorithms.
Owing to the stringent timing constraints, jets at the HLT are clustered from
particle candidates reconstructed using a simplified procedure, as compared to
the full offline reconstruction.

The integrated luminosity recorded by the available jet-related triggers is
given in Table~\ref{tab:trigger-luminosities}.
Several sets of triggers are deployed, which require the presence of at least
one jet (two jets) with a $\pt$ (average $\pt$) above certain predefined thresholds.
While distinct sets of single-jet triggers are deployed for anti-$\kt$ jets with
distance parameters of $R = 0.4$ and 0.8, only the former are used for
the dijet triggers.
The integrated luminosity delivered by each of these triggers depends on the
total time period during which it was deployed.
In addition, low-threshold triggers are prescaled by a factor that is
continually adjusted during data taking to optimize the data acquisition rate,
resulting in lower effective integrated luminosities.

\begin{table}[ht]
  \topcaption{Overview of the single-jet (dijet) triggers deployed
  for the different $\pt$ ($\ptaveNoIndices$) thresholds at the HLT,
  and the corresponding integrated luminosities.
  \label{tab:trigger-luminosities}}
  \centering
  \ifthenelse{\boolean{cms@external}}{
  \begin{tabular}{llllll}
    & \multicolumn{5}{l}{{Trigger threshold} (\GeVns{})}
    \\[1mm]
    & 40   & 60   & 80  & 140  & 200
    \\[1mm]\hline\\[-1mm]
    {Trigger set}
    & \multicolumn{5}{l}{{Int. luminosity} (\pbinvns{})}
    \\[1mm]
    Single-jet $R = 0.4$
    & 0.3  & 0.7  & 2.8 & 24.2 & 103.6
    \\
    Single-jet $R = 0.8$
    & 0.05 & 0.3  & 1.0 & 10.1 & 85.8
    \\
    Dijet      $R = 0.4$
    & 0.1 &  1.7  & 4.2 & 27.9 & 140.2
    \\[7mm]
    & \multicolumn{5}{l}{{Trigger threshold} (\GeVns{})}
    \\[1mm]
    & 260 & 320  & 400  & 450  & 500
    \\[1mm]\hline\\[-1mm]
    {Trigger set}
    & \multicolumn{5}{l}{{Int. luminosity} (\fbinvns{})}
    \\[1mm]
    Single-jet $R = 0.4$
    & 0.6 & 1.8  & 5.2  & 36.3 & 36.3
    \\
    Single-jet $R = 0.8$
    & 0.5 & 1.5  & 4.6  & 33.5 & 33.5
    \\
    Dijet      $R = 0.4$
    & 0.5 & 3.0  & 9.1  & \NA   & 29.6
  \end{tabular}
}{
  \begin{tabular}{p{35mm}llllp{12mm}lllll}
    \\[-2mm]
    & \multicolumn{10}{l}{{Trigger threshold} (\GeVns{})}
    \\[1mm]
    & 40   & 60   & 80  & 140  & 200 & 260 & 320  & 400  & 450  & 500
    \\[1mm]\cline{2-11}\\[-1mm]
    {Trigger set}
    & \multicolumn{5}{l}{{Int. luminosity} (\pbinvns{})}
    & \multicolumn{5}{l}{{Int. luminosity} (\fbinvns{})}
    \\[1mm]
    Single-jet $R = 0.4$
    & 0.3  & 0.7  & 2.8 & 24.2 & 103.6 & 0.6 & 1.8  & 5.2  & 36.3 & 36.3
    \\
    Single-jet $R = 0.8$
    & 0.05 & 0.3  & 1.0 & 10.1 &  85.8 & 0.5 & 1.5  & 4.6  & 33.5 & 33.5
    \\
    Dijet      $R = 0.4$
    & 0.1 &  1.7  & 4.2 & 27.9 & 140.2 & 0.5 & 3.0  & 9.1  & \NA   & 29.6
  \end{tabular}
}
\end{table}

To study the impact of the detector response on the measurement, samples of
simulated events produced using Monte Carlo~(MC) event generators are used.
Events are generated at LO in pQCD using \PYTHIAviii~\cite{Sjostrand:2014zea}
(version~8.212) with the CUETP8M1 tune~\cite{Khachatryan:2015pea}.
The matrix element calculation is matched to the parton shower and
takes multi-parton interactions and hadronization effects into account.
An alternative LO sample, generated using the \MGvATNLO{} program~\cite{Alwall:2014hca}
(version~2.2.2) and
interfaced with \PYTHIA for the simulation of parton showering and hadronization,
is used to estimate the dependence of results on the simulation model.

To simulate contributions from additional $\Pp\Pp$ collisions (pileup),
the particles emerging from the high-energy scattering are
overlaid with simulated \minbias events and propagated through a full
simulation of the CMS detector modeled using the \GEANTfour
package~\cite{Agostinelli:2002hh}.
The resulting signals are then processed using the same reconstruction
techniques used for collision data.
Differences between the simulated and measured pileup activity are accounted for
using a global reweighting of simulated events based on the mean number of pileup
interactions determined in data based on an estimated inelastic $\Pp\Pp$ collision
cross section of 69.2\unit{mb}. This number is obtained using the pileup counting
method described in the inelastic cross section measurement~\cite{CMS:2018mlc}.
About 23 pileup interactions occurred for each proton bunch collision during
the 2016 data taking~\cite{CMS:2020ebo}.

\section{Event reconstruction and selection\label{sec:selection}}

A global description of collision events
is achieved following the particle-flow approach~\cite{CMS-PRF-14-001},
which aims to identify and measure the kinematic properties of each individual
particle emerging from the collision using an optimized combination of
information from the various elements of the CMS detector.

The trajectories of charged particles, as well as their originating $\Pp\Pp$
interaction vertices are reconstructed from hits in the inner tracking
detectors.
The primary vertex is taken to be the vertex corresponding to the hardest
scattering in the event, evaluated using tracking information alone, as
described in Section 9.4.1 of Ref.~\cite{CMS-TDR-15-02}.

Muons
are identified as particle tracks in the inner detector layers that are
compatible with either a track or several hits in the muon system, and
are associated with calorimeter deposits consistent with the muon hypothesis.
The muon four-momentum is determined by fitting the muon trajectory using information
from both the inner tracker and the muon system.

Photons
are identified as ECAL energy clusters not linked to the extrapolation of any
charged-particle track to the ECAL\@.
Electrons
are identified by linking a primary charged-particle track to potential energy
deposits in the ECAL\@. The resulting energy clusters are required to be
spatially compatible with the extrapolated track to the ECAL, or consistent with
bremsstrahlung photons emitted in the tracker material.
While for photons the energy is obtained directly from the ECAL measurement, the
electron energy is determined from a combination of the track momentum at
the primary interaction vertex and the associated ECAL clusters.

Charged hadrons
are identified as particle tracks not identified as
electrons or muons, and neutral hadrons are identified as HCAL energy clusters
not linked to any charged-hadron trajectory, or as a combined ECAL and HCAL
energy excess with respect to the expected charged hadron energy deposit.
The energy of charged hadrons is determined by combining the track
momentum and the corresponding ECAL and HCAL energies, corrected for the
response function of the calorimeters to hadronic showers.
The energy of neutral hadrons is obtained from the corresponding
corrected ECAL and HCAL energies.

For each event, jets are clustered from the reconstructed particle candidates using the
infrared- and \collsafe anti-\kt algorithm~\cite{Cacciari:2008gp,
Cacciari:2011ma} with distance parameters of $R = 0.4$ and 0.8.
The jet momentum is determined as the vector sum of all particle momenta in the
jet, and is found from simulation to be, on average, within 5--10\% of the
true momentum over the entire \pt range and detector acceptance used in the analysis.
To mitigate the effect of pileup, which can contribute additional tracks and
calorimetric energy depositions to the jet momentum, charged particles identified
as originating from pileup vertices are discarded and an offset correction~\cite{CMS-PAS-JME-14-001}
is applied to account for the remaining contributions.

Jet energy corrections~\cite{Khachatryan:2016kdb} are derived from simulation studies so that the average
energy of reconstructed jets becomes identical to that of \parlevel jets.
The latter are defined as jets clustered from all stable particles produced in
the collision, excluding neutrinos.
In situ measurements of the momentum balance in dijet, $\text{photon} +
\text{jet}$, $\PZ + \text{jet}$, and multijet events are used to account for
any residual differences in the jet energy scale (JES) between data and
simulation.
The jet energy resolution (JER) typically amounts to 15--20\% at 30\GeV, 10\% at
100\GeV, and 5\% at 1\TeV~\cite{Khachatryan:2016kdb}.
It is measured in data using similar jet balancing approaches as for the JES,
and residual differences between data and simulation are corrected
by smearing the \pt of simulated jets accordingly.
Additional selection criteria~\cite{CMS-PAS-JME-16-003} are applied to each jet
to remove jets potentially dominated by spurious contributions from various
subdetector components or reconstruction failures.
Similarly, anomalous events caused by reconstruction failures, detector malfunctions,
or noncollision backgrounds are identified and rejected by dedicated
event filters.
These are designed to identify more than 85--90\% of anomalous events with
a misidentification rate of less than 0.1\%.
Further details can be found in Ref.~\cite{Sirunyan:2019kia}.

Events entering the 2D cross section measurements for both
$R = 0.4$ and 0.8 are required to have been accepted by at least one
single-jet trigger path operating on jets with the same distance parameter.
For the 3D measurements,
the dijet triggers are used on account of their lower overall prescale values.

To guarantee a high reconstruction efficiency at the trigger level, trigger
paths with different thresholds are assigned to mutually exclusive phase
space regions.
These are determined for single-jet (dijet) triggers based on measurements of
the trigger efficiency as a function of the leading jet \pt
(average \pt of the two leading jets), requiring
the trigger efficiency to remain above 99.5\% in each region.

During the 2016 data taking, a gradual shift in the timing of the inputs of the
ECAL first-level trigger in the region defined by $\abs{\eta} > 2.0$
caused the trigger signal to be
incorrectly associated to the previous bunch crossing (``prefiring''),
leading to a specific trigger inefficiency.
For events containing a jet with a \pt larger than $\approx$100\GeV, the efficiency
loss in the region $2.5 < \abs{\eta} < 3.0$ is $\approx$10--20\%, depending on
\pt, $\eta$, and time.
Correction factors were computed from data and applied to the acceptance
evaluated from simulation.

Further selection criteria are applied to events passing the trigger selection,
based on the kinematic properties of two jets with the highest \pt, denoted in the
following by the subscripts 1 and 2 for the \pt-leading and \pt-subleading jets,
respectively.
For the former, a requirement of $\pti{1}>100\GeV$ and $\abs{y_1}<3$ is imposed, while the
latter is required to satisfy $\pti{2}>50\GeV$ and $\abs{y_2}<3$.

\section{Cross section measurement\label{sec:measurement}}

The dijet production cross section is measured
both double- and triple-differentially
for anti-\kt jets with distance parameters of $R = 0.4$ and 0.8
in terms of the properties of
the system formed by the two \pt-leading jets.
The 2D spectra are reconstructed
as a function of $\mjj$
in five rapidity regions defined in terms of
the variable $\ymax = \abs{\ymaxNoAbs}$, where $\ymaxNoAbs$ corresponds to the
rapidity of the jet closer to the beam line (outermost jet), and is given by
\begin{linenomath*}
\ifthenelse{\boolean{cms@external}}{
\begin{multline}
  \ymaxNoAbs = \text{sign}(\abs{\text{max}(y_1, y_2)} - \abs{\text{min}(y_1, y_2)})\\
               \times\text{max}(\abs{y_1}, \abs{y_2}).\label{eq:def-ymax}
\end{multline}
}{
\begin{equation}
  \ymaxNoAbs = \text{sign}(\abs{\text{max}(y_1, y_2)} - \abs{\text{min}(y_1, y_2)})\,
               \text{max}(\abs{y_1}, \abs{y_2}).\label{eq:def-ymax}
\end{equation}
}
\end{linenomath*}

In the 3D case, the cross section is measured as a function of
$\mjj$ and $\ptave$ in 15 rapidity regions,
defined in terms of the dijet rapidity separation \ystar{}
and the total boost \yboost{} of the dijet system, as given by
\begin{linenomath*}
\begin{equation}
  \ystar = \frac{1}{2}\abs{y_1 - y_2},\quad
  \yboost = \frac{1}{2}\abs{y_1 + y_2}.\label{eq:def-ystar-yboost}
\end{equation}
\end{linenomath*}
The variables $\mjj$ and $\ptave$ are obtained as
\begin{linenomath*}
\ifthenelse{\boolean{cms@external}}{
\begin{equation}\begin{aligned}
  \mjj &= \sqrt{(E_1 + E_2)^2 - (\vec{p}_1 + \vec{p}_2)^2},\\
  \ptave &= \frac{1}{2}\left(\pti{1} + \pti{2}\right)\label{eq:def-ptave-mjj},
\end{aligned}\end{equation}
}{
\begin{equation}
  \mjj = \sqrt{(E_1 + E_2)^2 - (\vec{p}_1 + \vec{p}_2)^2},\quad
  \ptave = \frac{1}{2}\left(\pti{1} + \pti{2}\right)\label{eq:def-ptave-mjj},
\end{equation}
}
\end{linenomath*}
where the subscripts 1 and 2 refer to the \pt-leading and \pt-subleading jets,
respectively.

The differential dijet spectra are reconstructed from the effective event yield
$N_{\text{eff}}$ in
bins of the chosen observables, normalized to the integrated luminosity
$\intlumi$.
The effective event yield is calculated from the raw event yield, taking into
account the selection efficiency and subtracting background contributions.
In addition, events that enter the measurement are weighted according to the
prescale factor of the trigger path assigned to the corresponding phase space region.

The 2D cross section is obtained as a function of $\ymaxNoAbs$ and $\mjj$ as
\begin{linenomath*}
\begin{equation}\begin{aligned}
    \frac{\rd^2\sigma}{\rd y_{\text{max}}\rd\mjj} &=
        \frac{1}{\intlumi}\,\frac{N_{\text{eff}}}{(2\,\Delta\ymax)\Delta\mjj}.
    \label{eq:def-2d-cross-section}
\end{aligned}\end{equation}
\end{linenomath*}
Here, $\Delta\ymax$ and $\Delta\mjj$ denote the width of the bins in the
respective quantities.
The measurement is performed in five rapidity bins of equal size within
$0 < \ymax < 2.5$ and covers an invariant mass range of
$249 < \mjj < 10\,050\GeV$.
The measurement boundaries are chosen starting from a preliminary binning determined
using simulated samples based on the expected experimental resolution in $\mjj$, and
discarding bins at low $\mjj$ that do not meet the minimal trigger efficiency
requirement of 99.5\%, and bins at high $\mjj$ for which the statistical uncertainty
exceeds 50\%.

\begin{figure}[htb]
  \centering
    \includegraphics[width=0.49\textwidth]{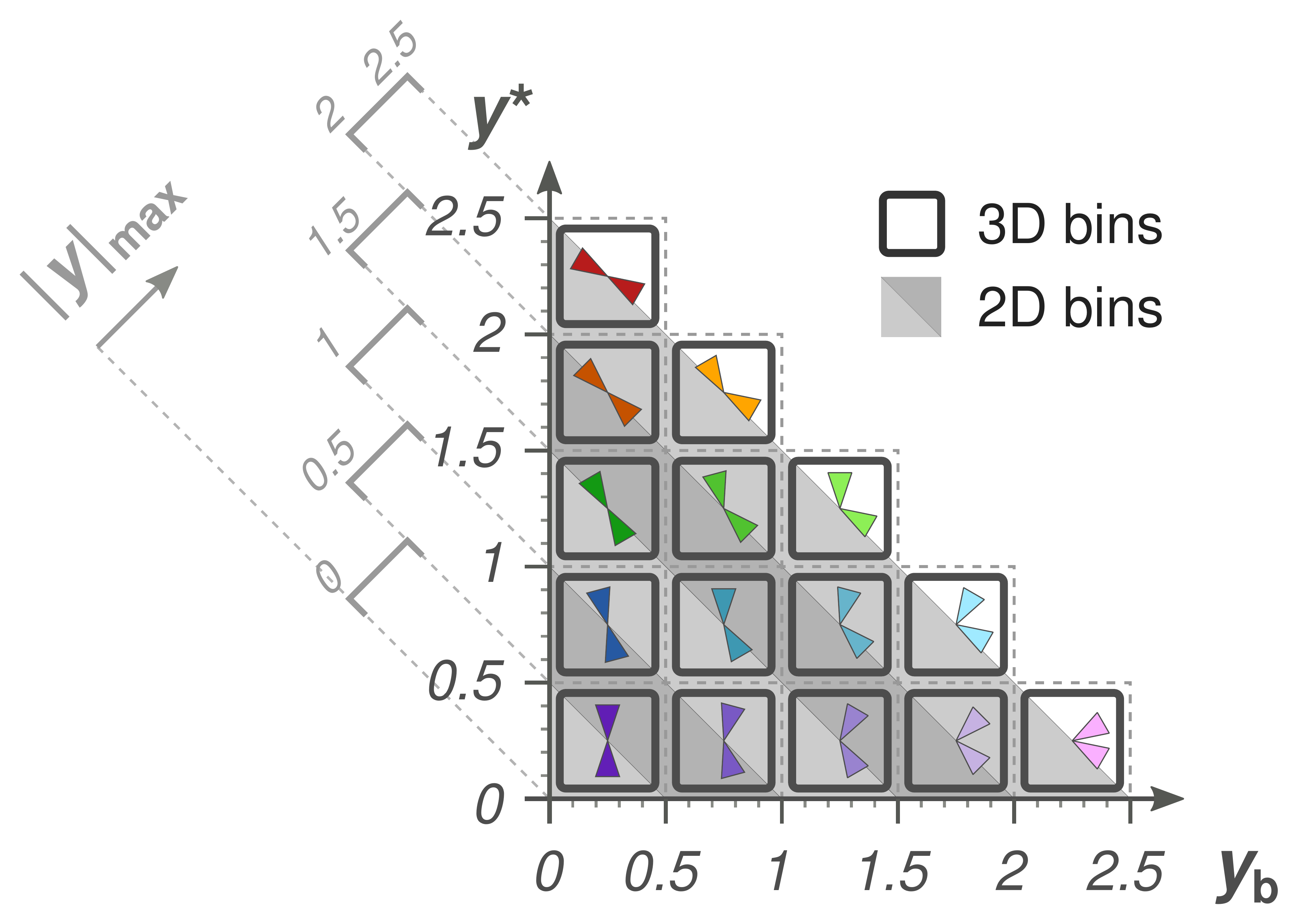}\hspace{0.1\textwidth}
    \caption{
    Illustration of the dijet rapidity phase space, highlighting the relationship 
    between the variables used for the 2D and 3D measurements. 
    The colored triangles are suggestive of the orientation of the two jets in the different phase 
    space regions in the laboratory frame, assuming that the beam line runs horizontally. 
    \label{fig:phase-space-illustration}}
\end{figure}

For the 3D measurement, the cross section is obtained as
\begin{linenomath*}
\begin{equation}\begin{aligned}
    \frac{\rd^3\sigma}{\rd\ystar\rd\yboost\rd x} &=
        \frac{1}{\intlumi}\,\frac{N_{\text{eff}}}{\Delta\ystar\Delta\yboost\Delta x}.
    \label{eq:def-3d-cross-section}
\end{aligned}\end{equation}
\end{linenomath*}
As in Eq.~\eqref{eq:def-2d-cross-section}, the event yield is normalized
to the observable bin widths $\Delta\ystar$, $\Delta\yboost$ and $\Delta x$,
where $x$ stands for either $\mjj$ or $\ptave$.
Fifteen rapidity regions are investigated, covering the range from
0 to 2.5 in each observable, as illustrated in
Fig.~\ref{fig:phase-space-illustration}.
Different invariant mass and average transverse momentum regions are measured
depending on the rapidity region, covering a range of
$306 < \mjj < 6094\GeV$ and $147 < \ptave < 2702\GeV$, respectively.
These ranges are obtained using an analogous procedure as for the 2D measurements.

\section{Unfolding\label{sec:unfolding}}

Because of the finite detector resolution and other experimental
effects, such as the reconstruction efficiency,
the properties of reconstructed jets differ from those of jets
defined at the particle level. This leads to a migration of dijet events within
the phase space spanned by the observables used for the cross section
measurement.
To enable a direct comparison of the measured cross sections to theoretical
\parlevel calculations or to other measurements, the effect of these
migrations is accounted for as part of a multi-dimensional unfolding procedure.

Using simulated event samples, the dijet observables of interest are computed
event by event based on both the two \pt-leading reconstructed jets and the jets
clustered directly from generated particles.
Response matrices are constructed to reflect the probability of bin-to-bin event
migrations between the particle and reconstruction levels, taking all the
observables used for a measurement into account simultaneously.

The measured event distributions are unfolded using the
\textsc{TUnfold} package~\cite{Schmitt:2012kp}, based on the
simulation-derived response matrices.
While no explicit regularization of the unfolded distributions is performed,
large fluctuations between neighboring bins stemming from an ill-conditioned
response matrix are avoided through an appropriate choice of bins.
These are chosen in such a way as to ensure that the bin sizes remain at least
twice as large as the resolution in these variables, and that the purity is at
the level of 50\% or above.
The latter is defined as the fraction of reconstructed events in each bin that
originate from genuine dijet events in the same bin at the particle level.

\begin{figure*}[t]
  \centering
    \includegraphics[width=\textwidth]{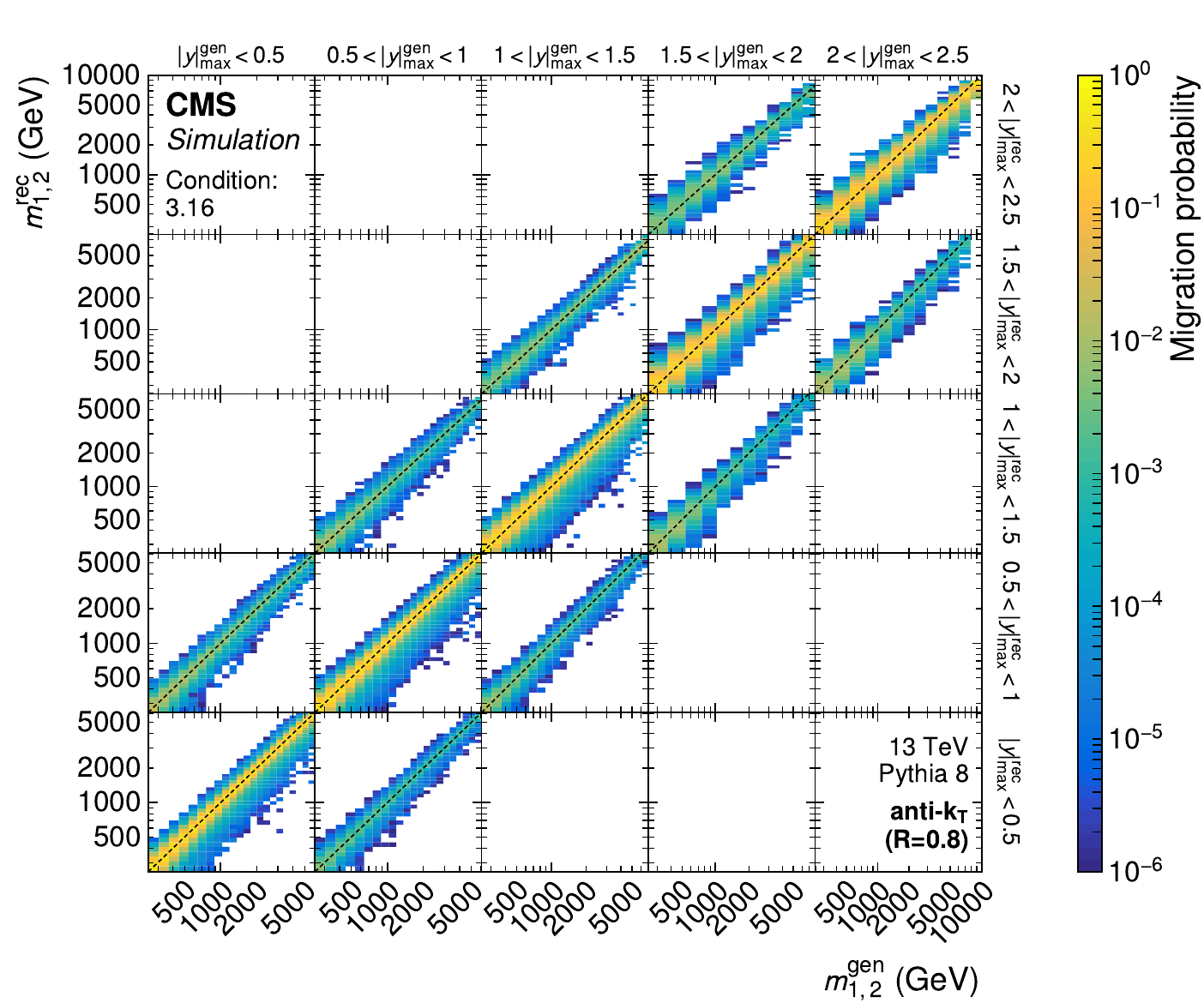}
    \caption{
    Response matrix for the 2D measurement as a function of $\mjj$ using jets with $R = 0.8$. 
    The entries represent the probability for a dijet event generated in the 
    phase space region (\mjjGen, \ymaxGen) indicated on the $x$ axis to be reconstructed in the 
    phase space region (\mjjRec, \ymaxRec) indicated on the $y$ axis. 
    Response matrices for all other jet sizes and observables can be found in 
    Appendix~\ref{app:supplementary-material}.
    \label{fig:response-matrices}}
\end{figure*}

To ensure that the unfolding problem is well-posed, a larger number of bins is
chosen for the reconstructed distributions than for the \parlevel
distributions.
Moreover, because of the larger resolution and the decrease in purity at outer
rapidities, a coarser \parlevel binning is chosen for the two outermost
$\ymax$ regions for the 2D measurements, and the corresponding nine outermost
$\ybys$ regions for the 3D measurements.
All response matrices obtained in this way exhibit condition numbers of ${\approx}3$
and are thus suitable for unfolding without regularization.
The condition number is defined as the absolute value of the ratio between the largest
and smallest matrix eigenvalue.
Figure~\ref{fig:response-matrices} shows the responses obtained for a representative
choice of jet distance parameters and dijet observables.

Aside from event migrations within the measurement phase space, contributions to
each bin from spurious jet reconstructions, pileup, changes in the \pt\ ordering
of jets, or migrations into the phase space, are evaluated in the simulation and
proportionally subtracted from the measured distributions prior to unfolding.
Similarly, to correct for event losses due to the finite reconstruction
efficiency, changes in the \pt\ ordering of jets, or migrations outside the phase
space, bin-by-bin correction factors are derived from simulation and applied to
the unfolded distributions.

\section{Experimental uncertainties\label{sec:uncertainties}}

Statistical fluctuations in the observed event counts and various systematic
effects give rise to experimental uncertainties
in the measured cross sections.
The statistical uncertainties are calculated from the event counts in each
bin assuming a Poisson distribution and the corresponding covariance
matrix is propagated through the unfolding procedure to yield a full set
of statistical uncertainties and correlations for the unfolded cross sections.
For both the 2D and 3D measurements, the statistical uncertainties remain below
2\% in most phase space regions, generally increasing to 2--5\% at outer
rapidities and reaching values of 20--40\% at large
$\mjj$ or $\ptave$.

The impact of systematic effects on the cross section is generally estimated by
varying experimental parameters within a $\pm1$ standard deviation interval around the
nominal value. The relative differences to the nominal result are used to
construct an asymmetric confidence interval for the unfolded cross sections in
each observable bin.

\begin{figure*}[htb]
  \centering
    \includegraphics[width=\cmsFigWidthTwo]{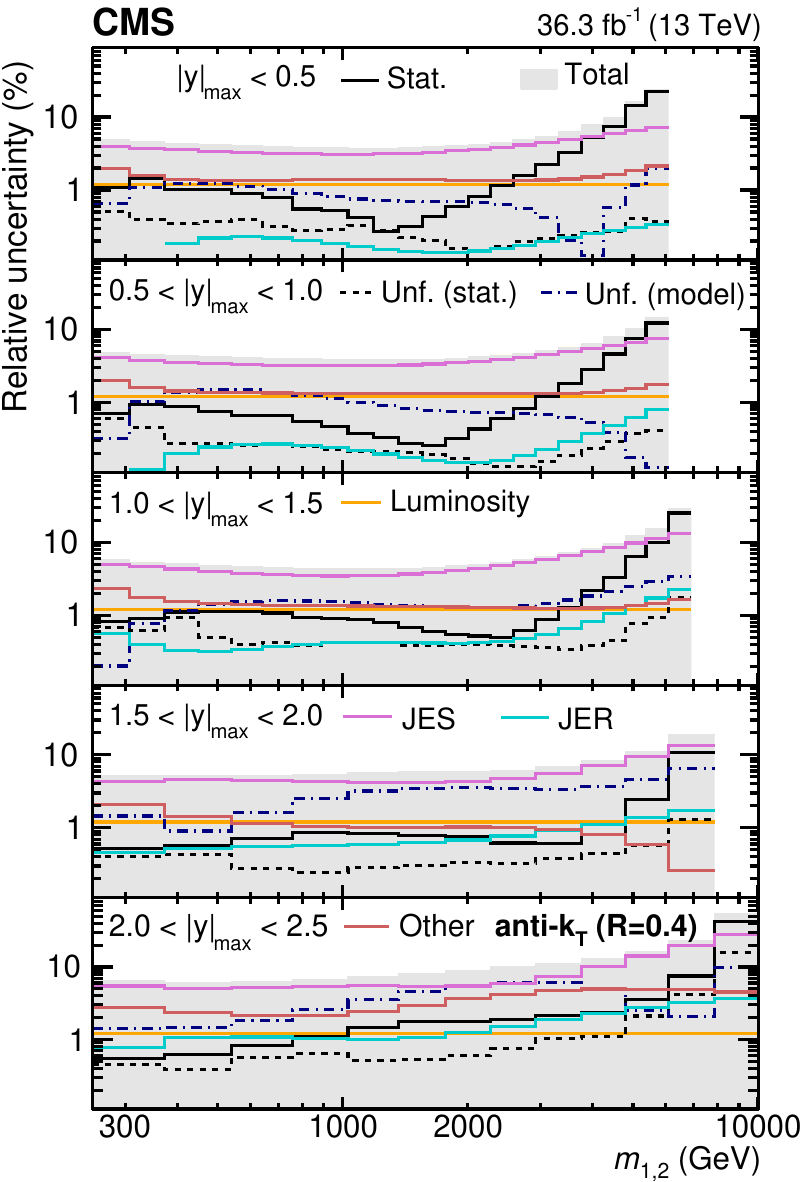}
    \includegraphics[width=\cmsFigWidthTwo]{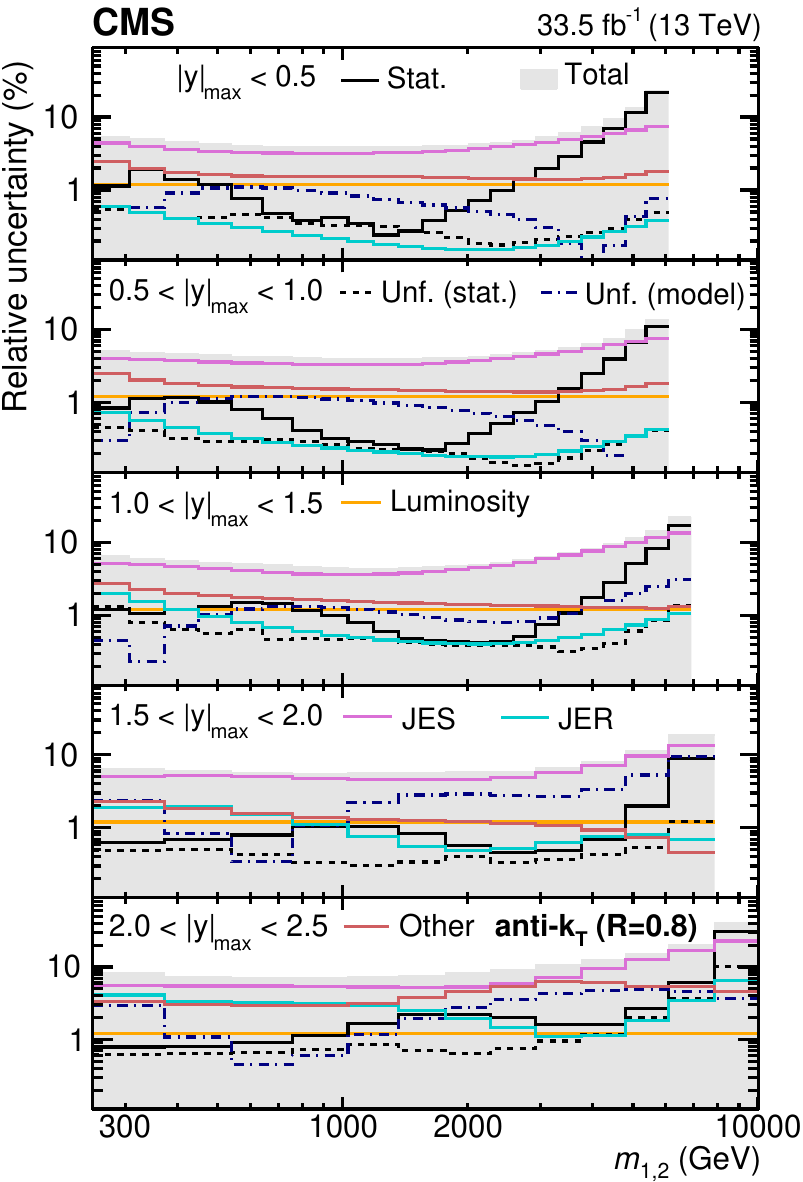}\\
    \caption{Breakdown of the experimental uncertainty  for the 2D measurements as a function of \mjj using jets with $R = 0.4$ (left) and 0.8 (right). The individual components and abbreviations are explained in Section~\ref{sec:uncertainties}. The abbreviation "Unf." refers to the unfolding uncertainties. The shaded area represents the sum in quadrature of all statistical and systematic uncertainty components. \label{fig:2d-uncertainties}}
\end{figure*}

\begin{figure*}[ht]
  \centering
    \includegraphics[width=\cmsFigWidthTwo]{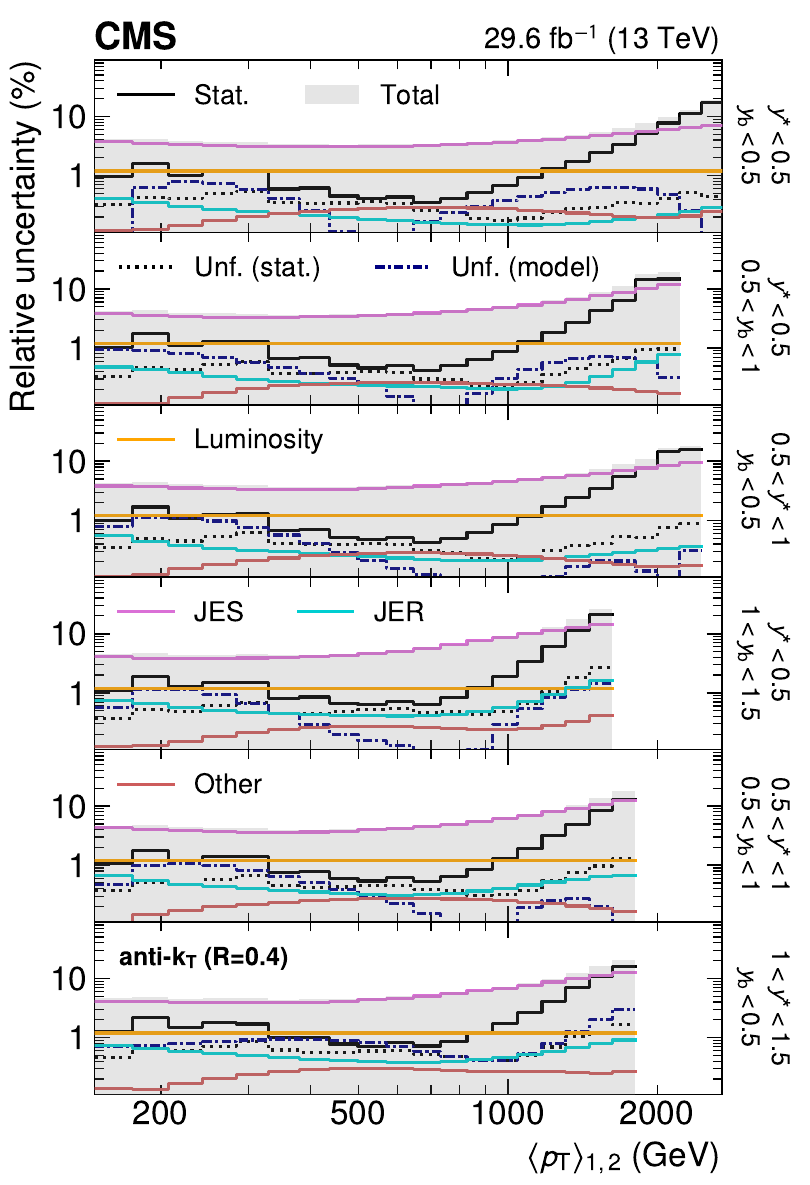}
    \includegraphics[width=\cmsFigWidthTwo]{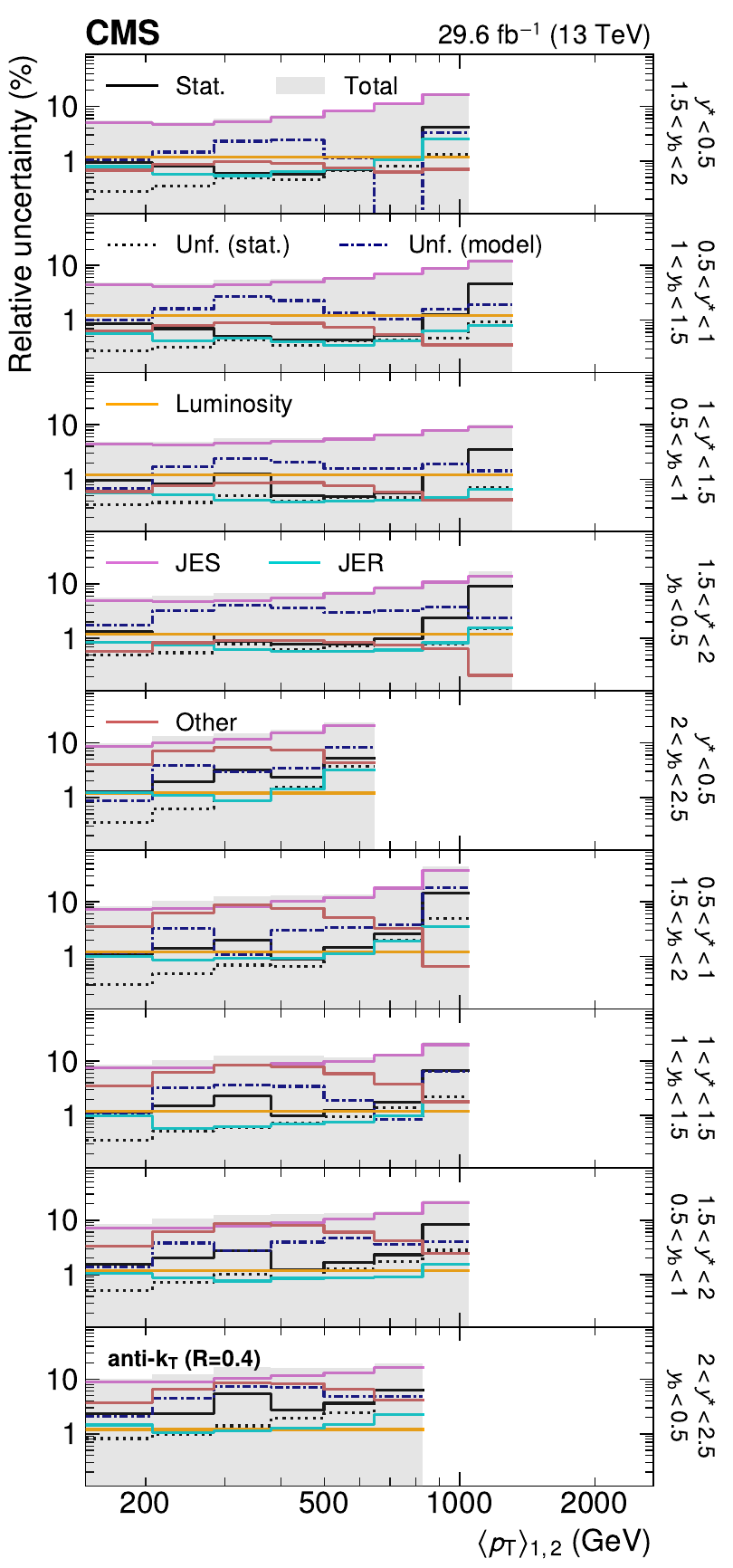}\\
 \caption{Breakdown of the experimental uncertainty for the 3D measurement as a function of \ptave\ using jets with $R = 0.4$. The individual components and abbreviations are explained in Section~\ref{sec:uncertainties}. The shaded area represents the sum in quadrature of all statistical and systematic uncertainty components. Similar plots for all other jet sizes and observables can be found in Appendix~\ref{app:supplementary-material}. \label{fig:3d-uncertainties}}
\end{figure*}

Figures~\ref{fig:2d-uncertainties} and~\ref{fig:3d-uncertainties}
show an overview of the main contributions to the experimental uncertainty
in the dijet cross section for the 2D measurement for both
values of $R$, and the 3D measurement as a function of \ptave
for $R=0.4$, respectively.
The main contributions to the systematic uncertainty are due to the
determination of the JES and JER, and the luminosity.
A further uncertainty results from the correction of the trigger prefiring
inefficiency, and is only significant in the outer rapidity regions
with contributions from jets with $\yabs > 2$.
Other contributions, which have an overall smaller impact on the cross section,
arise as a consequence of experimental methods such as unfolding.
The following sections describe the individual uncertainty contributions in
more detail.

\subsection{Jet energy scale uncertainty\label{sec:uncertainties:jes}}

The dominant contribution to the systematic uncertainty arises from the
determination of the JES\@.
Jets are calibrated in a multi-stage procedure to correct for experimental
effects, such as contributions from pileup collisions or shifts in the jet energy due to
detector or reconstruction effects.
The corrections depend on the $\pt$ and $\eta$ of the jet, and lead
to a total uncertainty in the energy scale of individual jets
of 1--2\% 
in the phase space considered here~\cite{CMS-DP-2020-019}.
Since the dijet spectrum decreases exponentially as a function of \mjj and
\ptave, the resulting uncertainty in the measured
differential dijet cross section is amplified by this exponent.
For the 2D cross sections, the JES uncertainty starts at 2--5\%,
reaching 30\% at higher values of $\mjj$.
For the 3D cross section the total JES uncertainty increases with \mjj (\ptave)
from about 3\% up to values between 8 and 60\% (40\%), depending on the rapidity region.

The total JES uncertainty is composed of 22 individual contributions
describing different systematic effects.
These include, in roughly descending order of their impact on the
cross section:
the change in experimental conditions over time,
the calibration of the relative and absolute JES as a function of $\eta$ and \pt,
the change in response for jets initiated by gluons and different quark flavors,
and pileup collisions.
Each contribution represents a fully correlated uncertainty across all data
points and is considered to be independent of the other contributions.

\subsection{Luminosity uncertainty\label{sec:uncertainties:lumi}}

The uncertainty due to the integrated luminosity measurement is evaluated to be
1.2\%~\cite{CMS-LUM-17-003} in all
phase space regions and is considered to be fully correlated across all bins.

\subsection{Jet energy resolution uncertainty\label{sec:uncertainties:jer}}

The effect of the finite JER on the cross section is modeled using
response matrices obtained from the simulation, where the effective
JER is increased by factors derived from control samples in data
to account for residual differences
between the detector simulation and the actual data-taking conditions. The
correction of the JER is applied as part of the unfolding procedure.
The JER uncertainty in the cross section is estimated by performing the
unfolding with response matrices derived from alternative samples, where the
jet energy was smeared by factors representing a $\pm$1 standard deviation shift in the JER
compared to the nominal value.
The resulting uncertainty values range from below 1\% at central rapidities
to at most 10\% in the outer rapidity regions, and are considered to be
correlated across all data points.

\subsection{Unfolding uncertainties\label{sec:uncertainties:unfolding}}

A further uncertainty arises as a consequence of the limited size of the
simulated samples used for deriving the response matrices as part of the
unfolding procedure.
These are thus subject to an intrinsic statistical uncertainty, which is
propagated analytically to the unfolded cross sections.
In most phase space regions, this uncertainty remains below 0.5\%, reaching
values of 5--10\% only in a small number of bins at the highest \mjj or \ptave.
As an estimate of the model dependence introduced by unfolding, the difference
in the cross sections unfolded with response matrices obtained from \PYTHIAviii
and \MGvATNLO{} is taken as an additional uncertainty, which is considered to
be correlated across all data points.
This uncertainty is typically at the level of 1\%, rising up to
at most 10\% at outer rapidities and high \mjj.

\subsection{Other uncertainties\label{sec:uncertainties:others}}

The correction applied to compensate for the trigger inefficiency due to
prefiring gives rise to an additional correlated uncertainty in the cross
section.
In general, this uncertainty is at the level of 1\% or below,
except in the outermost \ymax\ region and the five outermost \ybys\ regions,
where it rises to about 10\% (20\%) at the upper end of the \mjj (\ptave)
spectrum.

The uncertainty contribution from pileup interactions is determined by
varying the total inelastic $\Pp\Pp$ cross section used for reweighting the simulated
samples within its associated uncertainty of 4.6\%, as obtained using the pileup counting
method described in Ref.~\cite{CMS:2018mlc}.
The unfolding is then performed with the resulting response matrices, taking
the differences between the variations and the nominal value as
a fully correlated uncertainty in the unfolded cross section, which is
below 1\% in all phase space regions.

The normalization uncertainty in the background contribution from spurious
jet reconstructions or event migrations at the phase space boundaries
is estimated to be 5\% and propagated through the unfolding procedure.
A further contribution to the uncertainty in the unfolded cross section
is due to the correction of reconstruction inefficiencies and migrations
outside the measurement phase space and is estimated to be 5\% of the
corresponding correction factors.
Each of the above contributions is considered to be fully correlated across
all data points.

\preprintClearPage
\section{Theoretical predictions\label{sec:theory}}

{\tolerance=800
Fixed-order theoretical predictions for the 2D and 3D dijet cross sections are
obtained up to NNLO accuracy in pQCD with the \NNLOJET program
(revision~5918)~\cite{Gehrmann:2018szu}.
The \NNLOJET program is interfaced to \FASTNLO
(version~2.3)~\cite{Kluge:2006xs,Britzger:2012bs} via the \textsc{APPLfast}
interface (version~0.0.46)~\cite{Britzger:2019kkb,Britzger:2022lbf}
to provide interpolation grids that allow theoretical predictions to be
obtained for arbitrary PDFs and for different values of the renormalization
scale $\mur$, the factorization scale $\muf$, and the strong coupling constant
\asmz, without the need to repeat the full calculation.
\par}

\begin{figure*}[tb]
  \centering
  \includegraphics[width=\cmsFigWidthTwo]{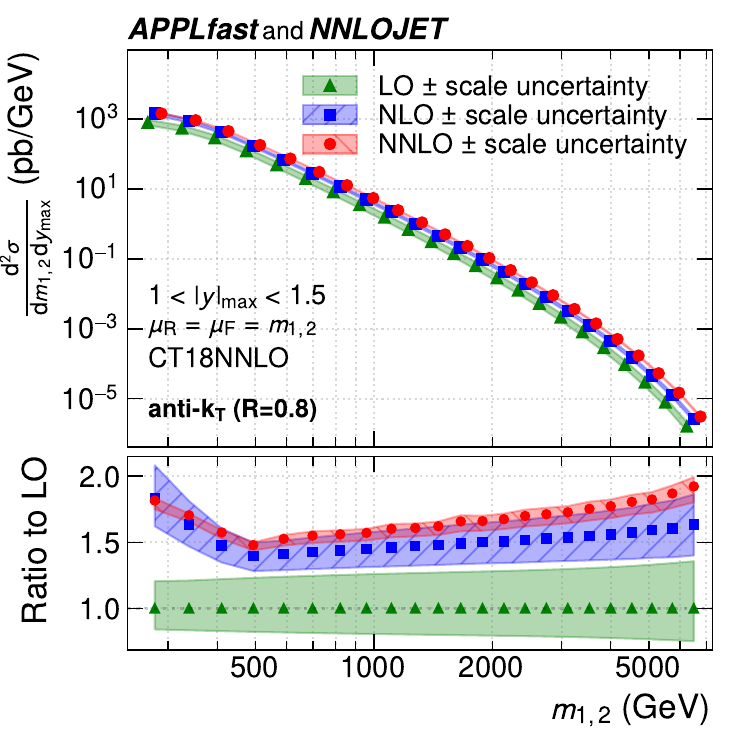}
  \includegraphics[width=\cmsFigWidthTwo]{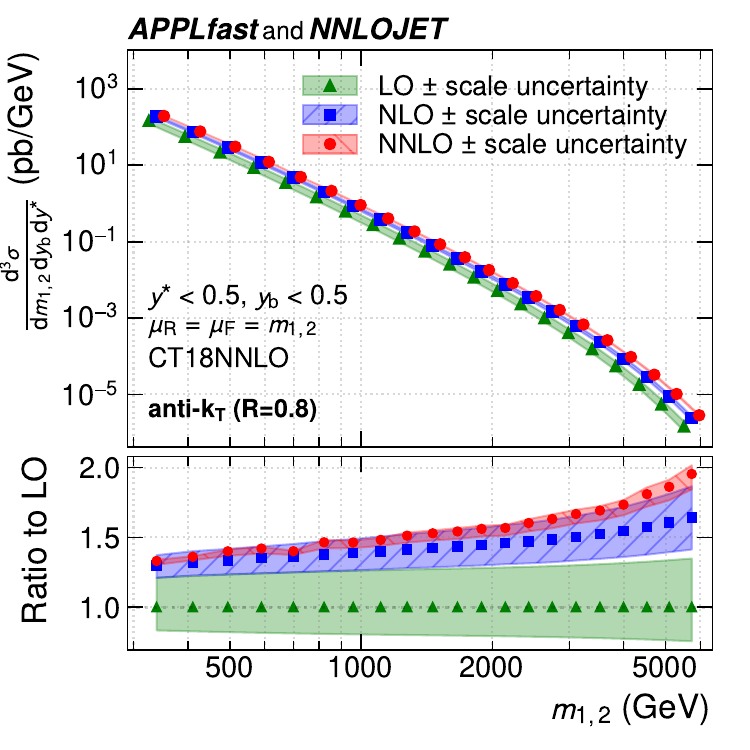}
  \caption{
  Theoretical predictions for the 2D (left) and 3D (right) cross sections,
  as a function of \mjj, 
  illustrated here in the rapidity regions $1.0<\ymax<1.5$ and
  $\yboost < 0.5$, $\ystar < 0.5$,
  together with the corresponding six-point scale uncertainty 
  for $\mur=\muf=\mjj$ using the CT18 NNLO PDF set. 
  In the upper panels, the curves and symbols are slightly shifted for better
  visibility.
  The lower panels show the ratio to the respective prediction at LO\@.
  The fluctuations in the NNLO predictions are due to the limited statistical precision
  of the calculation.
  \label{fig:theo:scaleunc}}
\end{figure*}

Following recommendations outlined in Ref.~\cite{Currie:2017eqf}, $\mjj$ is
chosen as the central reference value for both $\mur$ and $\muf$.
To estimate the theoretical uncertainty due to missing higher-order terms in
perturbation theory, the conventional recipe~\cite{Cacciari:2003fi,Catani:2003zt,Banfi:2010xy}
of varying the \mur and \muf scales is applied. More precisely, the so-called
scale uncertainty is derived from the envelope of the theoretical predictions
obtained for the six scale variations corresponding to
($\mur/\mjj$, $\muf/\mjj$) = (1/2,~1/2), (1/2,~1), (1,~1/2), (2,~1), (1,~2),
and (2,~2). As an example, Fig.~\ref{fig:theo:scaleunc} shows the
resulting uncertainty for theoretical predictions of the 2D and 3D dijet cross sections
obtained at LO, next-to-leading order (NLO), and NNLO\@.
In most phase space regions, the NLO and NNLO scale uncertainty bands
overlap, indicating good perturbative convergence.
However, towards small values of $\mjj$ and large rapidity separations \ystar,
a steep rise in the ratio of the higher-order predictions with respect to LO,
referred to here as the $K$ factors, is observed,
leading to a reduced overlap and an increased scale uncertainty.
For an ideal dijet event with two jets of equal \pt produced in a
back-to-back configuration, the dijet invariant mass is given by
$\mjj = 2\pt\cosh(\ystar)$.
The rise in the $K$ factors then is understood to be caused by the minimum
\pt requirements imposed on the two leading jets, which at small dijet
mass restrict the phase space accessible to LO processes in favor of
higher-order contributions.

The NNLO contribution is based on the leading-color and
leading-flavor-number approximation~\cite{Currie:2017eqf,Gehrmann-DeRidder:2019ibf}.
Subleading-color contributions have been shown to be at the percent level
at NLO and are expected to be even smaller in comparison to
the leading-color result at NNLO~\cite{Currie:2016bfm}.
It is worth noting, however, that in a recent investigation~\cite{Chen:2022tpk}
a significant impact of subleading-color contributions was found for the NNLO prediction
of the \ptave-dependent 3D CMS dijet measurement at $\sqrt{s} = 8\TeV$
with a jet distance parameter $R = 0.7$~\cite{Sirunyan:2017skj}.
The reported effect can lead to a decrease in the cross section of up to 5\%
for small \ptave and an increase of up to 3\% for large \ptave, which is
beyond the size of the scale uncertainty at NNLO\@.
For the CMS inclusive jet measurement at $\sqrt{s} = 13\TeV$~\cite{CMS:2016jip}
with jet distance parameters of $R = 0.4$ and~0.7, the effect was determined to be
much smaller and to be covered by scale uncertainty estimates, except towards
small jet \pt for the smaller jet size.
Predictions for the 2D dijet measurement performed as a function of \mjj and \ystar
at $\sqrt{s} = 7\TeV$ by the ATLAS Collaboration~\cite{ATLAS:2013jmu}
were also less affected even with $R = 0.4$.
The effect on the dijet observables under examination here is not yet known.
The 2D and 3D measurements presented here for two jet distance parameters, $R=0.4$ and 0.8,
and for the two dijet observables \mjj and \ptave,
provide an ideal set of measurements to further study the impact of subleading-color corrections
in comparison to data.

To compare with CMS data unfolded to the particle level, the fixed-order
predictions are complemented by nonperturbative (NP) correction factors
\cNP, which are defined as the ratio between the nominal cross sections
with and without multiple parton interactions (MPI) and hadronization (HAD)
effects, as given by a chosen MC event generator,
\begin{linenomath*}
\begin {equation}
  \cNP = \frac{\sigma^\text{PS+MPI+HAD}}{\sigma^\text{PS}},
  \label{eq:def-np-corr}
\end{equation}
\end{linenomath*}
where the parton shower (PS) is considered to be a perturbative component.

The model dependence of the NP corrections is evaluated by comparing
results from several MC event generators.
Leading-order \parlevel predictions are obtained from \PYTHIA (version 8.240),
using the tunes CUETP8M1~\cite{Khachatryan:2015pea} and CUETP8M2T4~\cite{CMS-PAS-TOP-16-021},
and \HERWIGpp{}~\cite{Bahr:2008pv} (version 2.7.1) using the EE5C tune~\cite{Seymour:2013qka}.
These generators are interfaced to
\POWHEG{}~\cite{Nason:2004rx,Frixione:2007vw,Alioli:2010xd,Alioli:2010xa} (version 2J V2\_Mar2016)
to provide NLO predictions.
An additional set of predictions is obtained from \HERWIG{}~7~\cite{Bellm:2015jjp}
(version 7.2.2) with the CH3 tune~\cite{CMS:2020dqt} at both LO and NLO.

To mitigate statistical fluctuations, the corrections
are parametrized by a
smooth function\linebreak$f(x) = a / x^b + c$, where $x$ is either $\mjj$ or $\ptave$.
The parameters $a$, $b$, and $c$ are obtained in a least-squares fit to the
binwise correction factors
$\cNP$ obtained from Eq.~\eqref{eq:def-np-corr}
in each rapidity region.
The fits provide a good description of the correction factors in most phase space regions.
For a number of low-\mjj bins, where the phase space is constrained by
the minimum \pt\ requirements on the two leading jets, the value
of $\cNP$ is taken directly as the correction factor.
The final correction factor in each bin is obtained as the midpoint between the largest
and smallest value of $\cNP$ obtained across all MC configurations, and half
the difference between the largest and smallest value is assigned as an uncertainty.

\begin{figure*}[p]
  \centering
  \includegraphics[width=\cmsFigWidthTwo]{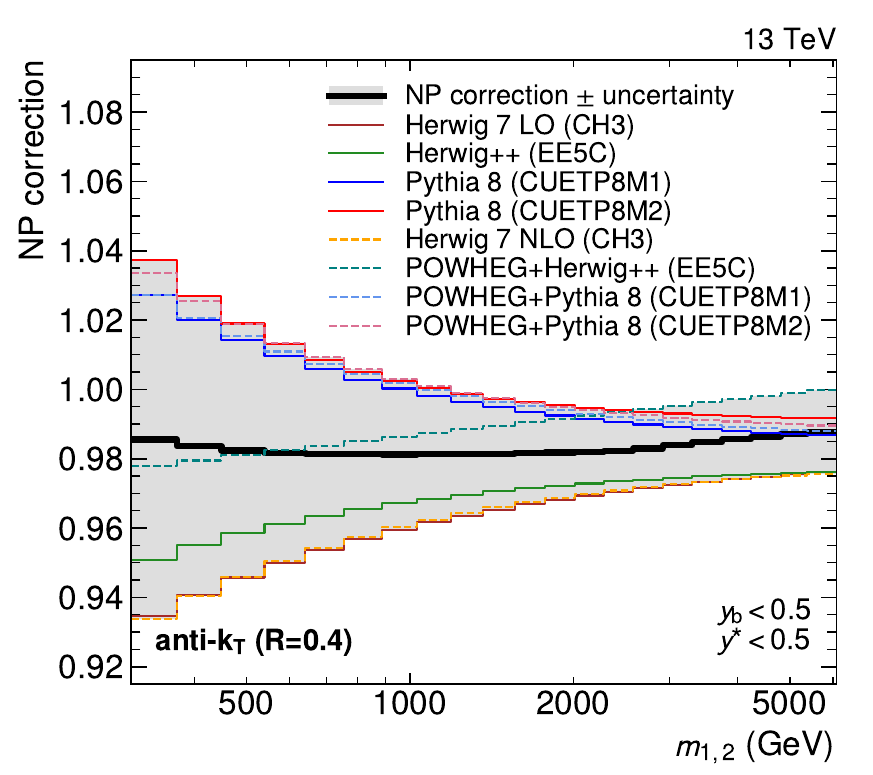}
  \includegraphics[width=\cmsFigWidthTwo]{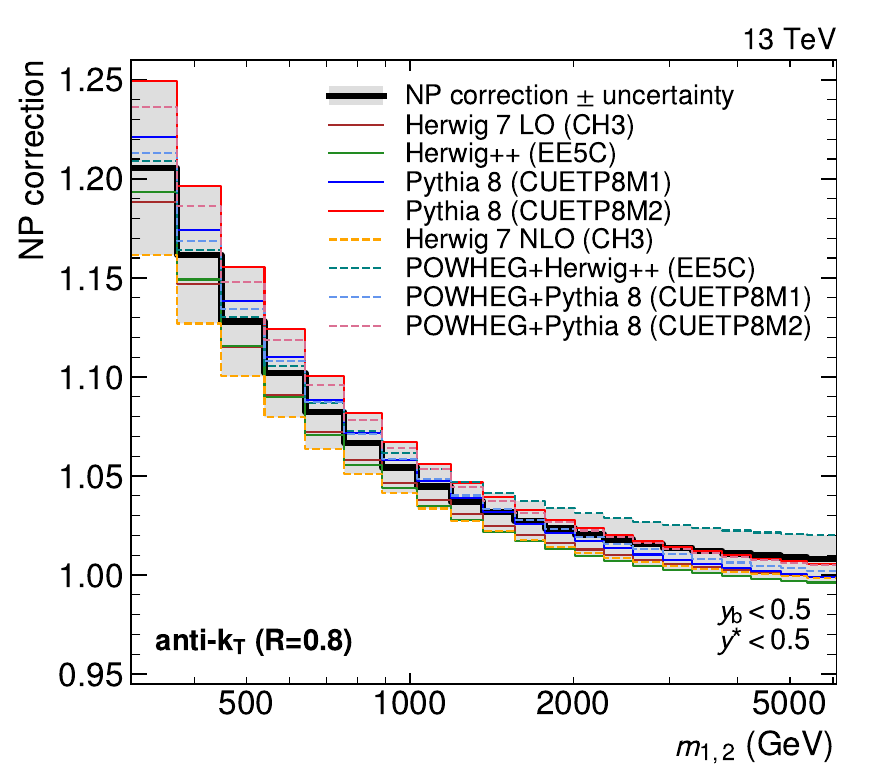}
  \caption{
  Nonperturbative correction factors obtained for jets with $R = 0.4$ (left) and 0.8 (right)
  as a function of \mjj, 
  illustrated here in the rapidity region ($\yboost < 0.5$, $\ystar < 0.5$). 
  Individual correction factors are first derived from simulation using eight different MC configurations. 
  The largest and smallest value obtained in each observable bin is then used to define the final correction 
  factor and its associated uncertainty. 
  The correction values are larger for jets with $R = 0.8$, increasing to over 20\% in the lowest $\mjj$ bin.
  \label{fig:np-corrections}}
\end{figure*}

The resulting NP corrections are illustrated in Fig.~\ref{fig:np-corrections}.
For jets with $R = 0.4$, the contributions from hadronization and MPI largely
cancel, leading to NP corrections compatible with unity within their uncertainty.
In contrast, the MPI contribution dominates for jets with $R = 0.8$, resulting in
significantly larger NP corrections of $\approx$20\% at low values of \mjj.
The size of the uncertainty is similar for both jet sizes.

It is also observed that the NP corrections obtained with \PYTHIAviii are in general larger
than those from \HERWIGpp or \HERWIG{}~7, such that this difference is the dominant
contribution to the NP correction uncertainty.
While some dependence on the tune is observed when comparing the predictions from
CUETP8M1 and CUETP8M2T4, the impact is typically small.
In most cases, the values obtained at NLO are seen to be comparable to those obtained
at LO from the same generator, with the notable exception of \HERWIGpp{}, where the
NLO result obtained using \POWHEG{} is consistently higher than the LO result.

\begin{figure*}[p]
  \centering
  \includegraphics[width=\cmsFigWidthTwo]{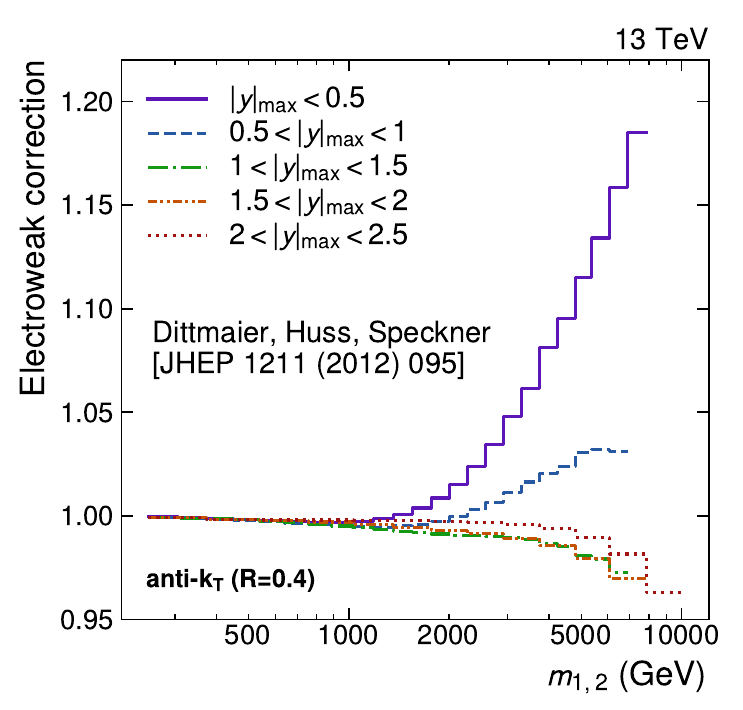}
  \includegraphics[width=\cmsFigWidthTwo]{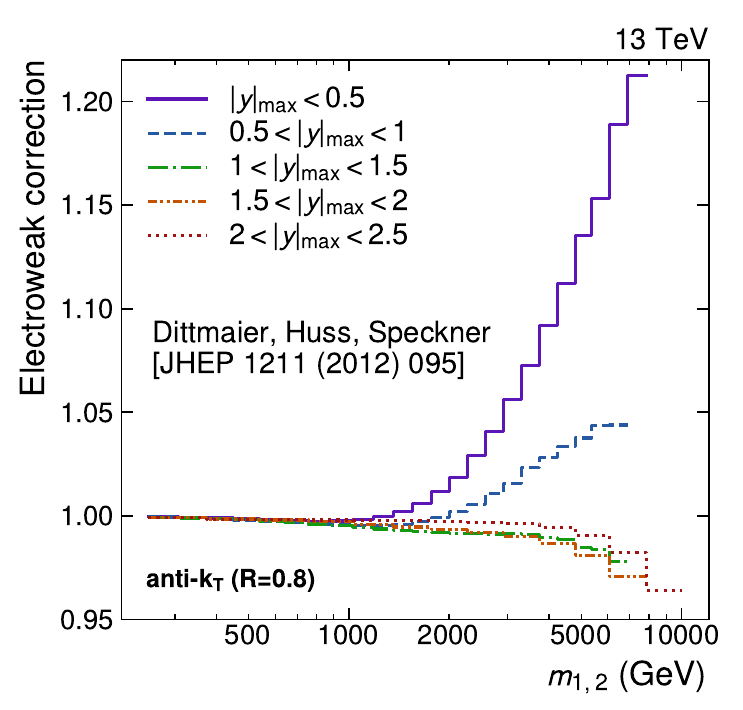}
  \caption{
  Electroweak correction factors obtained for jets with 
  $R = 0.4$ (left) and 0.8 (right)
  as a function of \mjj 
  in the five different $\ymax$ regions. 
  The corrections depend strongly on the kinematic properties of the jets and are 
  observed to be largest at central rapidities for $\mjj > 1\TeV$. 
  \label{fig:ew-corrections}}
\end{figure*}

For jet transverse momenta in the \TeVns{} range, electroweak contributions to the
differential dijet cross section become important and must be considered in
addition to the NNLO pQCD calculation~\cite{Dittmaier:2012kx}.
These effects, which arise from the virtual exchange of soft or collinear
\PW or \PZ bosons, are accounted for by applying a multiplicative correction
factor to the pQCD prediction.
As shown in Fig.~\ref{fig:ew-corrections}, these factors exhibit a strong
dependence on the rapidity and \pt of the jets.
Particularly at small \ymax or \ystar, the electroweak correction reaches 10--20\%
for dijet masses beyond 5\TeV, where experimental uncertainties become large as well.
The uncertainty on this correction is therefore considered to be negligible with
respect to other large uncertainties.

\preprintClearPage
\section{Comparison to theory\label{sec:results}}

\begin{figure*}[htb]
  \centering
    \includegraphics[width=0.47\textwidth]{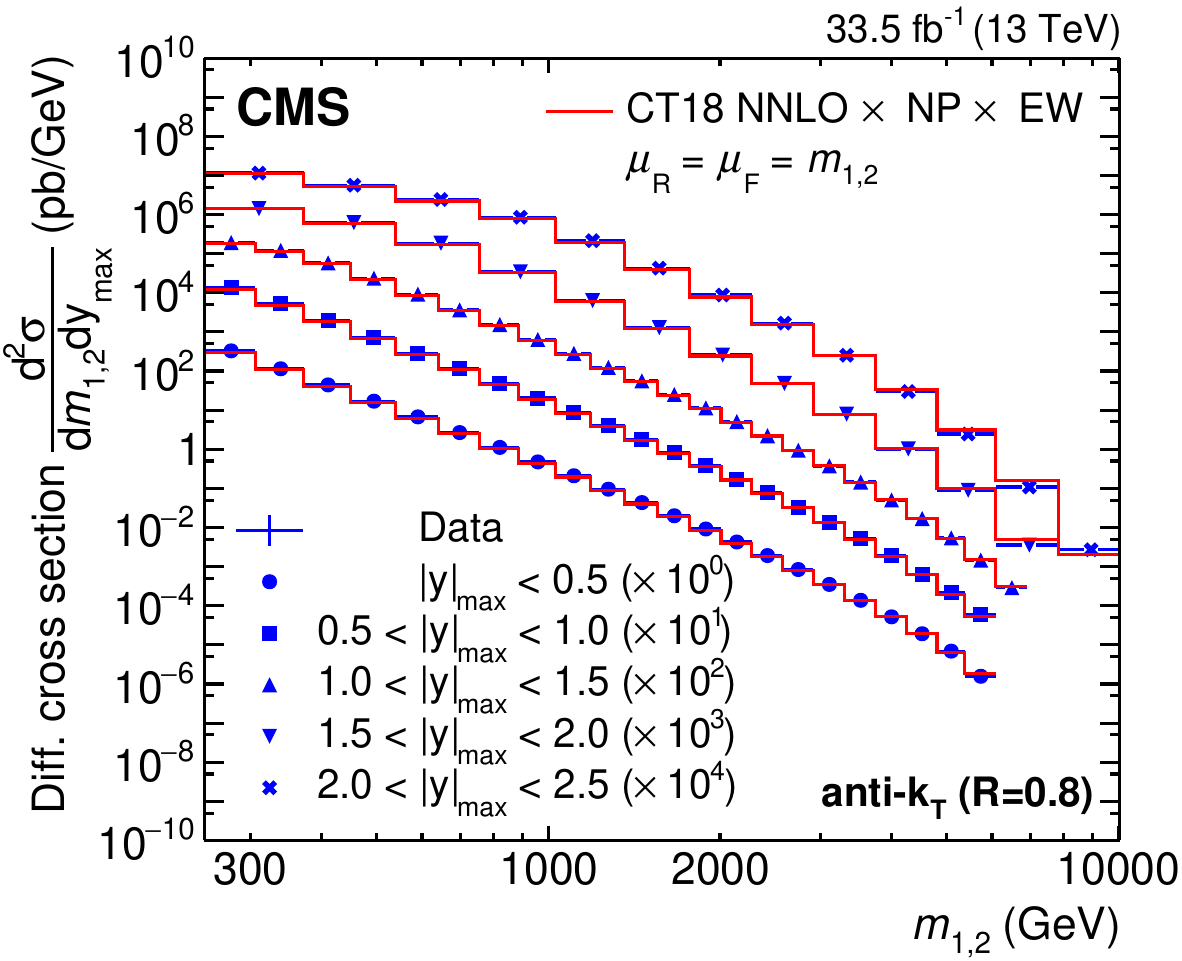}
    \includegraphics[width=0.51\textwidth]{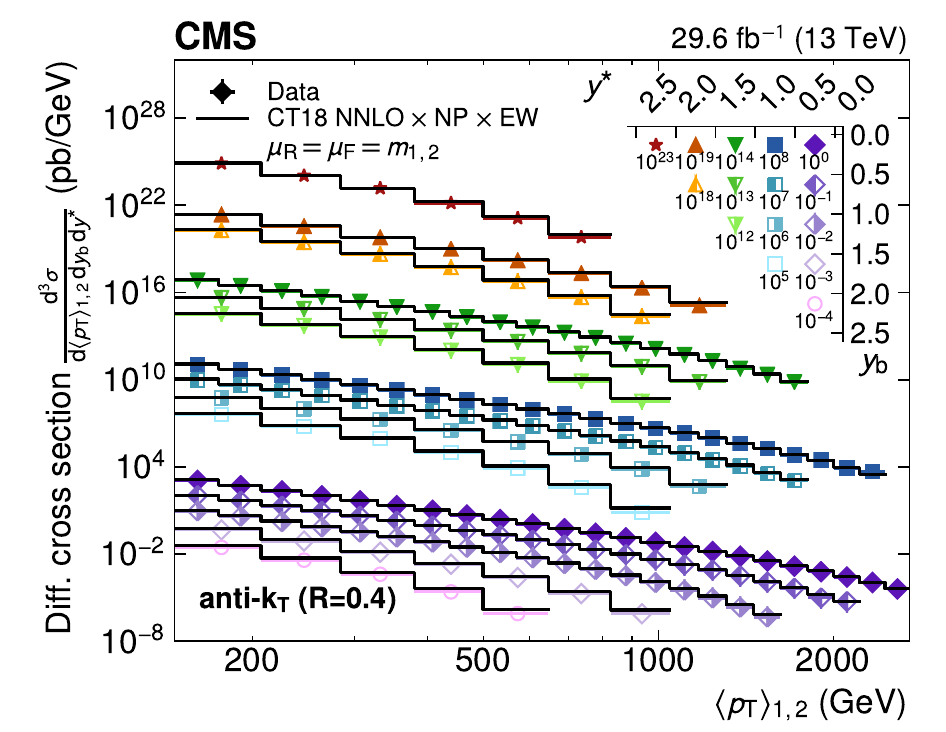}\\
    \caption{
    Differential dijet cross sections, illustrated here for
    the 2D measurement as a function of $\mjj$
    using jets with $R = 0.8$ (left), 
    and the 3D measurement as a function of $\ptave$ 
    using jets with $R = 0.4$ (right). 
    The markers and lines indicate the measured unfolded cross sections and
    the corresponding NNLO predictions, respectively.
    For better visibility, the values are scaled by a factor depending 
    on the rapidity region, as indicated in the legend. 
    Analogous plots for all other jet sizes and observables can be found in 
    Appendix~\ref{app:supplementary-material}.
    \label{fig:cross-sections}}
\end{figure*}

\begin{figure*}[htb]
  \centering
    \includegraphics[width=0.49\textwidth]{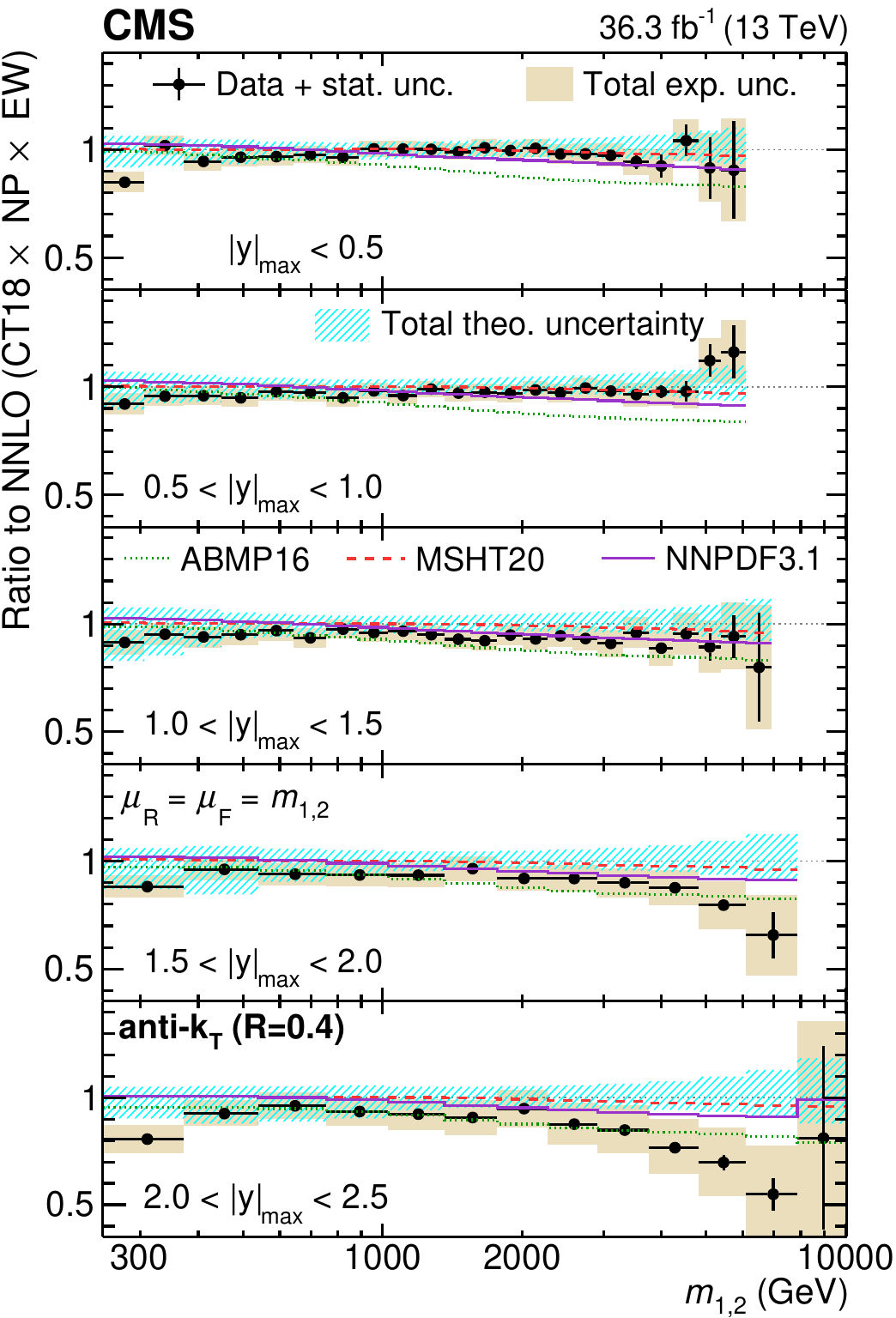}
    \includegraphics[width=0.49\textwidth]{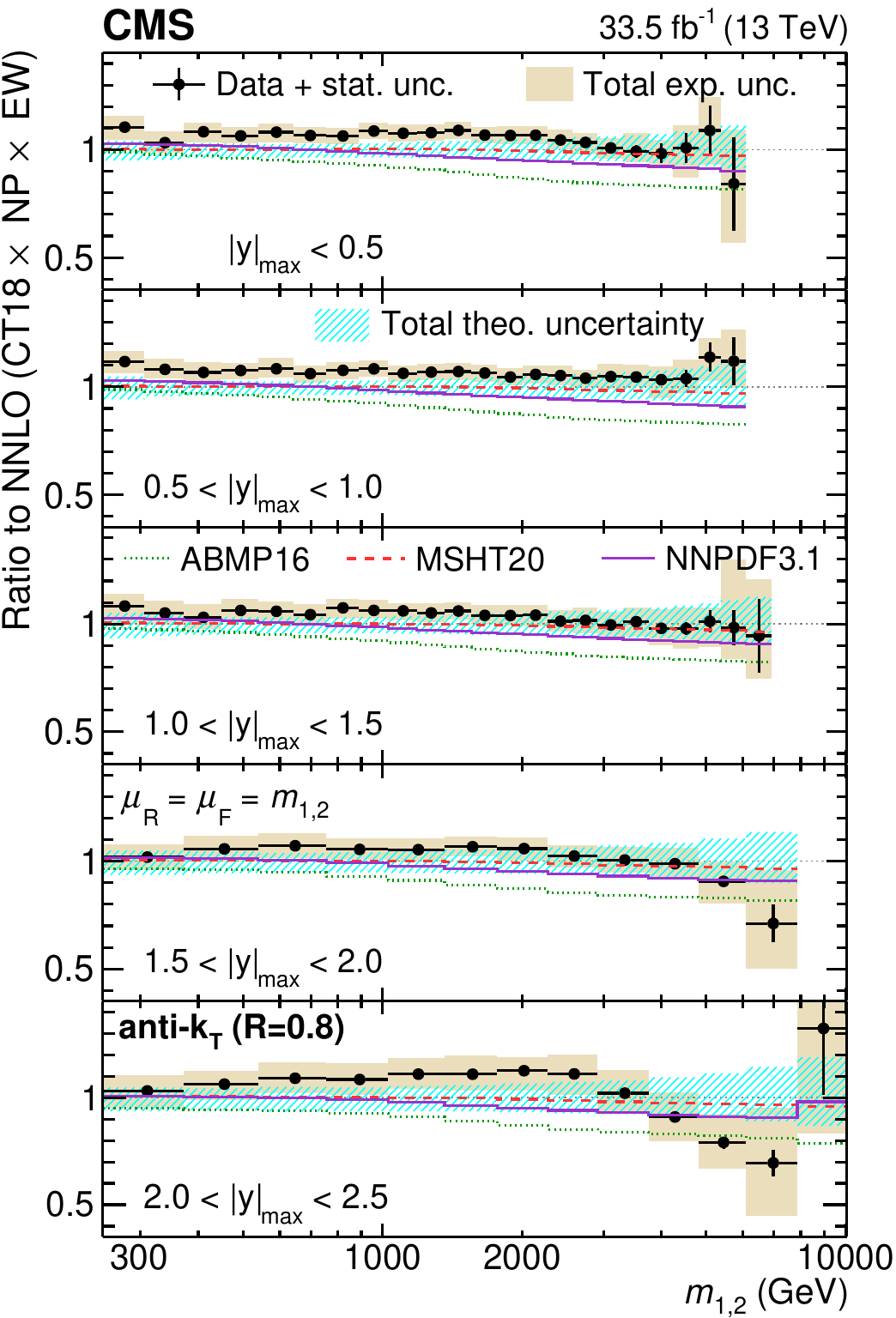}\\
    \caption{
    Comparison of the 2D dijet cross section 
    as a function of $\mjj$ 
    to fixed-order theoretical calculations at NNLO, 
    using jets with $R = 0.4$ (left) and 0.8 (right). 
    Shown are the ratios of the measured cross sections (markers)
    to the predictions obtained using the CT18 NNLO PDF set.
    The error bars and shaded yellow regions indicate the statistical and the
    total experimental uncertainties of the data, respectively,
    and the hatched teal band indicates the
    sum in quadrature of the PDF, NP, and scale uncertainties.
    Alternative theoretical predictions obtained using other global PDF sets are shown 
    as colored lines.
    Similar plots for the individual rapidity regions can be found in 
    Appendix~\ref{app:supplementary-material}.
    \label{fig:2d-ratios}}
\end{figure*}

\begin{figure*}[!htb]
  \centering
    \includegraphics[width=\cmsFigWidthTwo]{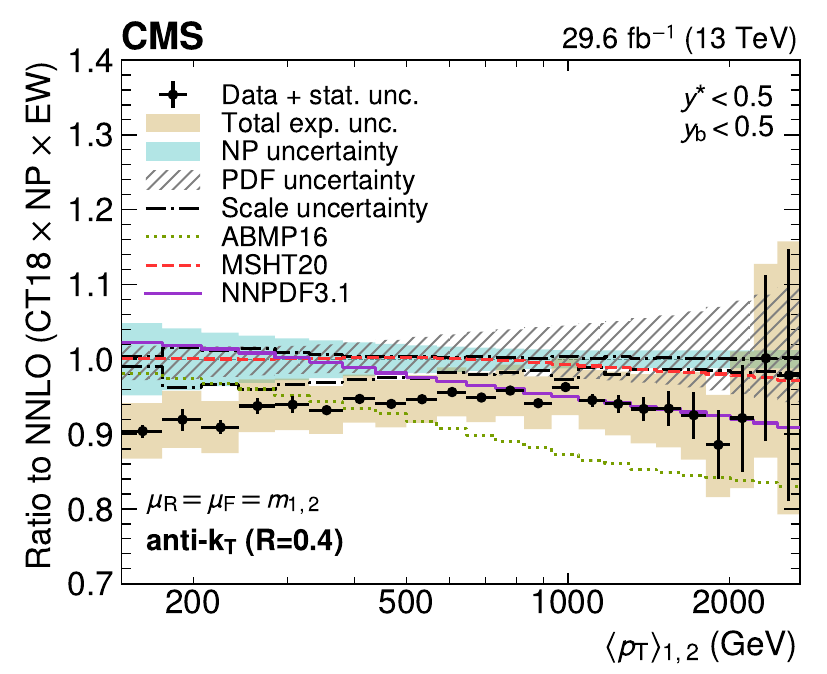}
    \includegraphics[width=\cmsFigWidthTwo]{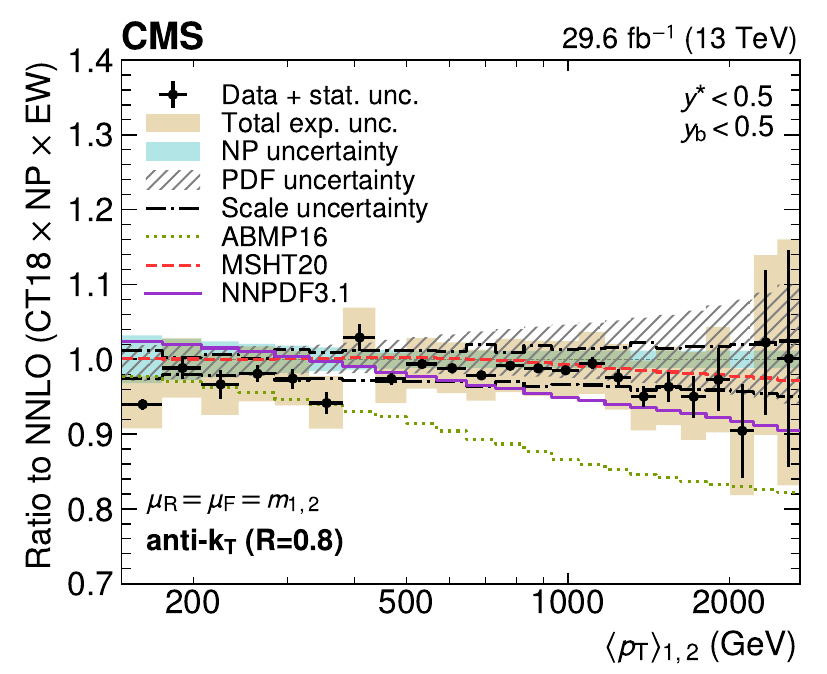}\\
    \includegraphics[width=\cmsFigWidthTwo]{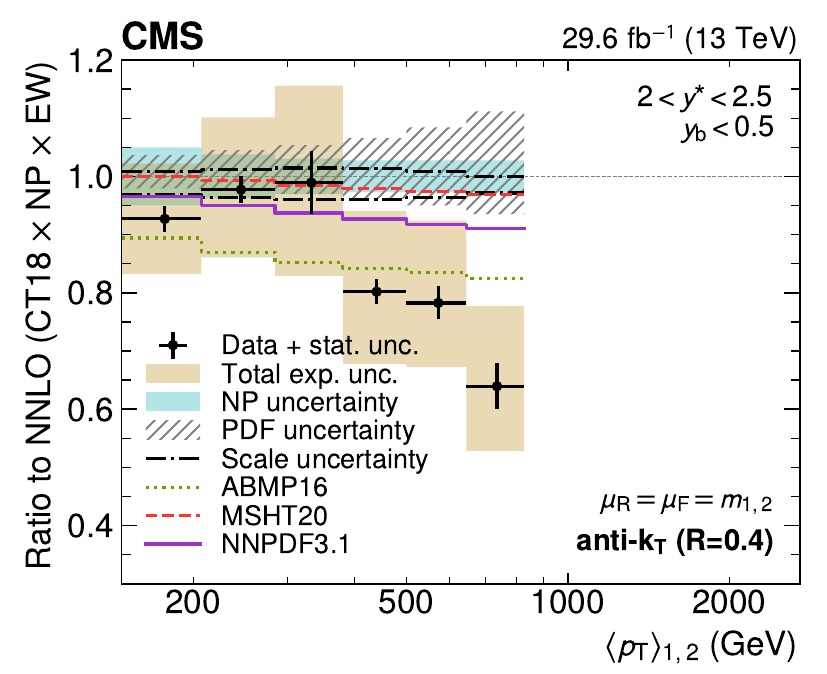}
    \includegraphics[width=\cmsFigWidthTwo]{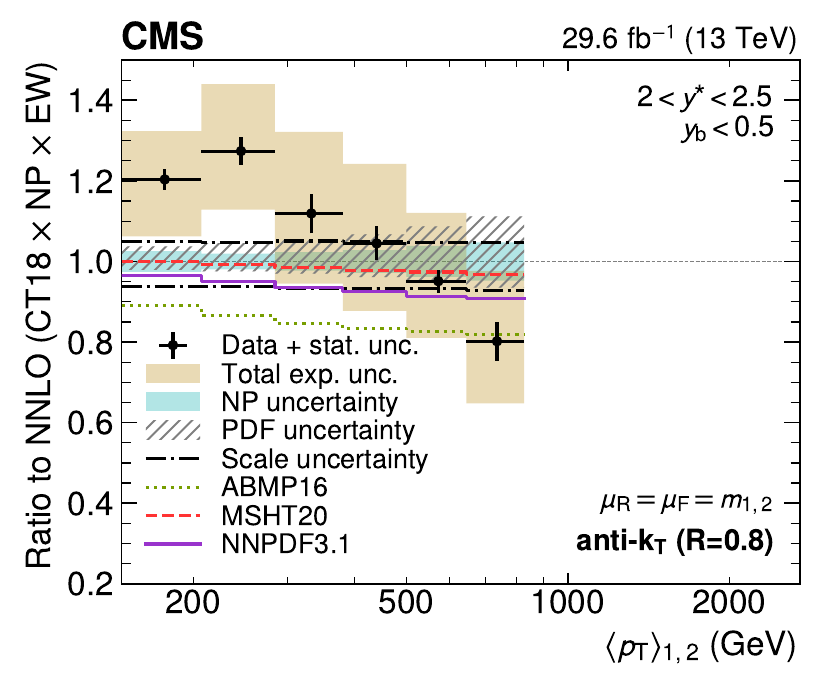}\\
    \includegraphics[width=\cmsFigWidthTwo]{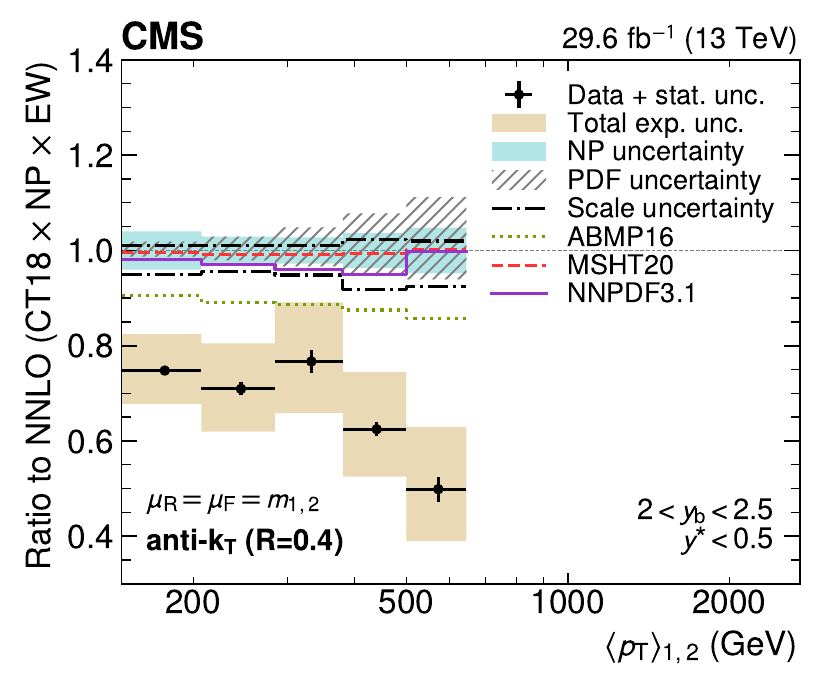}
    \includegraphics[width=\cmsFigWidthTwo]{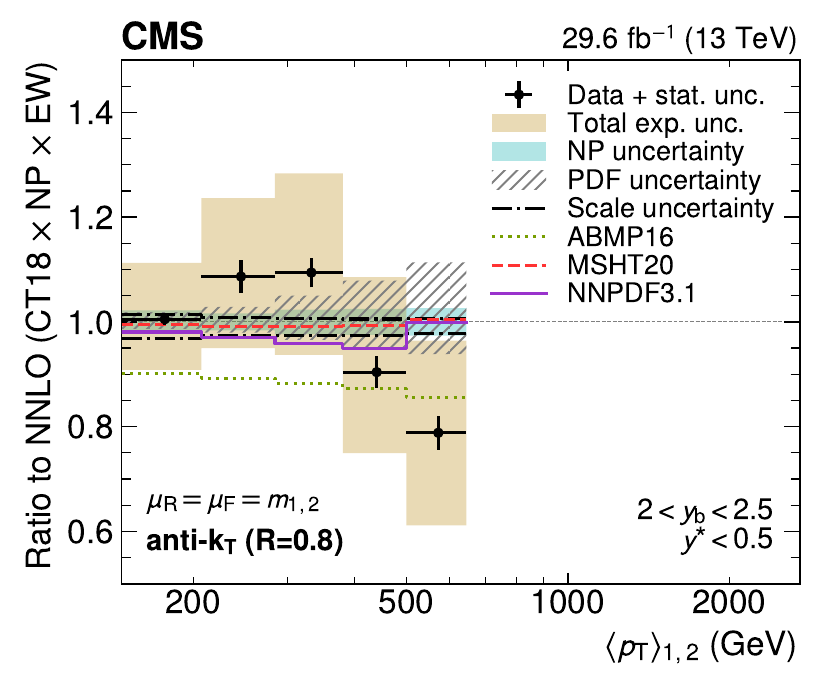}\\
    \caption{
    Comparison of the 3D dijet cross section 
    for jets with $R = 0.4$ (left) and 0.8 (right)
    as a function of $\ptave$ 
    to fixed-order theoretical calculations at NNLO, 
    shown here for three out of the total of 15 rapidity regions.
    The data points and predictions for alternative PDFs are 
    analogous to those in 
    Fig.~\ref{fig:2d-ratios}. 
    In addition, the separate contributions to the theory uncertainty due to the 
    CT18 PDFs, NP corrections, and six-point scale variations 
    are shown explicitly.
    Similar plots for all rapidity regions and observables can be found in 
    Appendix~\ref{app:supplementary-material}.
    \label{fig:3d-ratios}}
\end{figure*}

An overview of the unfolded cross sections obtained for the 2D and 3D
measurements and the corresponding fixed-order theoretical predictions at NNLO,
complemented by NP and electroweak corrections,
is presented in Fig.~\ref{fig:cross-sections}.
For a more detailed comparison, ratios of the measured cross sections
to the theoretical predictions are shown in
Figs.~\ref{fig:2d-ratios} and~\ref{fig:3d-ratios}.

The theoretical predictions are obtained using recent NNLO PDF sets available via
the \LHAPDF~\cite{Buckley:2014ana}
library (version 6.3.0), namely
ABMP16~\cite{Alekhin:2017kpj},
CT18~\cite{Hou:2019efy},
MSHT20~\cite{Bailey:2020ooq}, and
{NN\-PDF}3.1~\cite{Ball:2017nwa}.
All PDF sets are derived in global fits to data from multiple
experiments while fixing the value of the strong coupling constant \asmz to
0.118, except for
ABMP16, where $\asmz = 0.1147$ is determined in the fit together with all other
parameters.
The uncertainties in the cross section predictions due to the PDFs are
calculated as 68\% confidence intervals following the prescriptions given in
the respective references.
The PDF uncertainty bands shown in Fig.~\ref{fig:3d-ratios} are obtained using
the CT18 PDF set and do not account for the finite precision of \asmz.

The predictions for different PDFs are generally in agreement with each other
within the PDF uncertainties, except for the AMBP16 PDF, for which the predicted
cross sections are generally smaller than
those for other PDFs.
At large $\mjj$ or $\ptave$, the predictions obtained for the different
PDF sets show a diverging trend, while still remaining compatible
within the PDF uncertainties.

The level of agreement between the theoretical predictions and the data is observed
to be good in most phase space regions,
with some deviations at the lower ends of the spectra and in the outer
rapidity regions.
In general, the theoretical predictions for $R = 0.8$ are observed to provide a
better description of the data than for $R = 0.4$, which is consistent with
past observations~\cite{CMS:2014nvq,CMS:2016jip,ATLAS:2017kux,CMS:2021yzl,CMS:2021yzl_addendum}.

\journalClearPage

\section{The QCD analysis\label{sec:fits}}

To evaluate the impact of the present measurements on determinations of the proton PDFs and the
strong coupling constant, a QCD analysis is performed following the approach taken by
earlier HERAPDF analyses~\cite{H1:2009pze,H1:2015ubc,H1:2021xxi}.
The data used in the QCD analysis comprise DIS measurements~\cite{H1:2009pze,H1:2015ubc}
obtained in $\Pe^\pm\Pp$ collisions at the HERA collider experiments H1 and ZEUS
as a function of the momentum transfer $Q^2$, supplemented by the present measurements of the dijet
cross section.

The HERA measurements correspond to charged-current (CC) DIS data collected in $\Pem\Pp$ and
$\Pep\Pp$ collisions at a proton beam energy of $E_{\Pp} = 920\GeV$, and neutral-current (NC)
DIS data collected in $\Pep\Pp$ collisions at proton beam energies of $E_{\Pp} = 460$, 575,
820, and 920\GeV.
Only data points with $Q^2$ values above a threshold $Q^2_\text{min} = 10\GeV^2$ are included,
to ensure a good description of the measurements by the theoretical predictions.

It is well known that fixed-order predictions work most reliably for inclusive
observables, \ie, where the phase space for QCD radiation is not restricted.
Such restrictions---as for example the choice of a small distance parameter
$R$ for jet clustering---introduce disparities in the cancellation of
singularities between real emissions and virtual corrections,
leading to large logarithmic terms that have a negative impact on
the perturbative convergence and would require resummation~\cite{Bellm:2019yyh}.
Moreover, as discussed in Section~\ref{sec:theory},
there are indications that subleading-color corrections might have
a smaller impact for larger values of $R$, and for \mjj as compared to \ptave.
Thus, to profit most from the available predictions at NNLO, only the
dijet cross sections as a function of \mjj for the larger value of $R = 0.8$
are used in the QCD analysis.

Theoretical predictions for the dijet cross sections are obtained from \NNLOJET
and \FASTNLO
as interpolation grids at NNLO accuracy,
taking into account the full dependence of the NNLO cross sections
on the PDFs, the strong coupling constant, $\mur$, and $\muf$.
Following Ref.~\cite{Currie:2017eqf}, the dijet invariant mass $\mjj$ is chosen
as the central value for both $\mur$ and $\muf$.
The cross sections are corrected additionally for NP
and electroweak effects as described in Section~\ref{sec:theory}.

Simultaneous determinations of PDFs and the strong coupling constant at the scale of the
\PZ~boson mass, \asmz, are performed with the \textsc{xFitter} program
(version 2.0.1)~\cite{Alekhin:2014irh,xFitterDevelopersTeam:2017xal}, using the HERA data
together with either the 2D or 3D dijet cross sections as inputs.
Access to the theoretical predictions for the dijet cross sections is provided by \FASTNLO.
The evolution of PDFs following the DGLAP equations~\cite{Dokshitzer:1977sg,Gribov:1972ri,Altarelli:1977zs}
is performed using the \QCDNUM package (version 17-01/15)~\cite{Botje:2010ay}.
Contributions from heavy quarks are treated in the Thorne--Roberts optimal variable flavor number scheme
(RTOPT)~\cite{Thorne:1997ga,Thorne:2006qt,Thorne:2012az}, with the masses of the charm and bottom quarks
set to $m_\PQc = 1.43\GeV$ and $m_\PQb = 4.5\GeV$, respectively.

In the HERAPDF approach, the proton structure is expressed in terms of the gluon distribution
$\text{g}(x)$, the up and down valence quark distributions $\text{u}_\text{v}(x)$ and
$\text{d}_\text{v}(x)$, and the up- and down-type sea antiquark distributions $\UBar(x)$ and
$\DBar(x)$.
For each of these distributions $f\!$, the dependence on the proton momentum fraction $x$ carried
by a parton is parametrized at a
starting scale $\mufzero^2 = 1.9\GeV^2$ as
\begin{linenomath*}
\begin{equation}
  x\,{\!f}(x, \mufzero^2) = A_{\!f}\,x^{B_{\!f}}\,(1 - x)^{C_{\!f}}\,(1 + D_{\!f}\,x + E_{\!f}\,x^2).
  \label{eq:pdf-parametrization-general}
\end{equation}
\end{linenomath*}
The overall normalization of the PDFs is given by the $A$ parameters, with $A_{\text{u}_\text{v}}$,
$A_{\text{d}_\text{v}}$, and $A_{\text{g}}$ being constrained by the quark number and momentum sum rules.
The $B$ and $C$ parameters control the shape of the distribution
as $x$ approaches the edges of its domain at 0 and 1, respectively.
The $D$ and $E$ parameters represent additional degrees of freedom related to the functional forms.

The sea quark contributions are given by $\UBar(x) = \uBar(x)$ and $\DBar(x) = \dBar(x) + \sBar(x)$, where $\uBar(x)$,
$\dBar(x)$, and $\sBar(x)$ refer to the distribution of up, down, and strange antiquarks, respectively.
A fixed overall normalization ratio is imposed by the requirement $A_{\UBar} = A_{\DBar} (1 - \fs)$,
where the strangeness fraction is given by $\fs = \sBar/(\sBar + \dBar)$ and set to 0.4 following
Ref.~\cite{H1:2015ubc}.
To enforce a similar behavior of the quark sea as $x \to 0$, the requirement $B_{\UBar} = B_{\DBar}$
is imposed.
The total sea quark distribution is defined as $\Sigma(x) = 2\left(\UBar(x) + \DBar(x)\right)$.

The above constraints result in a total of ten $A$, $B$, and $C$ parameters
whose values are determined during the fit procedure.
Additional $D$ and $E$ parameters are included where it is found that these lead
to an improved fit quality in terms of the total $\chi^2$ value, following the
procedure outlined in Ref.~\cite{H1:2015ubc}.
Starting from the initial parametrization with no additional parameters, the change
in $\chi^2$ resulting from the inclusion of any of the remaining $D$ and $E$
parameters in the fit is evaluated.
The parameter resulting in the largest decrease in $\chi^2$ is added to the
parametrization, and the procedure is repeated until no further significant improvement
is observed.
The final parametrization obtained in this way for the fits including the CMS
dijet measurements is given by:
\begin{linenomath*}
  \begin{equation}\begin{aligned}
    x\,{\text{g}}(x, \mufzero^2)
        &= A_{\text{g}}\,x^{B_{\text{g}}}(1 - x)^{C_{\text{g}}},\\
    x\,{\text{u}_\text{v}}(x, \mufzero^2)
        &= A_{\text{u}_\text{v}}\,x^{B_{\text{u}_\text{v}}}(1 - x)^{C_{\text{u}_\text{v}}}(1 + D_{\text{u}_\text{v}}x + E_{\text{u}_\text{v}}x^2),\\
    x\,{\text{d}_\text{v}}(x, \mufzero^2)
        &= A_{\text{d}_\text{v}}\,x^{B_{\text{d}_\text{v}}}(1 - x)^{C_{\text{d}_\text{v}}},\\
    x\,{\UBar}(x, \mufzero^2)
        &= A_{\UBar}\,x^{B_{\UBar}}(1 - x)^{C_{\UBar}}\,(1 + D_{\UBar}\,x),\\
    x\,{\DBar}(x, \mufzero^2)
        &= A_{\DBar}\,x^{B_{\DBar}}(1 - x)^{C_{\DBar}}.
    \label{eq:pdf-parametrization-2d-3d}
  \end{aligned}\end{equation}
\end{linenomath*}

Uncertainties in the fitted PDFs are determined using a similar procedure as the
one described in Ref.~\cite{H1:2021xxi}. Separate contributions for
\textit{fit}, \textit{model}, \textit{scale}, and \textit{parametrization} uncertainties
are obtained as described in the following.

The fit uncertainty represents the propagation to the PDFs of the uncertainties in the
input measurements, theoretical predictions, and theory correction factors.
It is estimated following the MC method outlined in Refs.~\cite{Giele:1998gw,Giele:2001mr},
whereby a large number of alternative fits (MC replicas) are performed with random variations
of the input data according to their statistical and systematic uncertainties, taking the
standard deviation of the resulting PDFs as an estimate of the fit uncertainty.

An alternative estimate for the fit uncertainty is obtained via the Hessian
method~\cite{Pumplin:2001ct} and found to be comparable to the MC fit uncertainty in most
cases, apart from the $\text{u}_\text{v}$ distribution, where it is found that the fit
uncertainty is significantly underestimated by the Hessian method at $x < 0.1$.
This is understood to be a consequence of the more flexible parametrization of
$\text{u}_\text{v}$ resulting from the parametrization scan, which is driven by the
high-$x$ region where the input data are constraining.

The model uncertainty arises from the choices made for the values of certain non-PDF
parameters: the minimum $Q^2$ value used for restricting the HERA data, the strangeness
fraction $\fs$, the charm and bottom quark masses $m_\text{c}$ and $m_\text{b}$,
and the value of the starting scale $\mufzero$.
It is estimated by varying the values of these parameters up and down from their nominal
values as indicated in Table~\ref{tab:model-parameter-variations}, and adding the differences
to the nominal fit result in quadrature separately for each variation direction.

\begin{table}[htb]
  \topcaption{
    Nominal values and variations of parameters used to determine the PDF model uncertainty. 
    Variations marked with an asterisk are in conflict with the requirement $\mufzero < m_\PQc$ 
    and thus cannot be used directly for the uncertainty estimation. Following Ref.~\cite{H1:2021xxi}, 
    the results obtained for the opposite variation are symmetrized in these cases. 
    \label{tab:model-parameter-variations}}
  \centering
  \begin{tabular}{lccc}
    {Parameter}             &  {Nominal value}  &  \multicolumn{2}{c}{{Variations}} \\
                                     &                          &  \textit{down}  &  \textit{up}
    \\[1mm]\hline\\[-3mm]
    $Q^2_\text{min}$ ($\GeVns{}^2$)  &  10                      &  7.5            &  12.5 \\
    $\fs$                            &  0.4                     &  0.3            &  0.5 \\
    $m_\PQc$ (\GeVns{})              &  1.43                    &  \textit{1.37}* &  1.49 \\
    $m_\PQb$ (\GeVns{})              &  4.5                     &  4.25           &  4.75 \\
    $\mufzero^2$ ($\GeVns{}^2$)      &  1.9                     &  1.6            &  \textit{2.2}*  \\
  \end{tabular}
\end{table}

A further uncertainty arises because of the choice of PDF parametrization. It is estimated by
performing alternative fits that include one additional $D$ or $E$ parameter compared to the
nominal parametrization.
The maximum deviation between the nominal PDF and those obtained from the
alternative parametrizations is taken as an additional parametrization uncertainty.

Finally, a scale uncertainty is estimated to account for missing higher orders in perturbation theory
by varying $\mur$ and $\muf$ as described in Section~\ref{sec:theory}.
The envelope of the PDFs obtained with these alternative scale choices is defined as the
scale uncertainty.

As discussed in Section~\ref{sec:results}, the level of agreement between
the data and the theoretical predictions obtained with various global PDF sets
varies according to the phase space region and is generally worse
at outer rapidities.
For the PDF determinations performed using the present data, a poor fit quality
is observed in a small number of rapidity regions at high $\ymax$, $\ystar$ or
$\yboost$, with the partial $\chi^2$ divided by the number of data points reaching
values of $\approx$3.

The effect of including these regions in the PDF determinations is investigated
by comparing to fits performed with only a subset of
rapidity regions, in which the data are well described by the theoretical predictions.
While this results in an increased fit uncertainty,
a sizable reduction in the parametrization uncertainty---and to a lesser extent the scale
uncertainty---is achieved for the restricted fits.
Consequently, the fit results are derived with the chosen subset of rapidity
regions, which are indicated in Table~\ref{tab:partial-chi2} along with the total
and partial $\chi^2$ values, which are close to unity in most rapidity regions,
except for the bin $0.5 < \yboost < 1$, $\ystar < 0.5$.
The results of fits including all rapidity regions are provided for reference in
Appendix~\ref{app:fullrap}.

\begin{table*}[p]
  \topcaption{
    Goodness-of-fit values for the fits to the HERA DIS data alone, and together with the CMS dijet measurements,
    using the PDF parametrization given in Eq.~\ref{eq:pdf-parametrization-2d-3d}.
    The table shows the partial $\chi^2$ values divided by the number of data points for the HERA DIS datasets and
    each of the dijet rapidity regions. The total $\chi^2$ value, divided by the number of degrees of freedom, is
    given at the bottom of the table.
    \label{tab:partial-chi2}}
  \centering
  \begin{tabular}{lllccc}
    &                       &                                      &  \multicolumn{3}{c}{{Partial $\chi^2$ / $n_\text{data}$}} \\[1mm]\\[-3mm]
    &                       &                                      &   HERA DIS    &  \multicolumn{2}{c}{HERA DIS + CMS 13\TeV dijets} \\[1mm]
    \multicolumn{3}{l}{Data set}                                   &               &  \multicolumn{1}{c}{2D}  &  \multicolumn{1}{c}{3D} \\[1mm]
    \hline\\[-3mm]
    \multicolumn{3}{l}{CMS dijet 2D} \\[1mm]
    & \multicolumn{2}{l}{$\ymax < 0.5$}                            &               &      18 / 22  &               \\
    & \multicolumn{2}{l}{$0.5 < \ymax < 1$}                        &               &      15 / 22  &               \\
    & \multicolumn{2}{l}{$1 < \ymax < 1.5$}                        &               &      16 / 23  &               \\
    & \multicolumn{2}{l}{$1.5 < \ymax < 2$}                        &               &      15 / 12  &               \\[2mm]

    \multicolumn{3}{l}{CMS dijet 3D} \\[1mm]
    & $\yboost < 0.5$,      &  $\ystar < 0.5$                      &               &               &      22 / 21  \\[2mm]
    & $\yboost < 0.5$,      &  $0.5 < \ystar < 1$                  &               &               &      24 / 19  \\
    & $0.5 < \yboost < 1$,  &  $\ystar < 0.5$                      &               &               &      49 / 19  \\[2mm]
    & $0.5 < \yboost < 1$,  &  $0.5 < \ystar < 1$                  &               &               &      13 / 17  \\[2mm]
    & $0.5 < \yboost < 1$,  &  $1 < \ystar < 1.5$                  &               &               &      \phantom{0}8 / \phantom{0}7  \\
    & $1 < \yboost < 1.5$,  &  $0.5 < \ystar < 1$                  &               &               &      10 / \phantom{0}7  \\[2mm]
    & $0.5 < \yboost < 1$,  &  $1.5 < \ystar < 2$                  &               &               &      \phantom{0}9 / \phantom{0}6  \\
    & $1 < \yboost < 1.5$,  &  $1 < \ystar < 1.5$                  &               &               &      \phantom{0}4 / \phantom{0}6  \\
    & $1.5 < \yboost < 2$,  &  $0.5 < \ystar < 1$                  &               &               &      \phantom{0}8 / \phantom{0}5  \\[2mm]
    \multicolumn{3}{l}{HERA1+2} \\[1mm]
    & \multicolumn{2}{l}{CC $\Pem\Pp$, $E_\text{p} = 920\GeV$}  &    \phantom{0}51 / \phantom{0}42  &      \phantom{0}51 / \phantom{0}42  &      \phantom{0}50 / \phantom{0}42  \\
    & \multicolumn{2}{l}{CC $\Pep\Pp$, $E_\text{p} = 920\GeV$}  &    \phantom{0}37 / \phantom{0}39  &      \phantom{0}37 / \phantom{0}39  &      \phantom{0}38 / \phantom{0}39  \\
    & \multicolumn{2}{l}{NC $\Pem\Pp$, $E_\text{p} = 920\GeV$}  &    221 / 159  &    222 / 159  &    221 / 159  \\
    & \multicolumn{2}{l}{NC $\Pep\Pp$, $E_\text{p} = 460\GeV$}  &    198 / 177  &    197 / 177  &    198 / 177  \\
    & \multicolumn{2}{l}{NC $\Pep\Pp$, $E_\text{p} = 575\GeV$}  &    186 / 221  &    186 / 221  &    186 / 221  \\
    & \multicolumn{2}{l}{NC $\Pep\Pp$, $E_\text{p} = 820\GeV$}  &    \phantom{0}55 / \phantom{0}61  &      \phantom{0}55 / \phantom{0}61  &      \phantom{0}55 / \phantom{0}61  \\
    & \multicolumn{2}{l}{NC $\Pep\Pp$, $E_\text{p} = 920\GeV$}  &    359 / 317  &    364 / 317  &    362 / 317  \\[1mm]
    \hline\\[-3mm]
    \multicolumn{3}{l}{Total $\chi^2/n_\text{dof}$}                &  1161 / 1003  &  1232 / 1081  &  1339 / 1109  \\[1mm]
  \end{tabular}
\end{table*}

\begin{figure*}[!htb]
    \centering
    \includegraphics[width=0.49\textwidth]{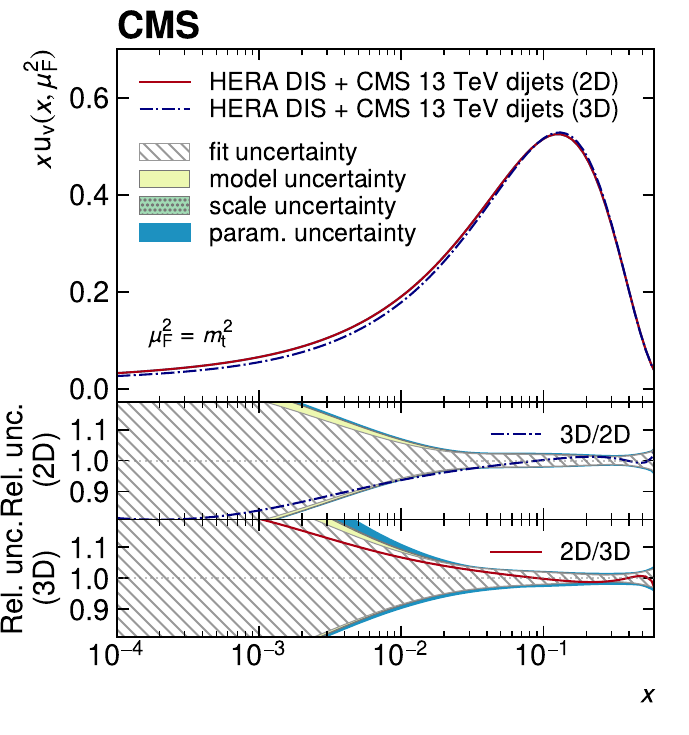}
    \includegraphics[width=0.49\textwidth]{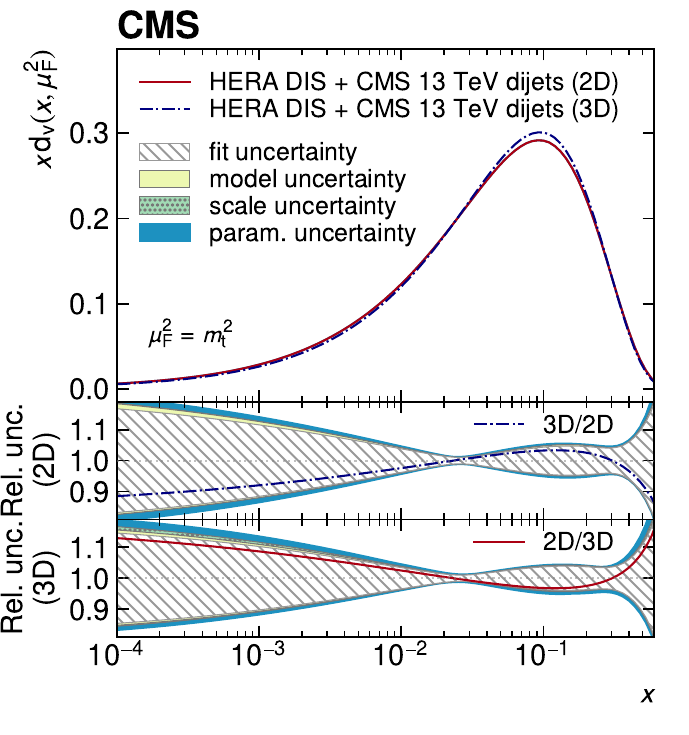}\\
    \includegraphics[width=0.49\textwidth]{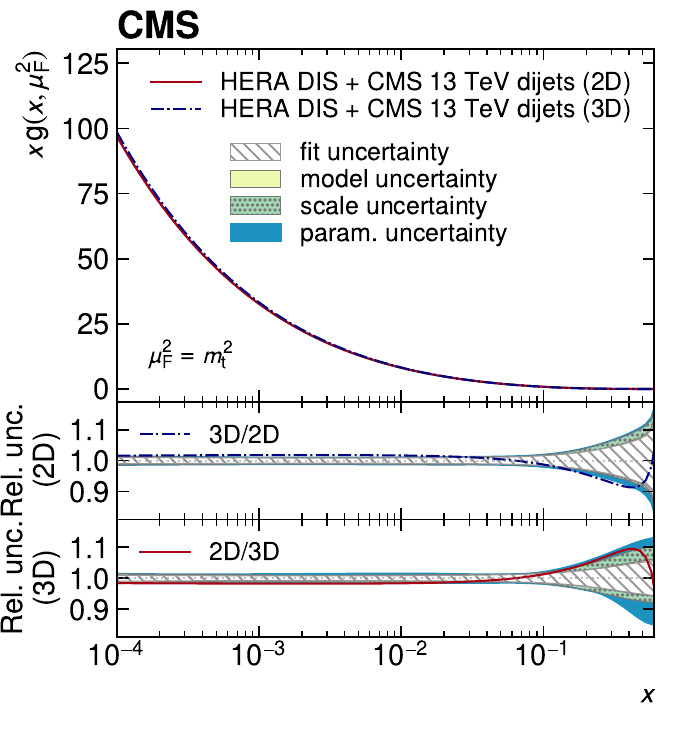}
    \includegraphics[width=0.49\textwidth]{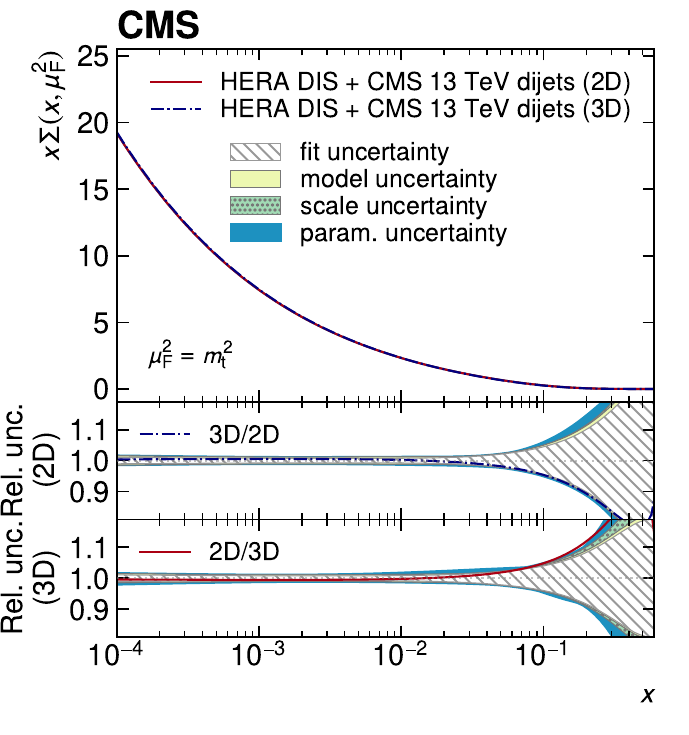}\\
    \caption{
    Parton distributions obtained in a fit to HERA DIS data together with the CMS
    2D or 3D dijet measurements. 
    The top panels show the PDFs of the up and down valence quarks (upper row), of the 
    gluon (lower left), and of the total sea quarks (lower right) as a function of the 
    fractional parton momentum $x$ at a factorization scale equal to the top quark mass. 
    The middle (lower) panels show the relative uncertainty contributions obtained for
    the 2D (3D) fit, as well as the ratios of the fitted central values.
    \label{fig:fits-pdf-uncertainties}}
\end{figure*}

The PDFs resulting from the fits including the CMS dijet measurements are shown in
Fig.~\ref{fig:fits-pdf-uncertainties},
along with the different uncertainty contributions.
The PDFs obtained with the inclusion of the 2D data are compatible with
those obtained from the 3D data within the total uncertainty, which is obtained by
adding together the parametrization uncertainty and the sum in quadrature of the fit,
model, and scale uncertainty contributions.
For most of the distributions, a smaller fit uncertainty is obtained in the 3D fit
compared to the 2D one, while the model uncertainty is of a similar size, and the
scale and parametrization uncertainties are slightly larger for the 3D fit in certain
$x$ regions.

\begin{figure*}[!htb]
    \centering
    \includegraphics[width=0.49\textwidth]{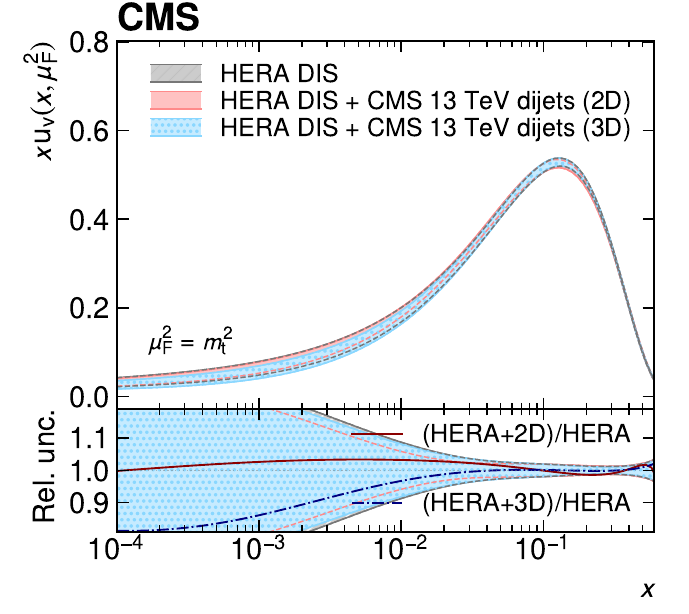}
    \includegraphics[width=0.49\textwidth]{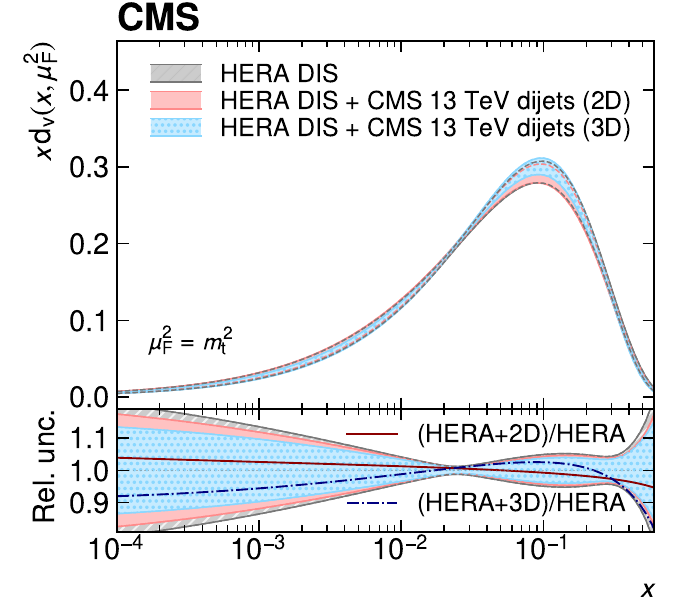}\\
    \includegraphics[width=0.49\textwidth]{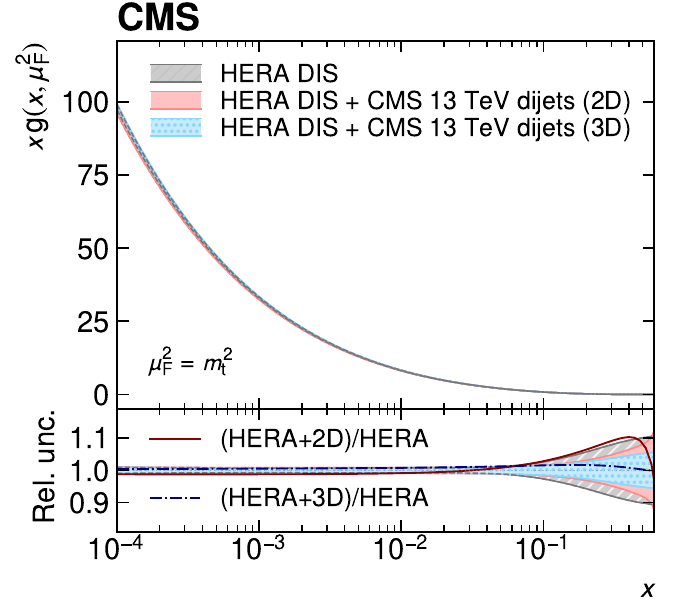}
    \includegraphics[width=0.49\textwidth]{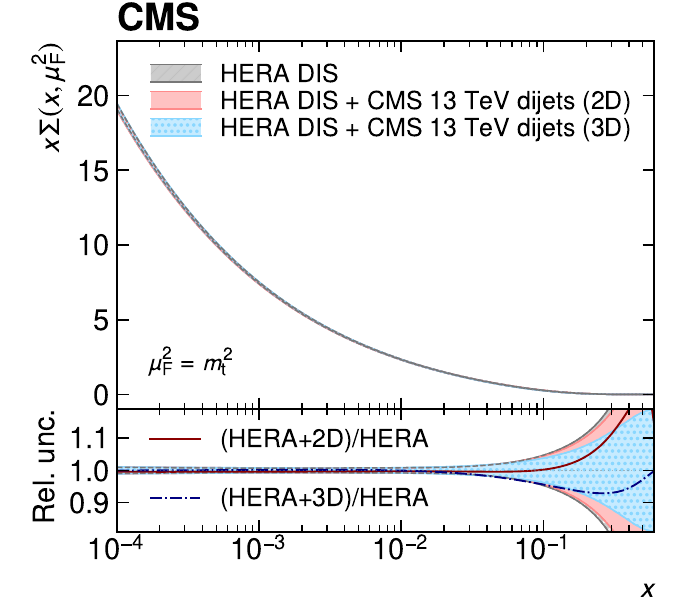}\\
    \caption{
    Parton distributions obtained in a fit to HERA DIS data together with the CMS dijet data, 
    compared to a fit to HERA DIS data alone. 
    Shown are the PDFs of the up and down valence quarks (upper row), of the gluon (lower left), 
    and of the total sea quarks (lower right) as a function of the fractional parton momentum $x$ 
    at a factorization scale equal to the top quark mass. 
    The bands indicate the fit uncertainty and are shown in the lower panels
    as a relative uncertainty with respect to the corresponding central values.
    The lines in the lower panels show the ratios between the fitted central values. 
    \label{fig:fits-compare-dis-dijets}}
\end{figure*}

To evaluate the impact of the present measurements on the PDF determination, fits are
performed using the HERA DIS data alone, using the same PDF parametrization as for
the fits including the dijet measurements.
Figure~\ref{fig:fits-compare-dis-dijets} shows a comparison between the PDFs obtained
using only the HERA DIS data, and those obtained when fitting the CMS dijet data in
addition, along with the respective fit uncertainties.
The distributions obtained with and without the inclusion of the dijet measurements are
observed to be compatible with each other, and a general reduction in the fit uncertainty
is observed when the CMS data is included in the fit.
In particular, the precision of the gluon PDF is improved for parton momentum fractions
$x > 0.1$, where the uncertainty is reduced by up to a factor of $\approx$2 by the
inclusion of the dijet measurements.
The 3D data are observed to constrain the gluon PDF to higher values of $x$ compared
to the 2D data.

For the PDF determinations presented above, the value of \asmz is extracted from the data
by including it in the fits as a free parameter, thus ensuring a consistent treatment of
correlations between \asmz and the PDF parameters, in particular those of the gluon
distribution.
The value of \asmz obtained in the fit to the 2D dijet cross sections is
\begin{linenomath*}
\begin{equation}
  \begin{aligned}
  \asmzResult{\asmz}{0.1179}{0.0015}{0.0008}{0.0008}{0.0001}{0.0019},
  \end{aligned}
  \label{eq:asmz-fitresult-2D}
\end{equation}
\end{linenomath*}
where the central value (fit uncertainty) is obtained as the average (standard deviation)
over the ensemble of MC replicas.
The remaining uncertainties are determined analogously to the PDFs, and in particular the
parametrization uncertainty contributes linearly to the total uncertainty while
the remaining contributions are added in quadrature.
For the 3D dijet measurement, the result obtained is
\begin{linenomath*}
\begin{equation}\begin{aligned}
  \asmzResult{\asmz}{0.1181}{0.0013}{0.0009}{0.0006}{0.0002}{0.0022},
  \label{eq:asmz-fitresult-3D}
\end{aligned}\end{equation}
\end{linenomath*}
which is in good agreement with the 2D result.

The values of $\asmz$ determined from the dijet measurements are in agreement
with the value of $0.1166\pm0.0017$ obtained in Ref.~\cite{CMS:2021yzl_addendum},
and with the world average value of
$0.1179\pm0.0009$~\cite{ParticleDataGroup:2022pth}.

Parton distributions obtained in previous analyses at $\sqrt{s}$~=~8 or 13\TeV of the
inclusive jet~\cite{CMS:2021yzl,CMS:2021yzl_addendum,CMS:2016lna} or the 3D dijet cross section~\cite{Sirunyan:2017skj}
are not easily comparable directly because of significant differences in the fit setup,
the PDF parametrizations, the model parameters, and particularly in the theoretical calculations
at 8\TeV, which were only available at NLO\@.
Taking the fit uncertainty in $\asmz$ obtained in a simultaneous fit with the PDFs
as a figure of merit, the 13\TeV results are more precise, which is consistent with
the increase in integrated luminosity.

\section{Summary\label{sec:summary}}

The dijet production cross section is measured based on $\Pp\Pp$ collision data
recorded by the CMS detector in 2016 at $\sqrt{s} = 13\TeV$,
corresponding to an integrated luminosity of up to 36.3\fbinv.

The measurements are performed double-differentially (2D)
as a function of the dijet invariant mass \mjj in five regions of
the maximal absolute rapidity \ymax\ of the two jets with the largest
transverse momenta,
and triple-differentially (3D)
as a function of either \mjj or the average transverse momentum \ptave{}
in 15 bins of the rapidity variables \ystar\ and \yboost.
The latter two variables correspond to
the rapidity separation of the two jets, and the total boost of the dijet
system, respectively.
All measurements are performed for jets clustered using the anti-\kt jet
algorithm with distance parameters $R=0.4$ and 0.8, and the cross sections
are unfolded in all measurement dimensions simultaneously to correct for
detector effects.

This is the first time that such a large set of multidifferential dijet measurements for two
observables, \ptave and \mjj, and two jet distance parameters, $R = 0.4$ and 0.8, is made available
for comparison to theory and use in fits of the parton distribution functions (PDFs) of the proton.
Predictions at next-to-next-to-leading order (NNLO) in perturbative quantum chromodynamics,
supplemented with electroweak and nonperturbative corrections are observed to describe the data better
for $R = 0.8$.

Using the measurement of \mjj for $R = 0.8$,
the PDFs of the proton
are determined simultaneously in fits to the dijet measurements together with
deep-inelastic scattering data from the HERA experiments following the approach
described in earlier HERAPDF analyses~\cite{H1:2009pze,H1:2015ubc,H1:2021xxi}.
The results obtained from the double- and triple-differential measurements
are compatible within the estimated uncertainties.
The inclusion of either of the dijet measurements leads to an improved
determination of the PDFs compared to fits to HERA data alone.
In particular, the uncertainty in the gluon distribution at fractional
proton momenta $x>0.1$ is reduced, with the 3D dijet data providing
tighter constraints at higher values of $x$ compared to the 2D data.
The strong coupling constant at the \PZ boson mass is
determined simultaneously with the PDFs, yielding
consistent results between the 2D and 3D dijet measurements, with
the former resulting in the slightly more precise value of $\asmz = 0.1179 \pm 0.0019$ at NNLO\@.

The impact of subleading-color contributions to the leading-color NNLO calculation
used here is not yet known~\cite{Chen:2022tpk}. Apart from being useful as inputs to
PDF fits or studies of jet size dependence, the present 2D and 3D measurements for two jet size
parameters, $R=0.4$ and 0.8, and for the two dijet observables \mjj and \ptave, provide
an ideal testing ground for further investigations.

\begin{acknowledgments}
 We congratulate our colleagues in the CERN accelerator departments for the excellent performance of the LHC and thank the technical and administrative staffs at CERN and at other CMS institutes for their contributions to the success of the CMS effort. In addition, we gratefully acknowledge the computing centers and personnel of the Worldwide LHC Computing Grid and other centers for delivering so effectively the computing infrastructure essential to our analyses. Finally, we acknowledge the enduring support for the construction and operation of the LHC, the CMS detector, and the supporting computing infrastructure provided by the following funding agencies: SC (Armenia), BMBWF and FWF (Austria); FNRS and FWO (Belgium); CNPq, CAPES, FAPERJ, FAPERGS, and FAPESP (Brazil); MES and BNSF (Bulgaria); CERN; CAS, MoST, and NSFC (China); MINCIENCIAS (Colombia); MSES and CSF (Croatia); RIF (Cyprus); SENESCYT (Ecuador); MoER, ERC PUT and ERDF (Estonia); Academy of Finland, MEC, and HIP (Finland); CEA and CNRS/IN2P3 (France); SRNSF (Georgia); BMBF, DFG, and HGF (Germany); GSRI (Greece); NKFIH (Hungary); DAE and DST (India); IPM (Iran); SFI (Ireland); INFN (Italy); MSIP and NRF (Republic of Korea); MES (Latvia); LAS (Lithuania); MOE and UM (Malaysia); BUAP, CINVESTAV, CONACYT, LNS, SEP, and UASLP-FAI (Mexico); MOS (Montenegro); MBIE (New Zealand); PAEC (Pakistan); MES and NSC (Poland); FCT (Portugal); MESTD (Serbia); MCIN/AEI and PCTI (Spain); MOSTR (Sri Lanka); Swiss Funding Agencies (Switzerland); MST (Taipei); MHESI and NSTDA (Thailand); TUBITAK and TENMAK (Turkey); NASU (Ukraine); STFC (United Kingdom); DOE and NSF (USA).
 
 \hyphenation{Rachada-pisek} Individuals have received support from the Marie-Curie program and the European Research Council and Horizon 2020 Grant, contract Nos.\ 675440, 724704, 752730, 758316, 765710, 824093, and COST Action CA16108 (European Union); the Leventis Foundation; the Alfred P.\ Sloan Foundation; the Alexander von Humboldt Foundation; the Science Committee, project no. 22rl-037 (Armenia); the Belgian Federal Science Policy Office; the Fonds pour la Formation \`a la Recherche dans l'Industrie et dans l'Agriculture (FRIA-Belgium); the Agentschap voor Innovatie door Wetenschap en Technologie (IWT-Belgium); the F.R.S.-FNRS and FWO (Belgium) under the ``Excellence of Science -- EOS" -- be.h project n.\ 30820817; the Beijing Municipal Science \& Technology Commission, No. Z191100007219010 and Fundamental Research Funds for the Central Universities (China); the Ministry of Education, Youth and Sports (MEYS) of the Czech Republic; the Shota Rustaveli National Science Foundation, grant FR-22-985 (Georgia); the Deutsche Forschungsgemeinschaft (DFG), under Germany's Excellence Strategy -- EXC 2121 ``Quantum Universe" -- 390833306, and under project number 400140256 - GRK2497; the Hellenic Foundation for Research and Innovation (HFRI), Project Number 2288 (Greece); the Hungarian Academy of Sciences, the New National Excellence Program - \'UNKP, the NKFIH research grants K 124845, K 124850, K 128713, K 128786, K 129058, K 131991, K 133046, K 138136, K 143460, K 143477, 2020-2.2.1-ED-2021-00181, and TKP2021-NKTA-64 (Hungary); the Council of Science and Industrial Research, India; ICSC -- National Research Center for High Performance Computing, Big Data and Quantum Computing, funded by the EU NexGeneration program (Italy); the Latvian Council of Science; the Ministry of Education and Science, project no. 2022/WK/14, and the National Science Center, contracts Opus 2021/41/B/ST2/01369 and 2021/43/B/ST2/01552 (Poland); the Funda\c{c}\~ao para a Ci\^encia e a Tecnologia, grant CEECIND/01334/2018 (Portugal); the National Priorities Research Program by Qatar National Research Fund; MCIN/AEI/10.13039/501100011033, ERDF ``a way of making Europe", and the Programa Estatal de Fomento de la Investigaci{\'o}n Cient{\'i}fica y T{\'e}cnica de Excelencia Mar\'{\i}a de Maeztu, grant MDM-2017-0765 and Programa Severo Ochoa del Principado de Asturias (Spain); the Chulalongkorn Academic into Its 2nd Century Project Advancement Project, and the National Science, Research and Innovation Fund via the Program Management Unit for Human Resources \& Institutional Development, Research and Innovation, grant B37G660013 (Thailand); the Kavli Foundation; the Nvidia Corporation; the SuperMicro Corporation; the Welch Foundation, contract C-1845; and the Weston Havens Foundation (USA).  
\end{acknowledgments}

\bibliography{auto_generated}

\appendix

\section{Full-rapidity fit results\label{app:fullrap}}

This section documents the results obtained for the fits described in Section~\ref{sec:fits}
when all dijet rapidity regions are included. The partial $\chi^2$ values indicating the
goodness-of-fit in each rapidity region are given in Table~\ref{tab:partial-chi2-fullrap},
and Eqs.~\eqref{eq:asmz-fitresult-2D-fullrap} and~\eqref{eq:asmz-fitresult-3D-fullrap} show
the values of \asmz obtained in fits including the 2D and 3D dijet cross sections,
respectively:

\begin{table*}[ht]
  \topcaption{
    Goodness-of-fit values for the fits to the HERA DIS data alone, and together with the CMS dijet measurements,
    including all rapidity regions. 
    The table shows the partial $\chi^2$ values divided by the number of data points for the HERA DIS datasets and
    each of the dijet rapidity regions. The total $\chi^2$ value, divided by the number of degrees of freedom, is
    given at the bottom of the table.
    \label{tab:partial-chi2-fullrap}
  }
  \centering
  \begin{tabular}{lllcc}
    &                       &                                      &  \multicolumn{2}{c}{{Partial $\chi^2$ / $n_\text{data}$}} \\[1mm]\\[-3mm]
    &                       &                                      &  \multicolumn{2}{c}{HERA DIS + CMS 13\TeV dijets} \\[1mm]
    \multicolumn{3}{l}{Data set}                                   &  \multicolumn{1}{c}{2D}  &  \multicolumn{1}{c}{3D} \\[1mm]
    \hline\\[-3mm]
    \multicolumn{3}{l}{CMS dijet 2D} \\[1mm]
    & \multicolumn{2}{l}{$\ymax < 0.5$}                            &      24 / 22  &               \\
    & \multicolumn{2}{l}{$0.5 < \ymax < 1$}                        &      14 / 22  &               \\
    & \multicolumn{2}{l}{$1 < \ymax < 1.5$}                        &      22 / 23  &               \\
    & \multicolumn{2}{l}{$1.5 < \ymax < 2$}                        &      15 / 12  &               \\
    & \multicolumn{2}{l}{$2 < \ymax < 2.5$}                        &      30 / 12  &               \\[2mm]

    \multicolumn{2}{l}{CMS dijet 3D} \\[1mm]
    & $\yboost < 0.5$,      &  $\ystar < 0.5$                      &               &       32 / 21 \\[2mm]
    & $\yboost < 0.5$,      &  $0.5 < \ystar < 1$                  &               &       23 / 19 \\
    & $0.5 < \yboost < 1$,  &  $\ystar < 0.5$                      &               &       40 / 19 \\[2mm]
    & $\yboost < 0.5$,      &  $1 < \ystar < 1.5$                  &               &       45 / 17 \\
    & $0.5 < \yboost < 1$,  &  $0.5 < \ystar < 1$                  &               &       18 / 17 \\
    & $1 < \yboost < 1.5$,  &  $\ystar < 0.5$                      &               &       44 / 17 \\[2mm]
    & $\yboost < 0.5$,      &  $1.5 < \ystar < 2$                  &               &        15 / \phantom{0}7 \\
    & $0.5 < \yboost < 1$,  &  $1 < \ystar < 1.5$                  &               &        \phantom{0}7 / \phantom{0}7 \\
    & $1 < \yboost < 1.5$,  &  $0.5 < \ystar < 1$                  &               &        \phantom{0}9 / \phantom{0}7 \\
    & $1.5 < \yboost < 2$,  &  $\ystar < 0.5$                      &               &        20 / \phantom{0}6 \\[2mm]
    & $\yboost < 0.5$,      &  $2 < \ystar < 2.5$                  &               &        19 / \phantom{0}6 \\
    & $0.5 < \yboost < 1$,  &  $1.5 < \ystar < 2$                  &               &        16 / \phantom{0}6 \\
    & $1 < \yboost < 1.5$,  &  $1 < \ystar < 1.5$                  &               &        \phantom{0}6 / \phantom{0}6 \\
    & $1.5 < \yboost < 2$,  &  $0.5 < \ystar < 1$                  &               &        \phantom{0}1 / \phantom{0}5 \\
    & $2 < \yboost < 2.5$,  &  $\ystar < 0.5$                      &               &        15 / \phantom{0}4 \\[2mm]
    \multicolumn{3}{l}{HERA1+2} \\[1mm]
    & \multicolumn{2}{l}{CC $\Pem\Pp$, $E_\text{p} = 920\GeV$}  &      50 / 42  &       48 / 42 \\
    & \multicolumn{2}{l}{CC $\Pep\Pp$, $E_\text{p} = 920\GeV$}  &      37 / 39  &       41 / 39 \\
    & \multicolumn{2}{l}{NC $\Pem\Pp$, $E_\text{p} = 920\GeV$}  &    222 / 159  &     227 / 159 \\
    & \multicolumn{2}{l}{NC $\Pep\Pp$, $E_\text{p} = 460\GeV$}  &    197 / 177  &     201 / 177 \\
    & \multicolumn{2}{l}{NC $\Pep\Pp$, $E_\text{p} = 575\GeV$}  &    186 / 221  &     187 / 221 \\
    & \multicolumn{2}{l}{NC $\Pep\Pp$, $E_\text{p} = 820\GeV$}  &      55 / 61  &       55 / 61 \\
    & \multicolumn{2}{l}{NC $\Pep\Pp$, $E_\text{p} = 920\GeV$}  &    368 / 317  &     365 / 317 \\[1mm]
    \hline\\[-3mm]
    \multicolumn{3}{l}{Total $\chi^2/n_\text{dof}$}                &  1283 / 1094  &   1557 / 1167 \\[1mm]
  \end{tabular}
\end{table*}

\begin{linenomath*}
  \begin{equation}\begin{aligned}
    \asmzResultNoTotal{\asmz_{\text{2D}}}{0.1201}{0.0012}{0.0008}{0.0008}{0.0005},
    \label{eq:asmz-fitresult-2D-fullrap}
  \end{aligned}\end{equation}
  \begin{equation}\begin{aligned}
      \asmzResultNoTotal{\asmz_{\text{3D}}}{0.1201}{0.0010}{0.0005}{0.0008}{0.0006}.
      \label{eq:asmz-fitresult-3D-fullrap}
  \end{aligned}\end{equation}
\end{linenomath*}

\clearpage

\section{Additional figures \label{app:supplementary-material}}

\begin{figure*}[htb!]
  \centering
    \includegraphics[width=0.8\textwidth]{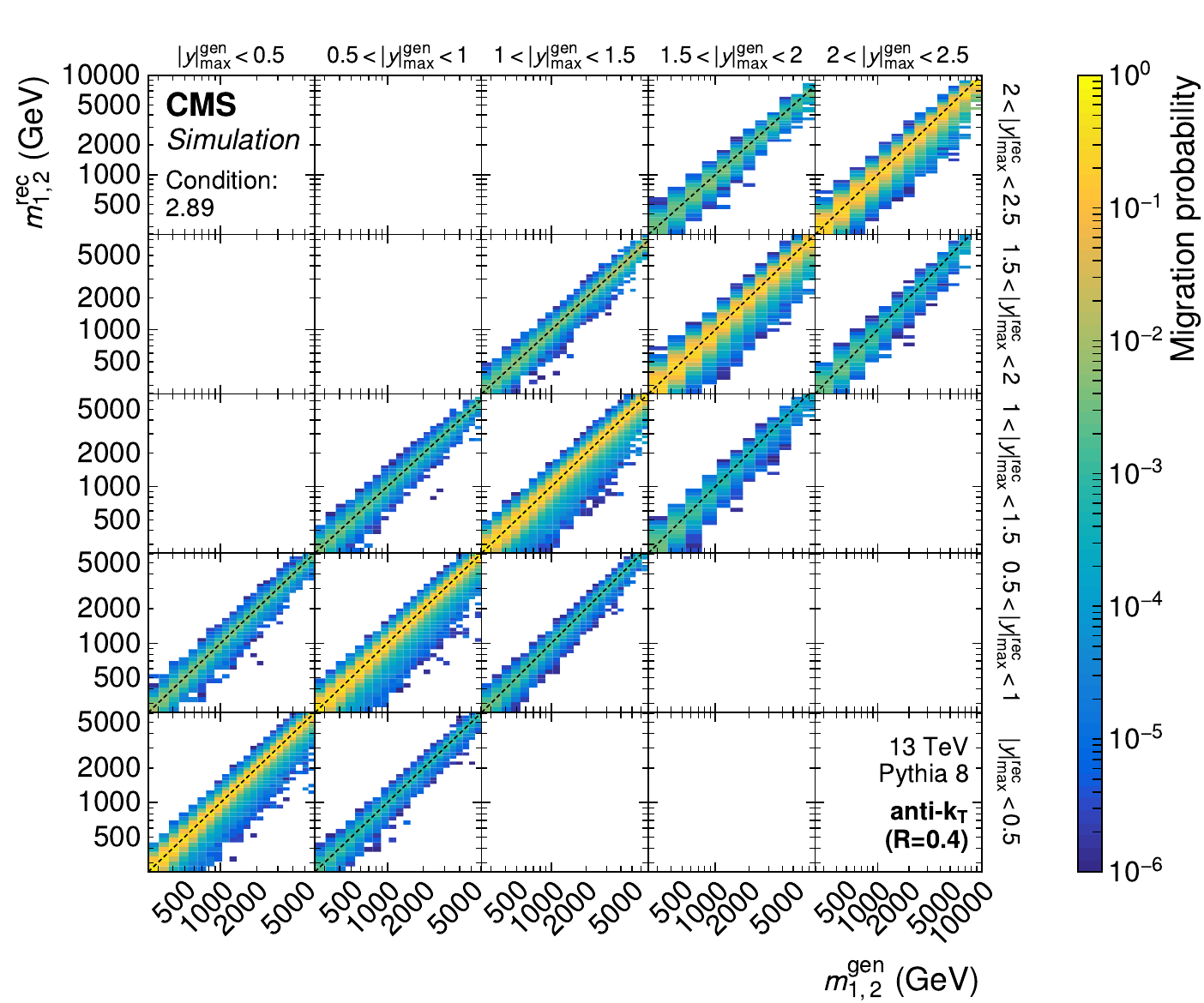}
    \caption{
    Response matrix for the 2D measurements
    as a function of $\mjj$
    for jets with $R = 0.4$. 
    The details correspond to those of 
    Fig.~\ref{fig:response-matrices}. 
    \label{fig:appendix-2d-matrix-mjj}}
\end{figure*}

\begin{figure*}[htb!]
  \centering
    \includegraphics[width=0.8\textwidth]{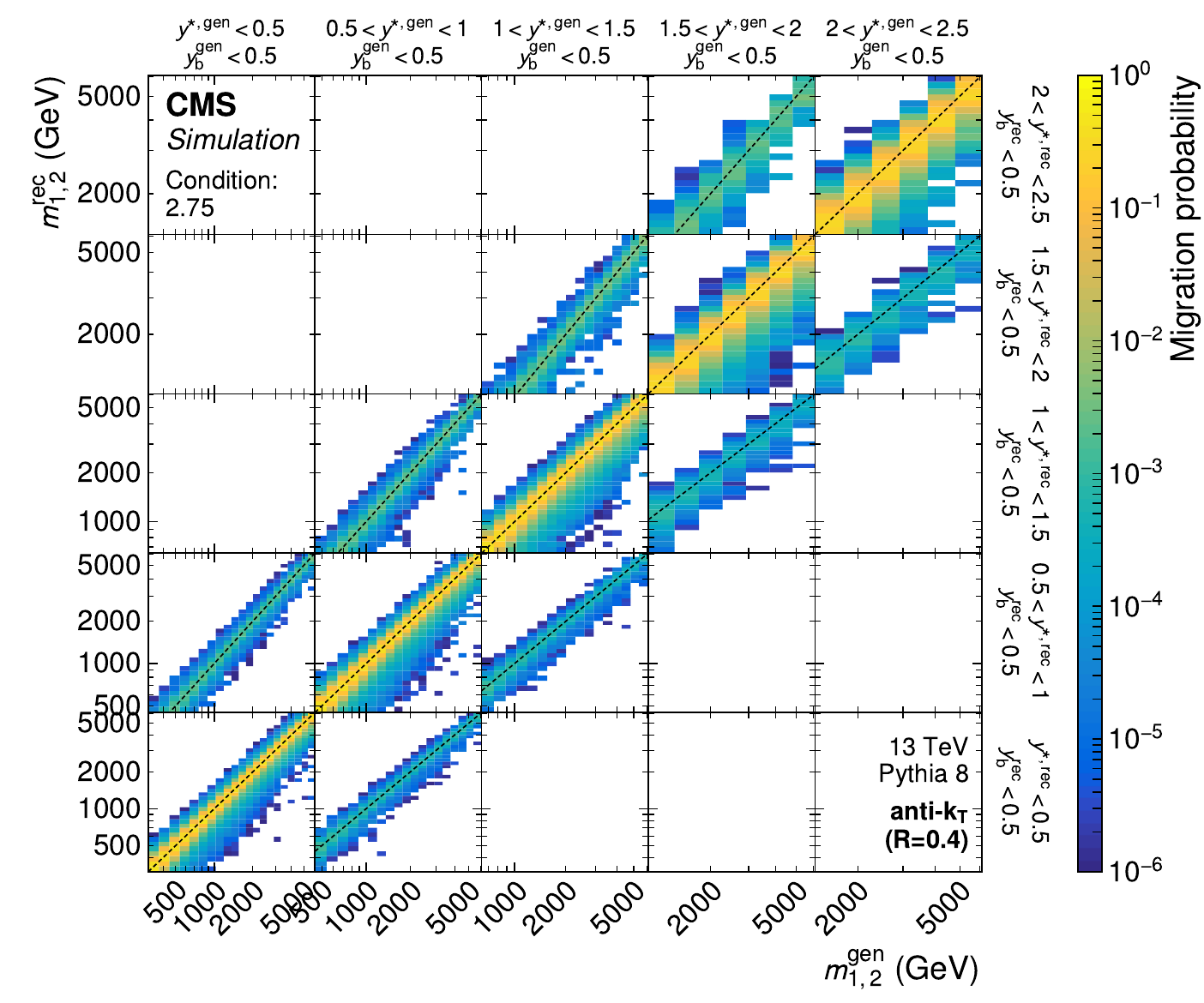}\\
    \includegraphics[width=0.8\textwidth]{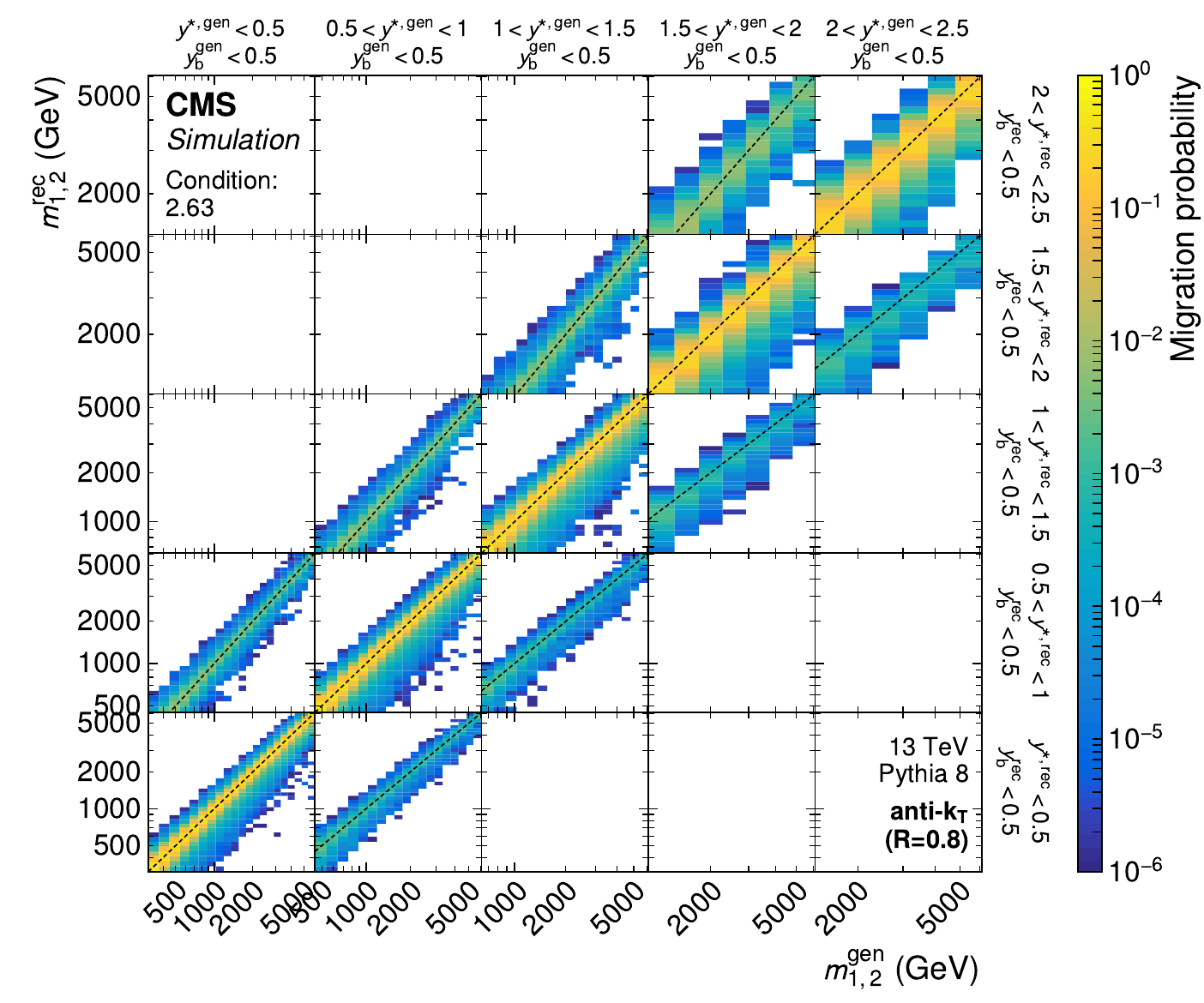}
    \caption{
    Partial response matrices for the 3D measurements 
    as a function of $\mjj$ 
    using jets with $R = 0.4$ (upper) and 0.8 (lower), 
    shown here for the five rapidity regions with $\yboost < 0.5$. 
    The details correspond to those of 
    Fig.~\ref{fig:response-matrices}. 
    \label{fig:appendix-3d-matrix-mjj-yb0}}
\end{figure*}

\begin{figure*}[!htb]
  \centering
    \includegraphics[width=0.8\textwidth]{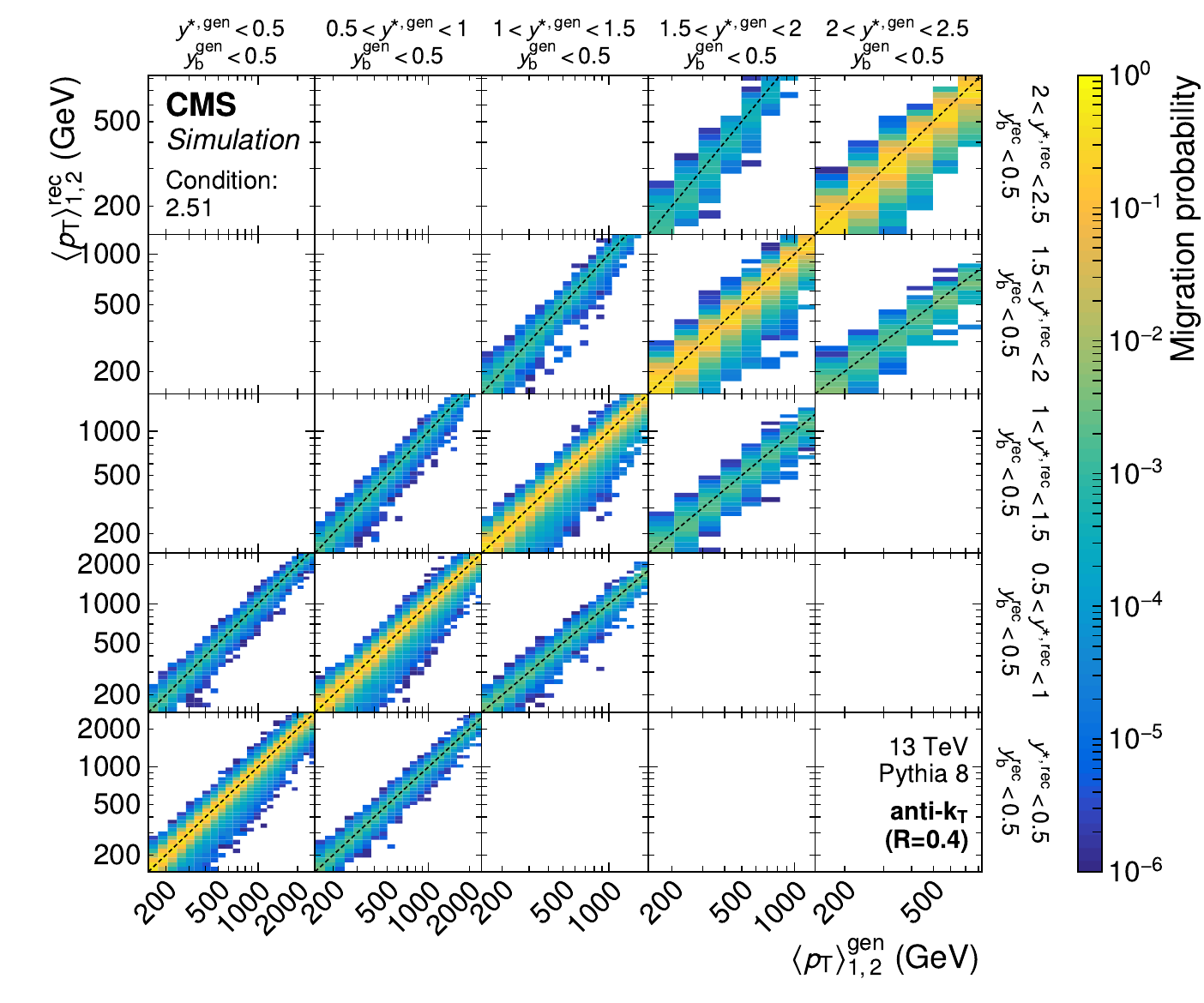}\\
    \includegraphics[width=0.8\textwidth]{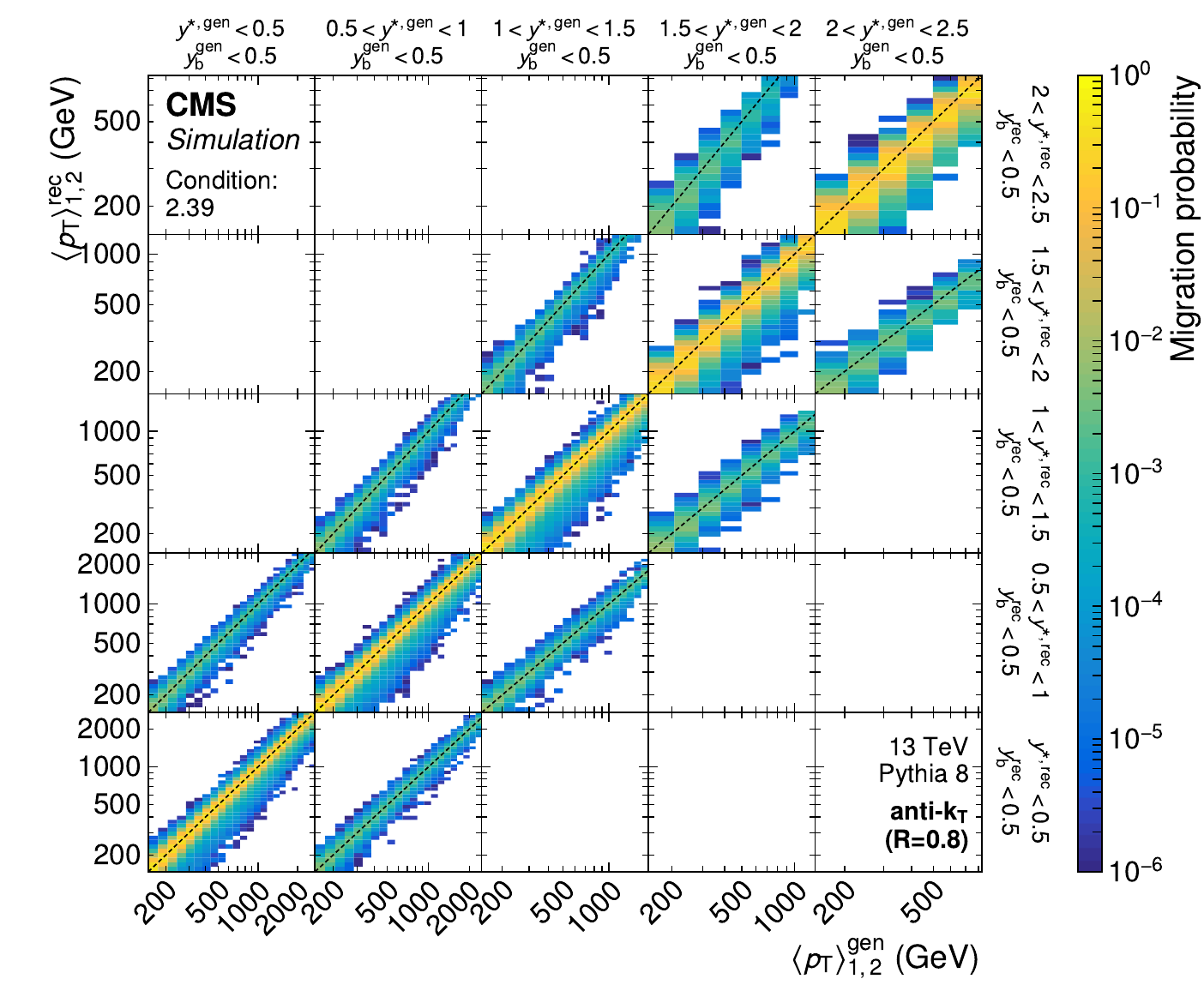}
    \caption{
    Partial response matrices for the 3D measurements 
    as a function of $\ptave$ 
    using jets with $R = 0.4$ (upper) and 0.8 (lower), 
    shown here for the five rapidity regions with $\yboost < 0.5$. 
    The details correspond to those of 
    Fig.~\ref{fig:response-matrices}. 
    \label{fig:appendix-3d-matrix-ptave-yb0}}
\end{figure*}

\begin{figure*}[htb]
  \centering
    \includegraphics[width=\cmsFigWidthTwo]{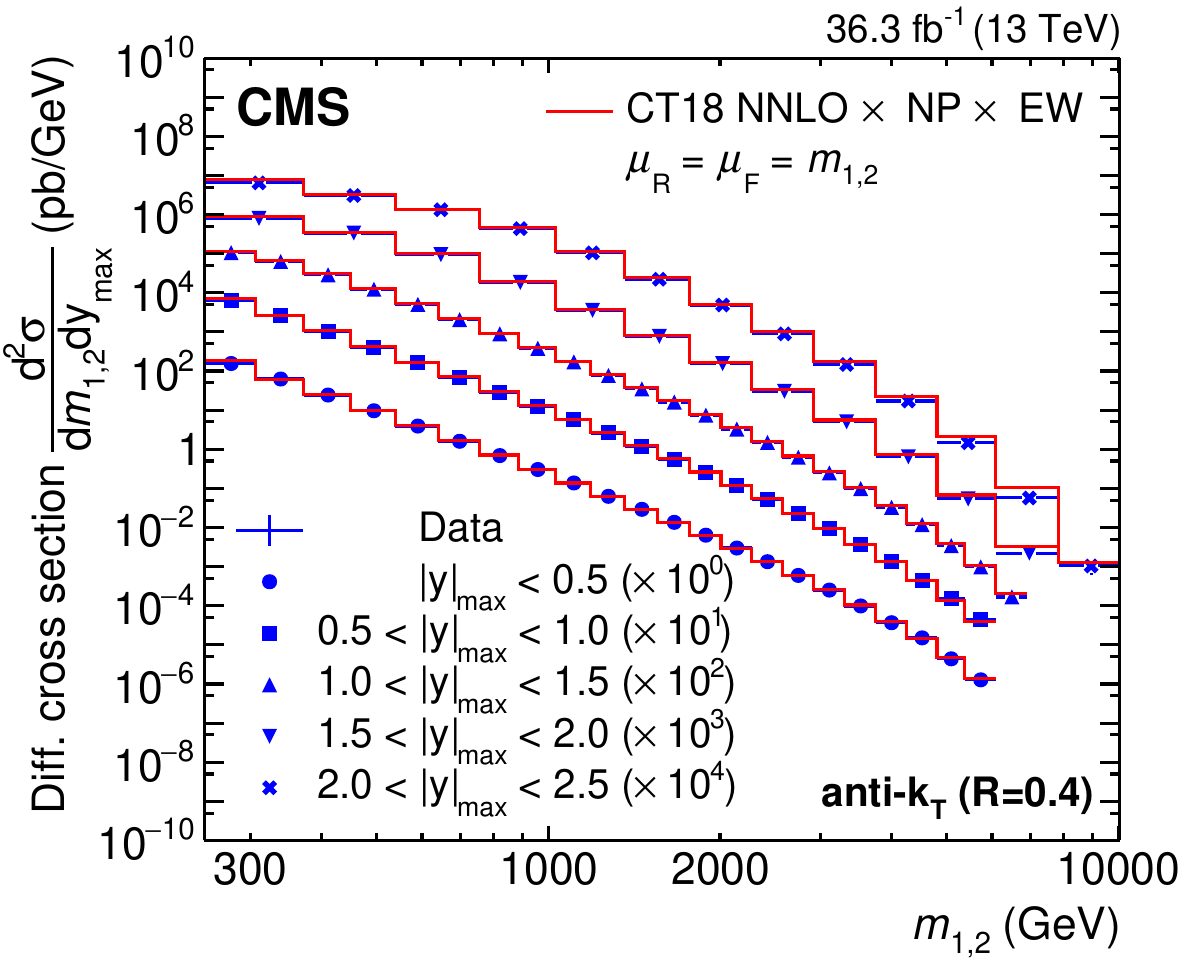}
    \caption{
    Overview of the 2D dijet cross section 
    as a function of $\mjj$ in all 5 $\ymax$ regions, 
    using jets with $R = 0.4$. 
    The details correspond to those of 
    Fig.~\ref{fig:cross-sections}. 
    \label{fig:appendix-2d-cross-sections}}
\end{figure*}

\begin{figure*}[htb]
  \centering
    \includegraphics[width=\cmsFigWidthTwo]{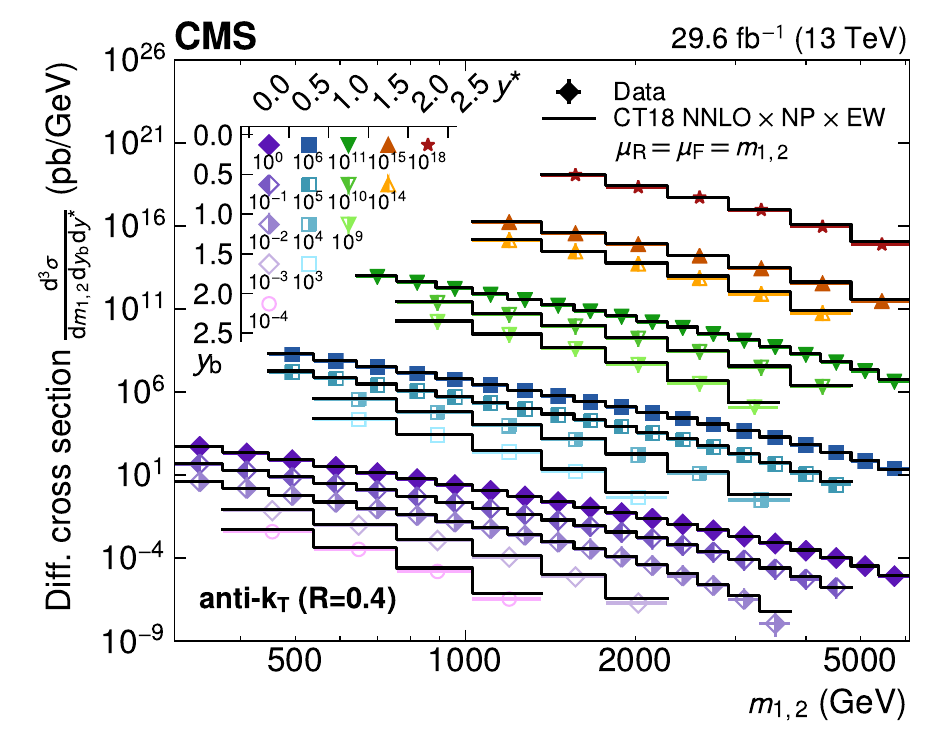}
    \includegraphics[width=\cmsFigWidthTwo]{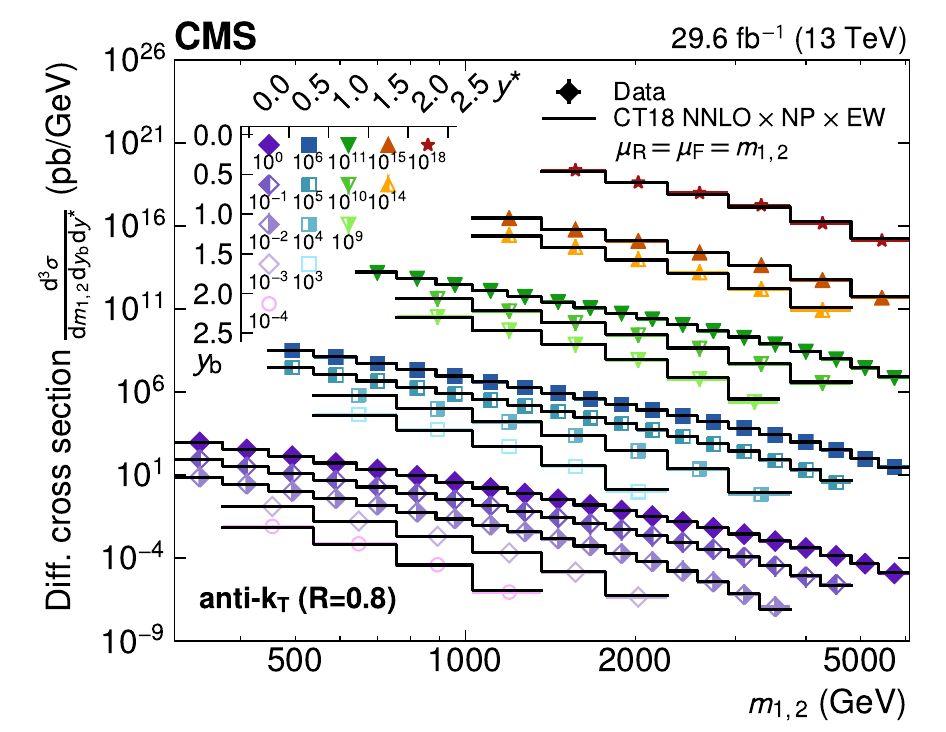}
    \caption{
    Overview of the 3D dijet cross section 
    as a function of $\mjj$ in all 15 $\ybys$ regions, 
    using jets with $R = 0.4$ (left) and 0.8 (right). 
    The details correspond to those of 
    Fig.~\ref{fig:cross-sections}. 
    \label{fig:appendix-3d-cross-sections-mjj}}
\end{figure*}

\begin{figure*}[htb]
  \centering
    \includegraphics[width=\cmsFigWidthTwo]{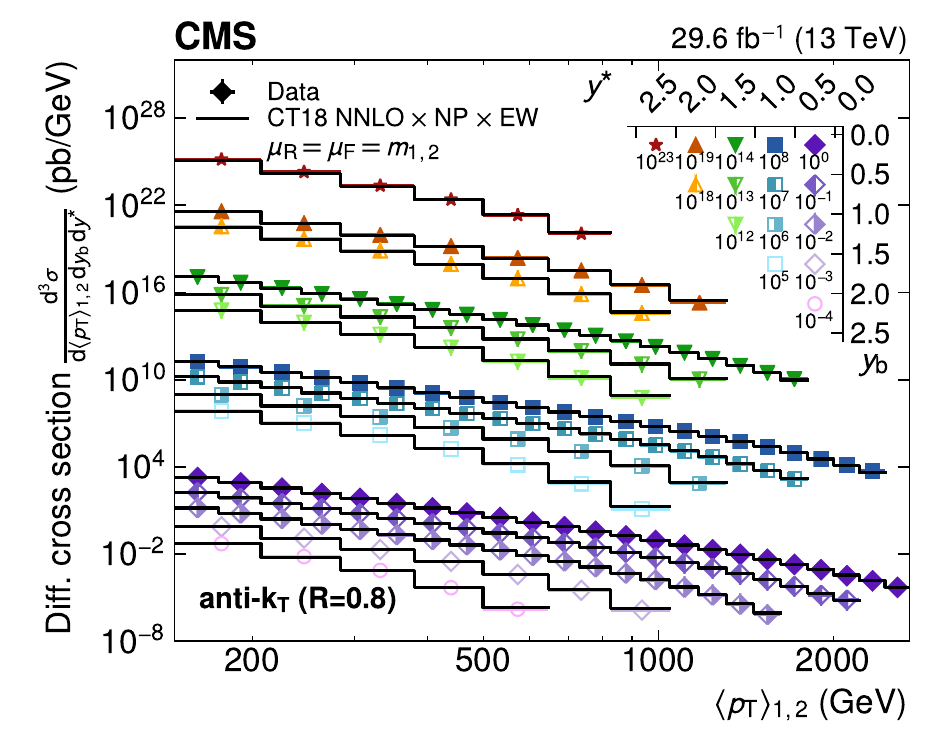}
    \caption{
    Overview of the 3D dijet cross section 
    as a function of $\ptave$ in all 15 $\ybys$ regions, 
    using jets with $R = 0.8$. 
    The details correspond to those of 
    Fig.~\ref{fig:cross-sections}. 
    \label{fig:appendix-3d-cross-sections-ptave}}
\end{figure*}

\begin{figure*}[htb]
  \centering
    \includegraphics[width=\cmsFigWidthTwo]{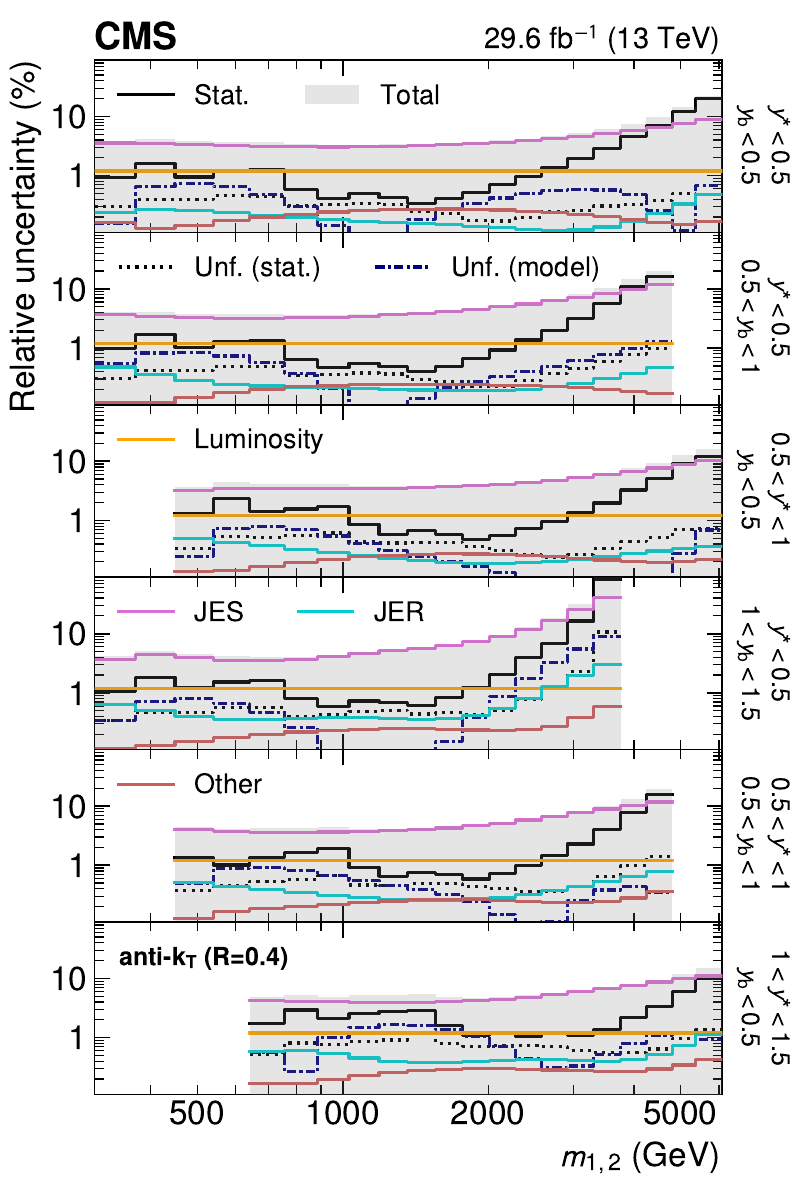}
    \includegraphics[width=\cmsFigWidthTwo]{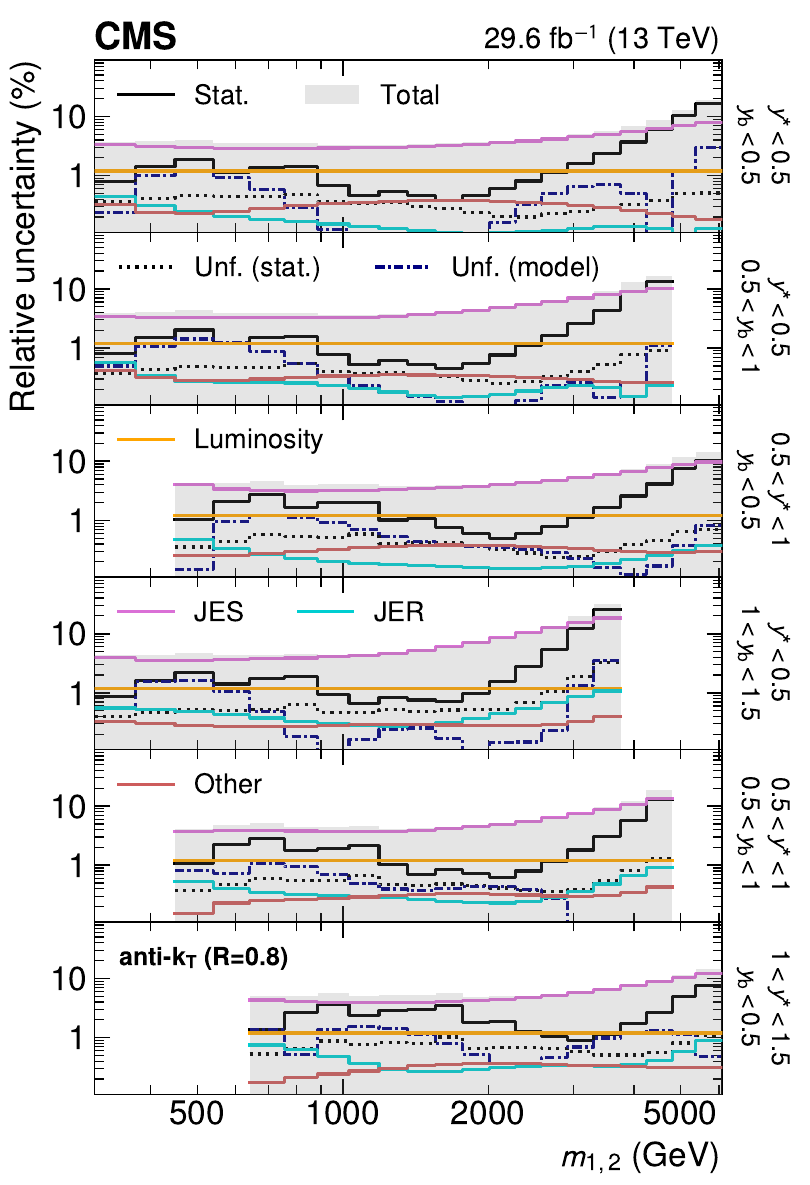}
    \caption{
    Breakdown of the experimental uncertainty 
    for the 3D measurements as a function of \mjj{} 
    using jets with $R = 0.4$ (left) and 0.8 (right), 
    in six out of 15 $\ybys$ bins. 
    The details correspond to those of Fig.~\ref{fig:3d-uncertainties}. 
    \label{fig:appendix-3d-uncertainties-mjj-part1}}
\end{figure*}

\begin{figure*}[htb]
  \centering
    \includegraphics[width=\cmsFigWidthTwo]{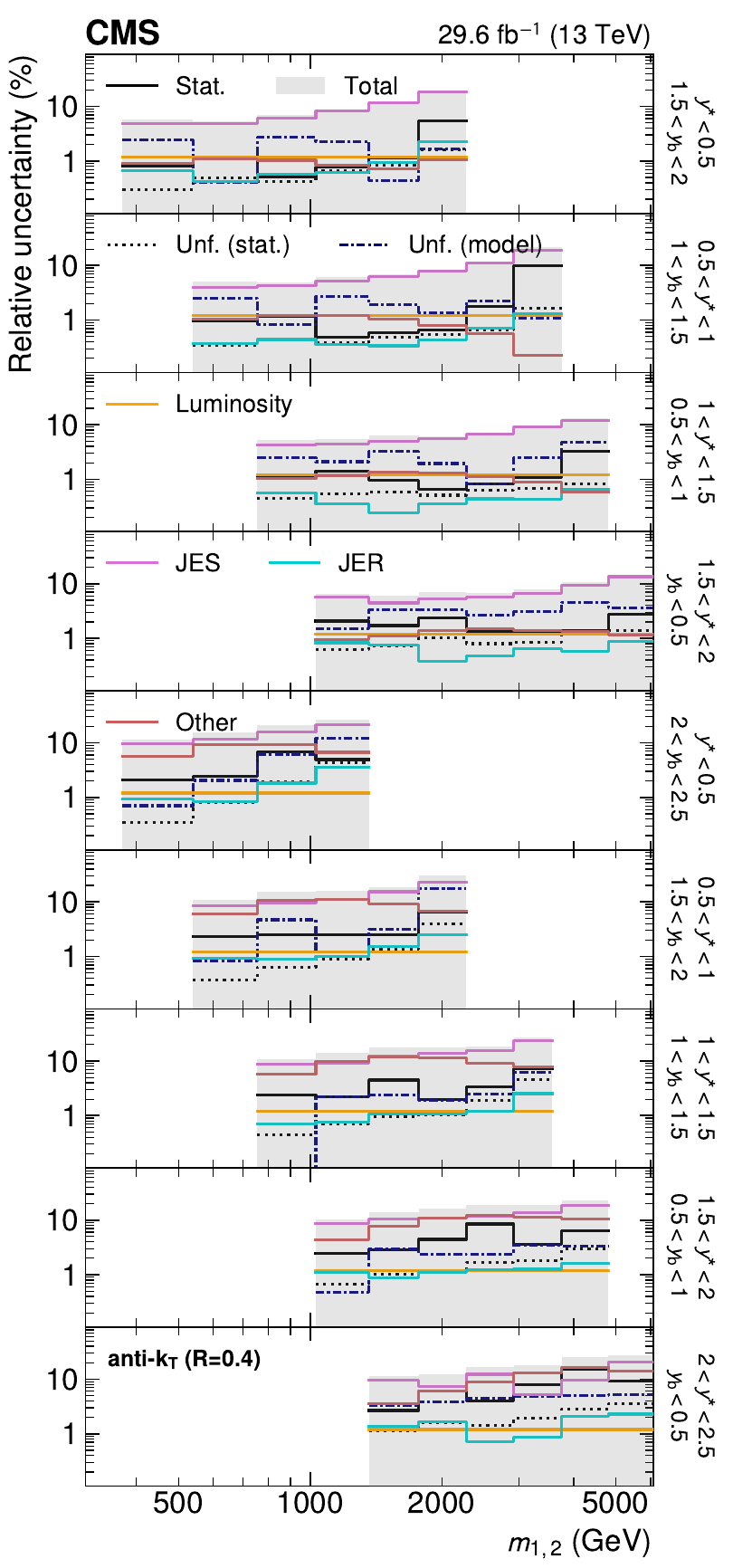}
    \includegraphics[width=\cmsFigWidthTwo]{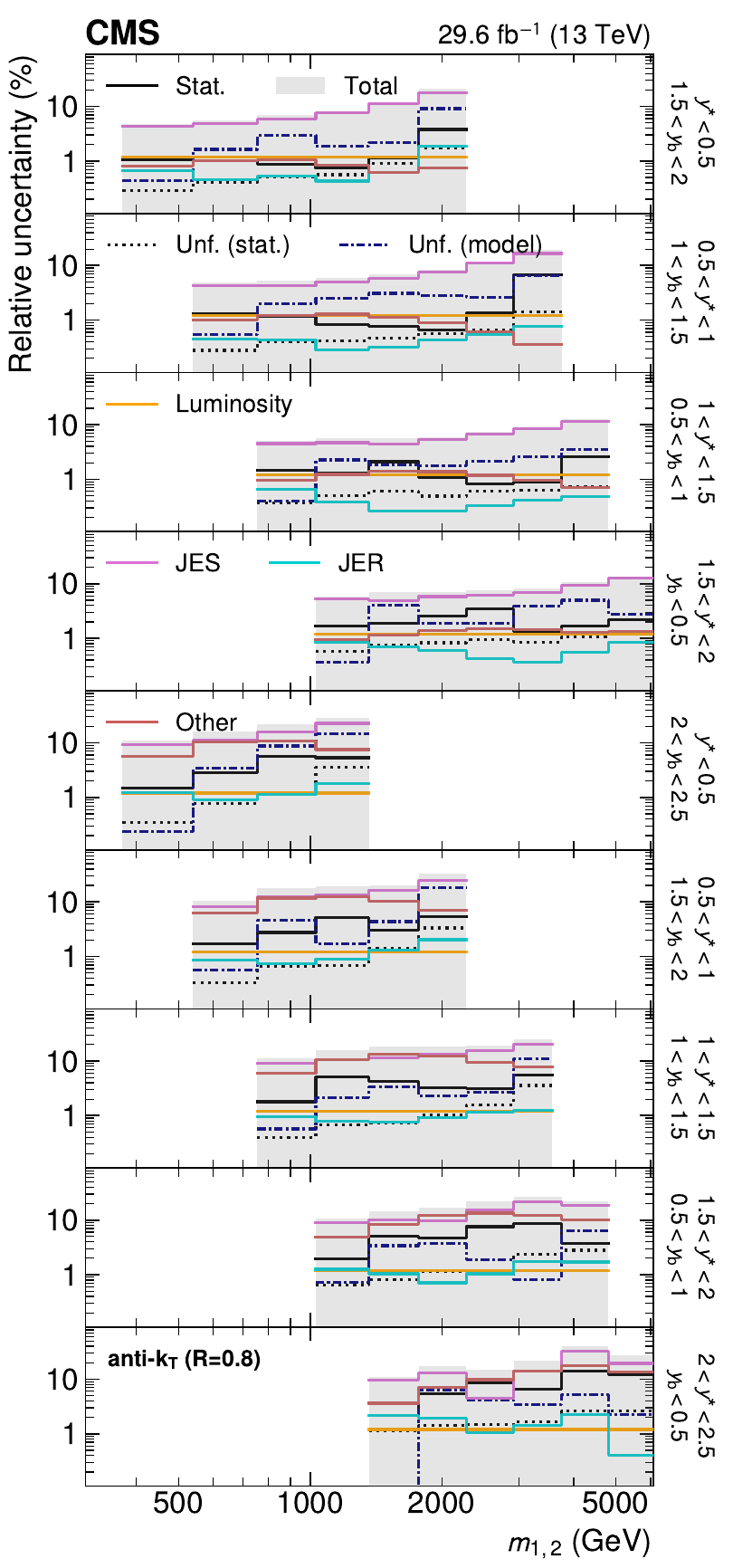}
    \caption{
    (\textit{continuation of Fig.~\ref{fig:appendix-3d-uncertainties-mjj-part1}}) 
    Breakdown of the experimental uncertainty 
    for the 3D measurements as a function of \mjj{} 
    using jets with $R = 0.4$ (left) and 0.8 (right), 
    in the remaining nine out of 15 $\ybys$ bins. 
    The details correspond to those of 
    Fig.~\ref{fig:3d-uncertainties}. 
    \label{fig:appendix-3d-uncertainties-mjj-part2}}
\end{figure*}

\begin{figure*}[htb]
  \centering
    \includegraphics[width=\cmsFigWidthTwo]{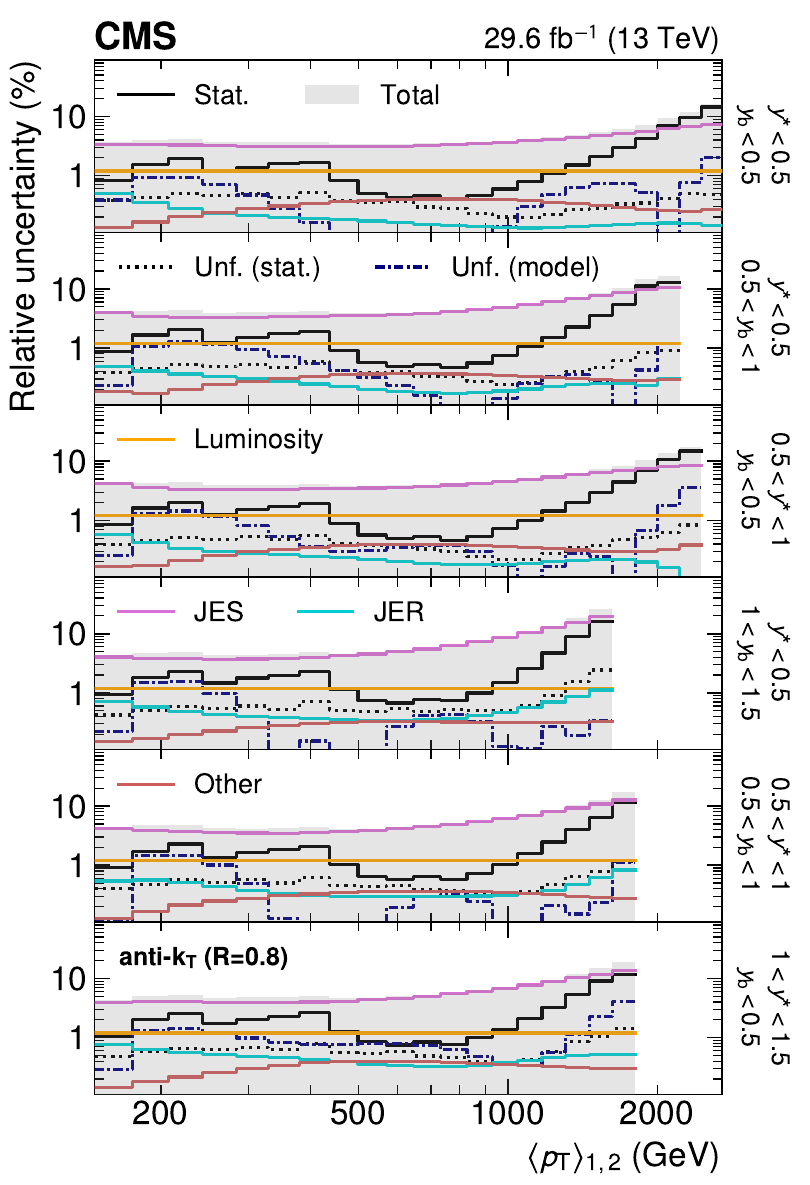}
    \caption{
    Breakdown of the experimental uncertainty 
    for the 3D measurements as a function of \ptave{} 
    using jets with $R = 0.8$, 
    in six out of 15 $\ybys$ bins. 
    The details correspond to those of 
    Fig.~\ref{fig:3d-uncertainties}. 
    \label{fig:appendix-3d-uncertainties-ptave-part1}}
\end{figure*}

\begin{figure*}[htb]
  \centering
    \includegraphics[width=\cmsFigWidthTwo]{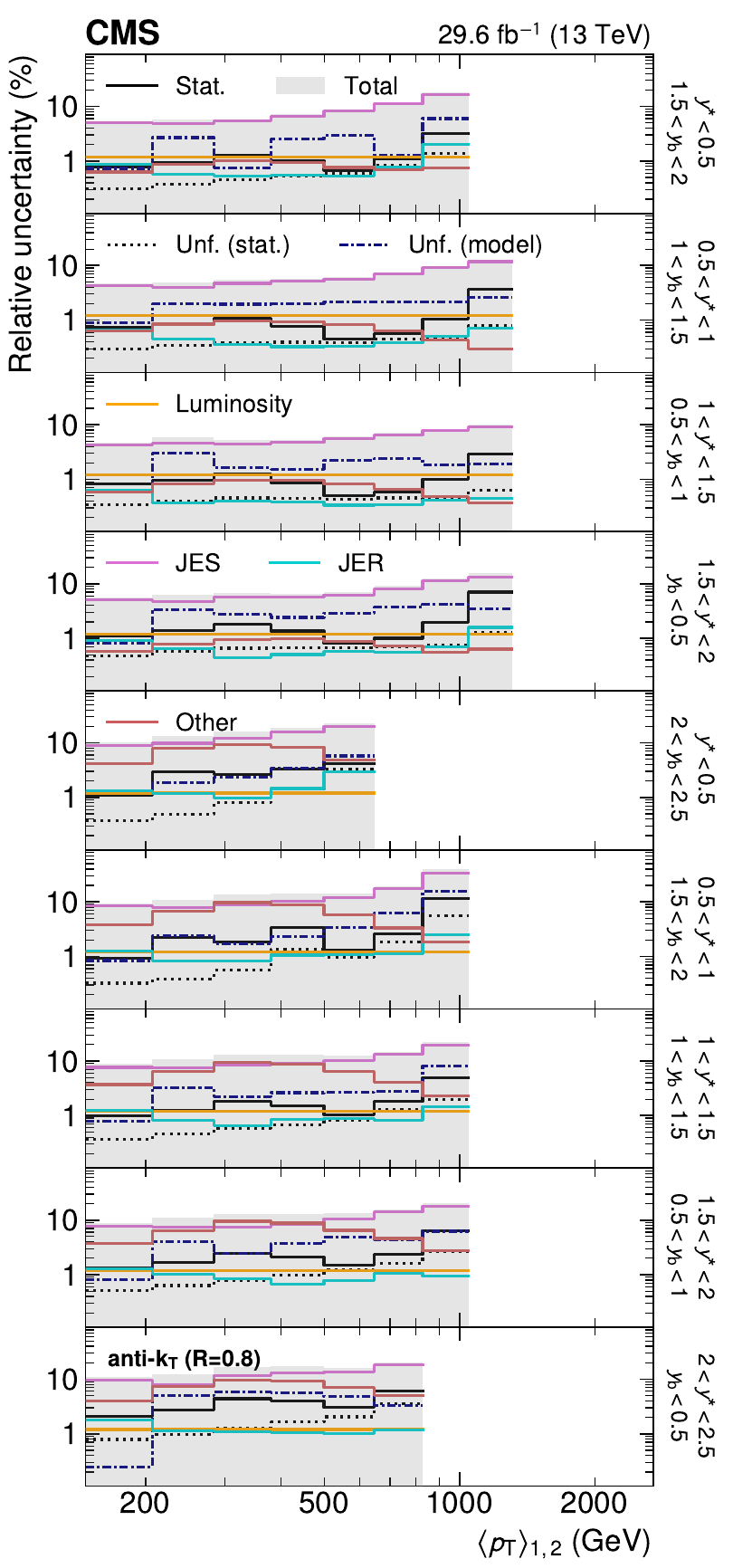}
    \caption{
    (\textit{continuation of Fig.~\ref{fig:appendix-3d-uncertainties-ptave-part1}}) 
    Breakdown of the experimental uncertainty 
    for the 3D measurements as a function of \ptave{} 
    using jets with $R = 0.8$, 
    in the remaining nine out of 15 $\ybys$ bins. 
    The details correspond to those of 
    Fig.~\ref{fig:3d-uncertainties}. 
    \label{fig:appendix-3d-uncertainties-ptave-part2}}
\end{figure*}

\begin{figure*}[htb]
  \centering
    \includegraphics[width=\cmsFigWidthTwo]{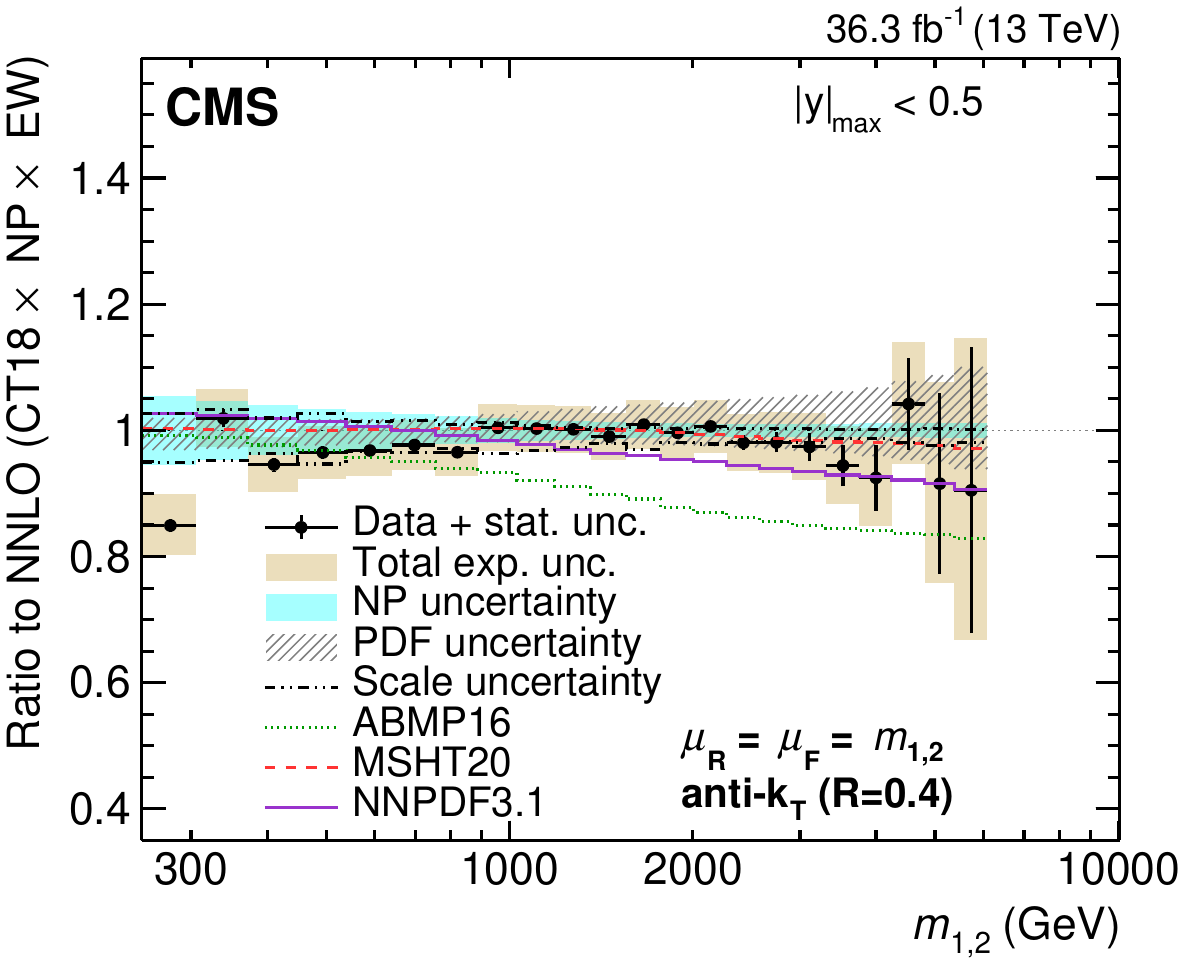}
    \includegraphics[width=\cmsFigWidthTwo]{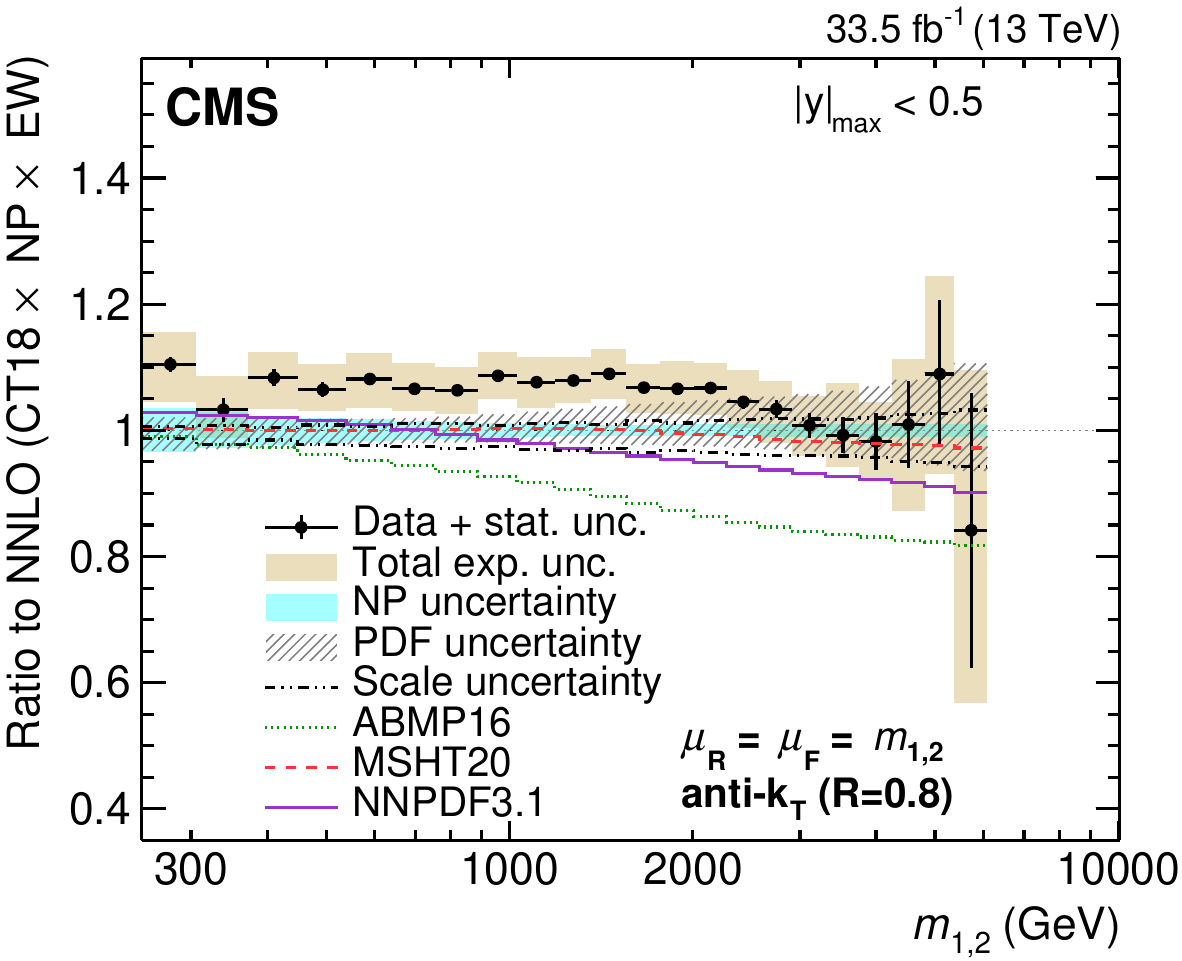}\\
    \includegraphics[width=\cmsFigWidthTwo]{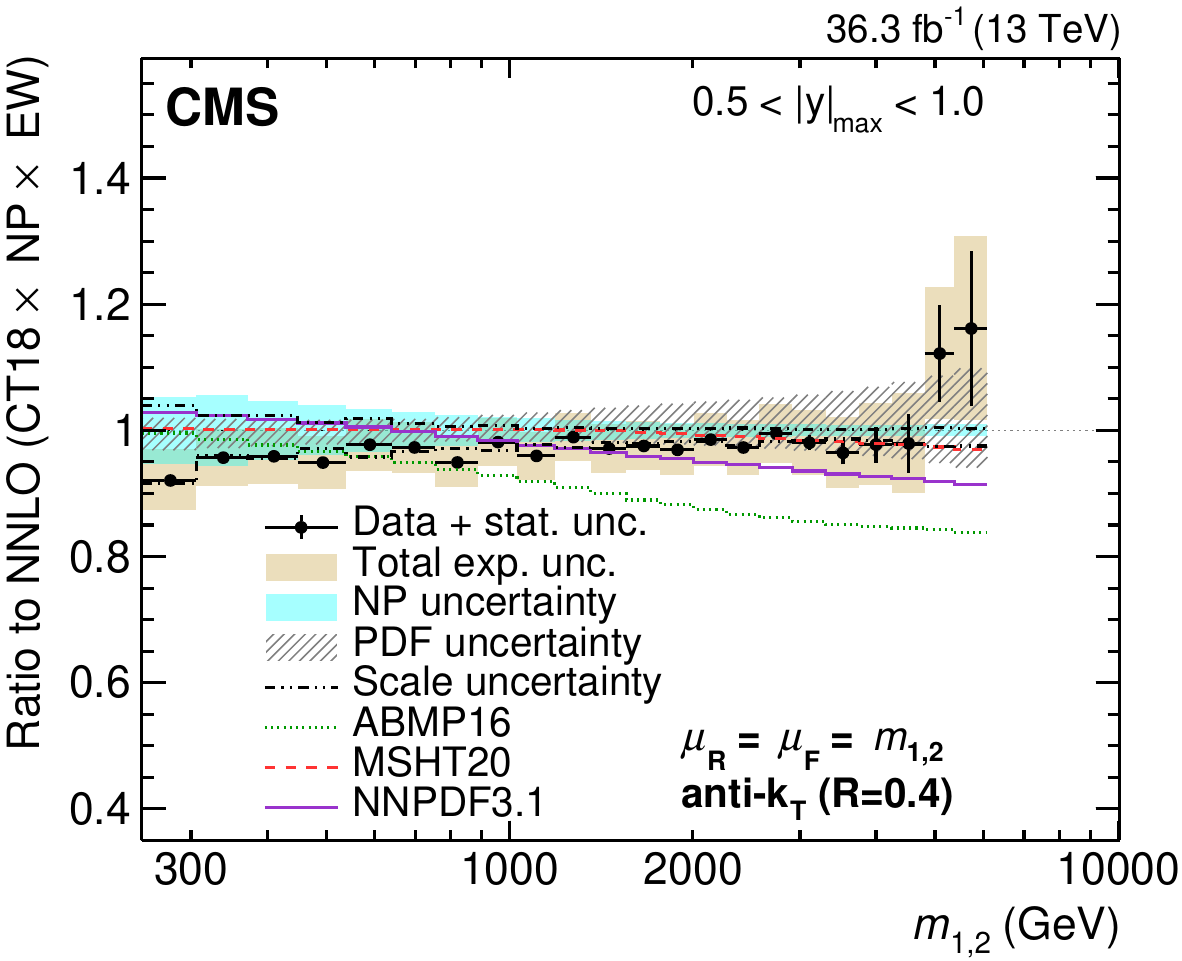}
    \includegraphics[width=\cmsFigWidthTwo]{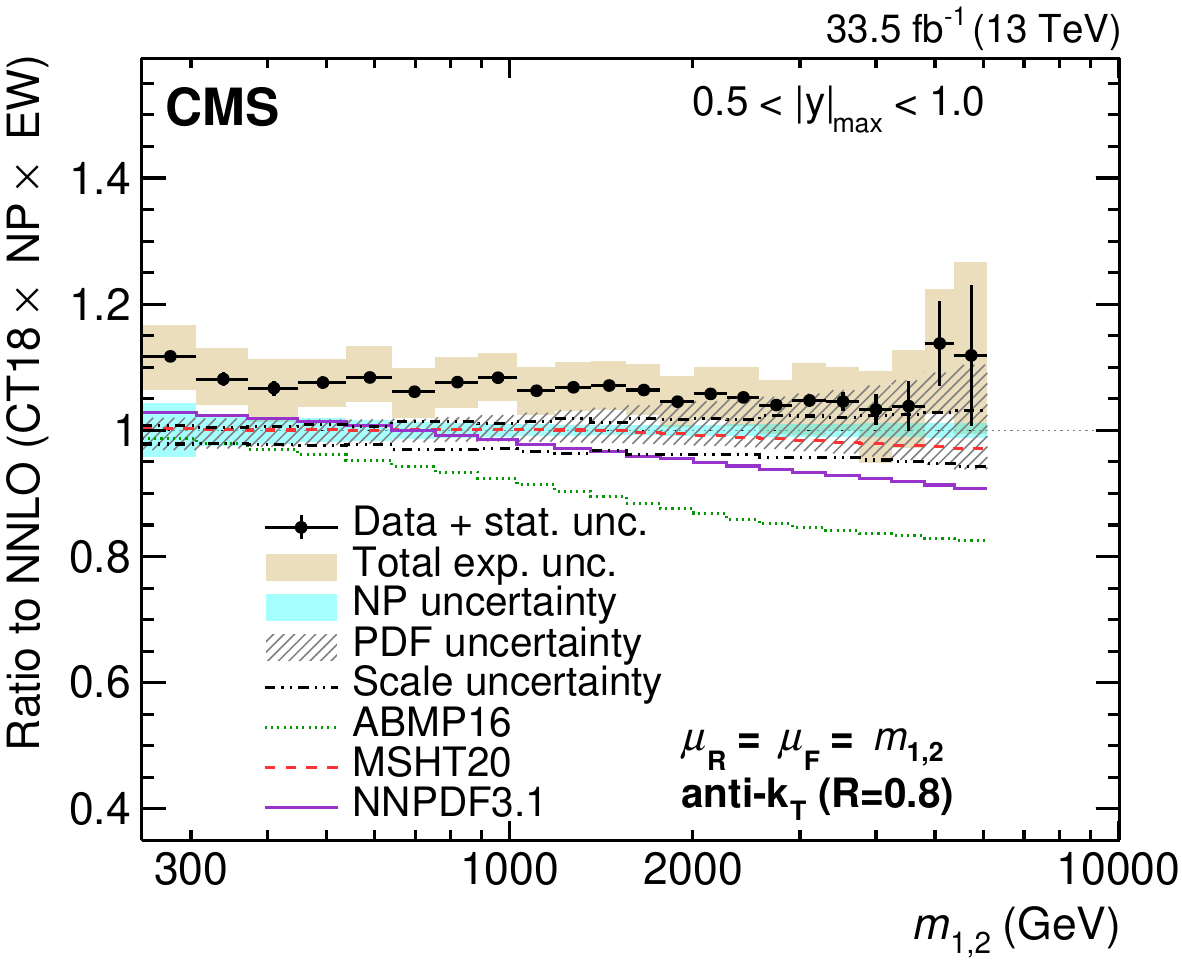}\\
    \includegraphics[width=\cmsFigWidthTwo]{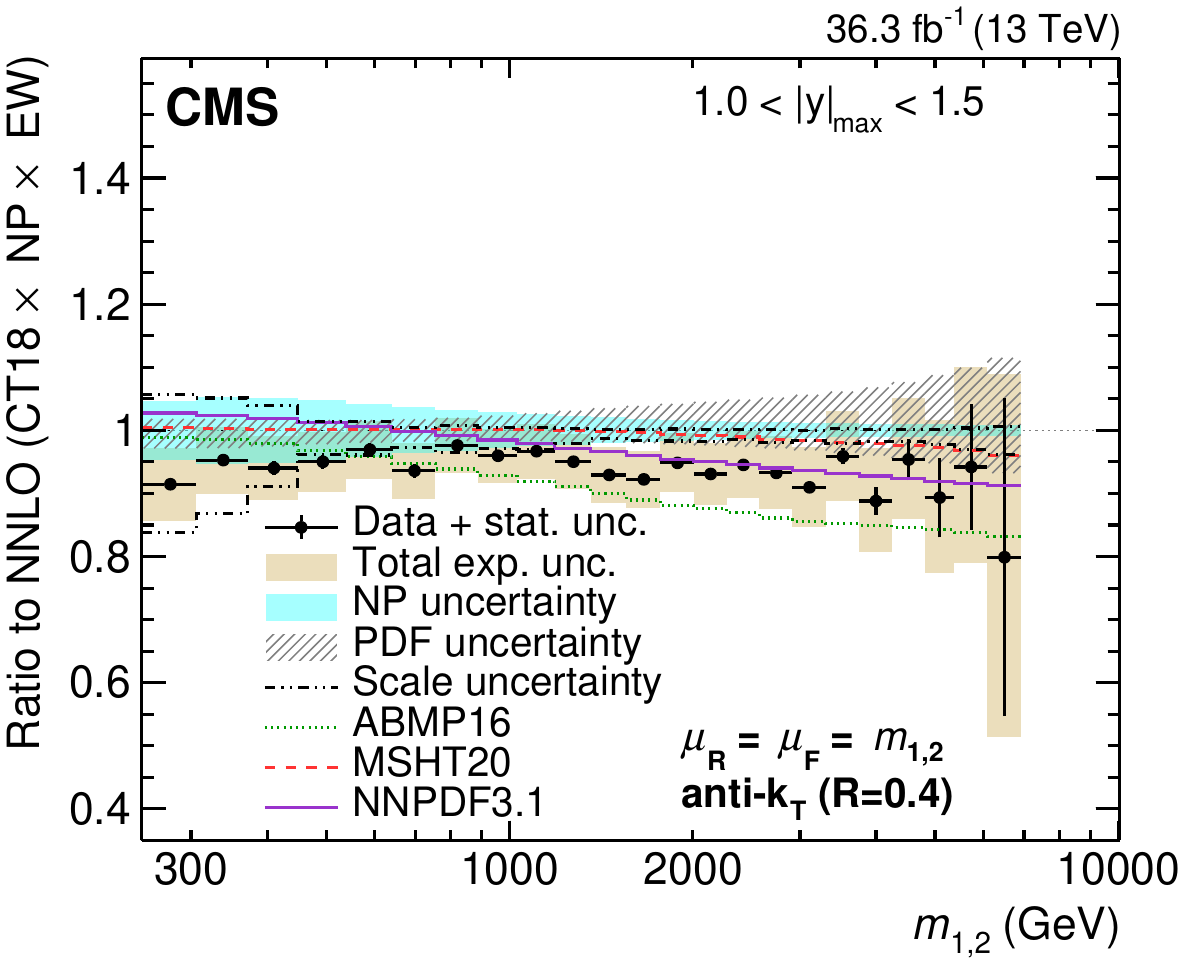}
    \includegraphics[width=\cmsFigWidthTwo]{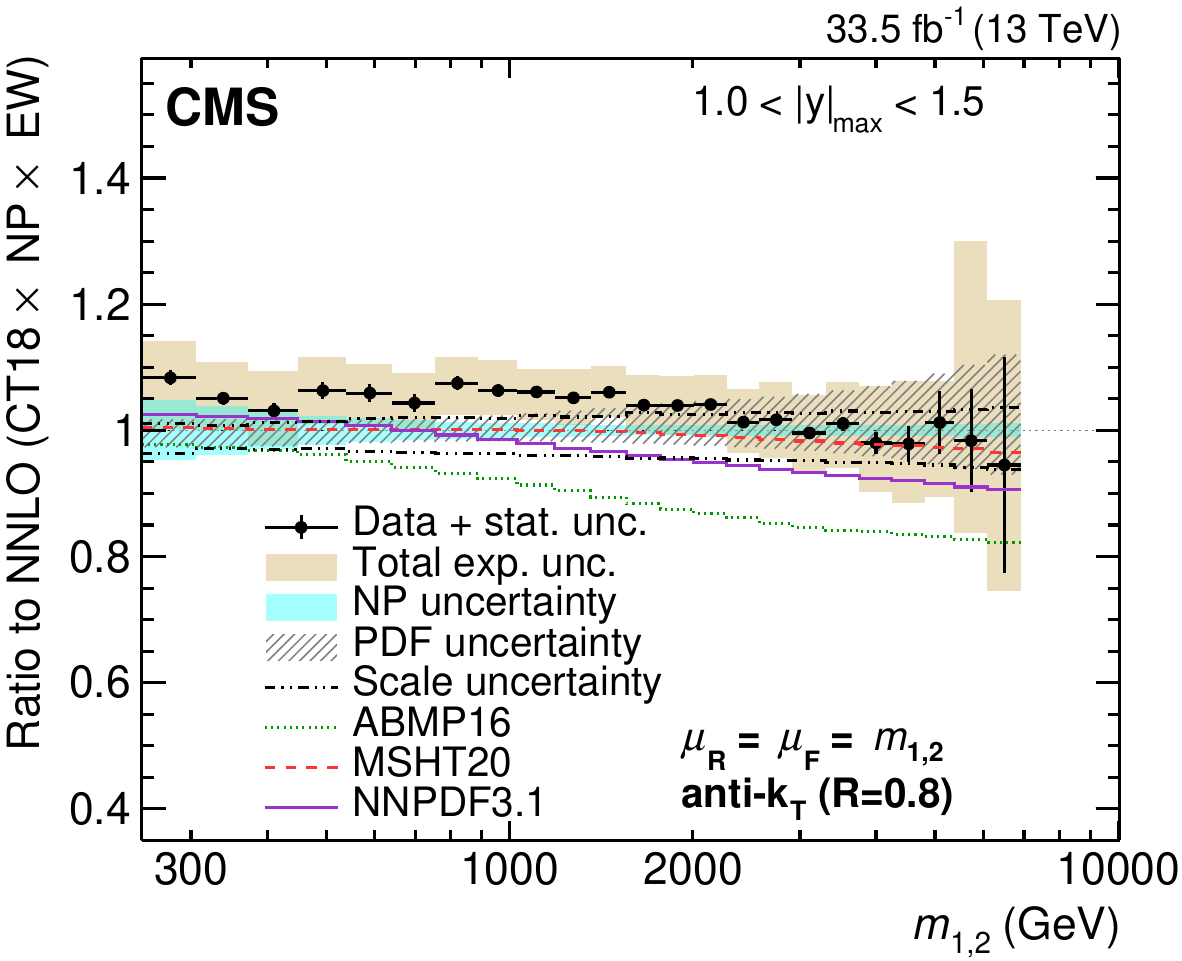}
    \caption{
    Comparison of the 2D dijet cross section 
    for jets with $R = 0.4$ (left) and 0.8 (right)
    as a function of $\mjj$ 
    to fixed-order theoretical calculations at NNLO, 
    shown here for three inner $\ymax$ regions. 
    The details correspond to those of 
    Fig.~\ref{fig:2d-ratios}. 
    \label{fig:appendix-2d-ratios-part1}}
\end{figure*}

\begin{figure*}[htb]
  \centering
    \includegraphics[width=\cmsFigWidthTwo]{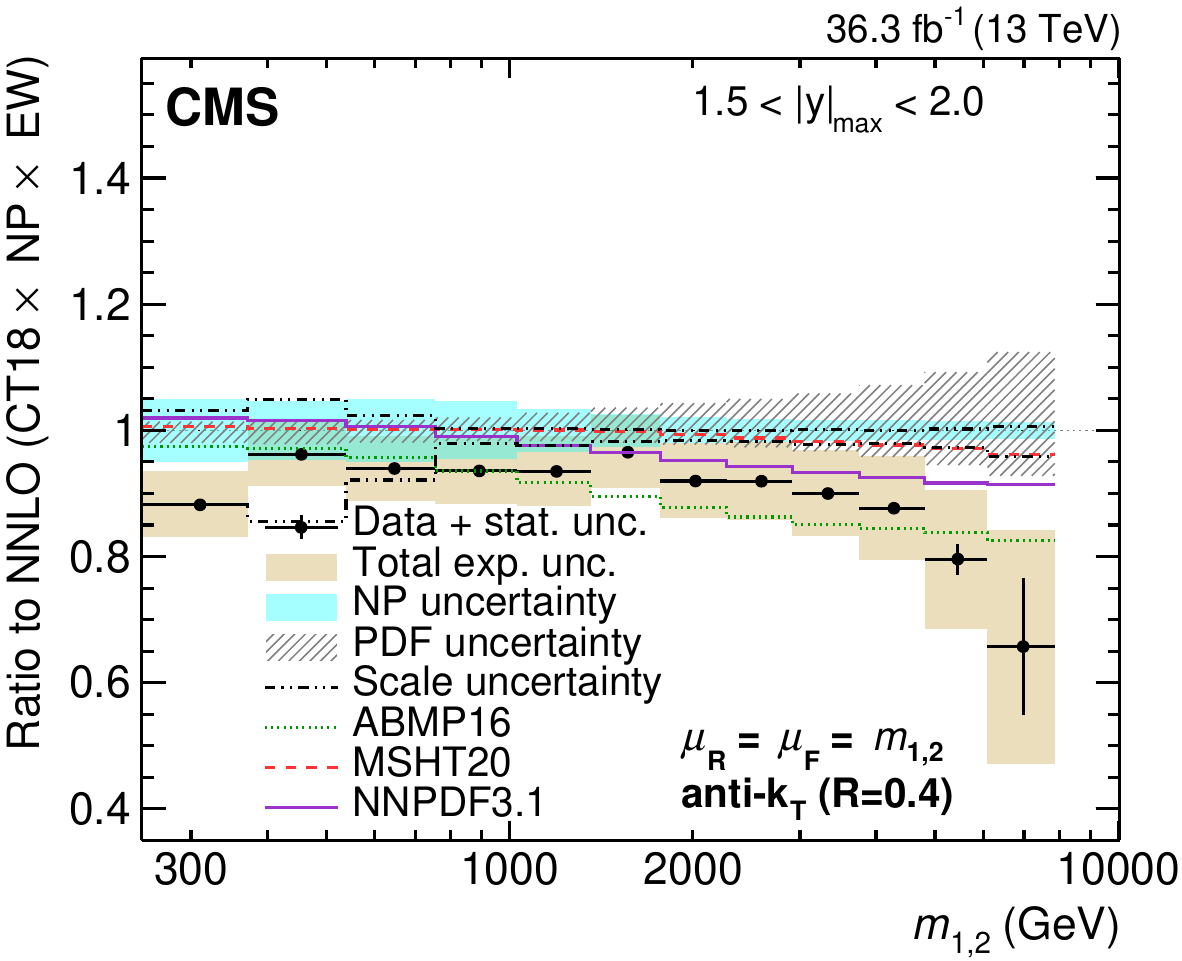}
    \includegraphics[width=\cmsFigWidthTwo]{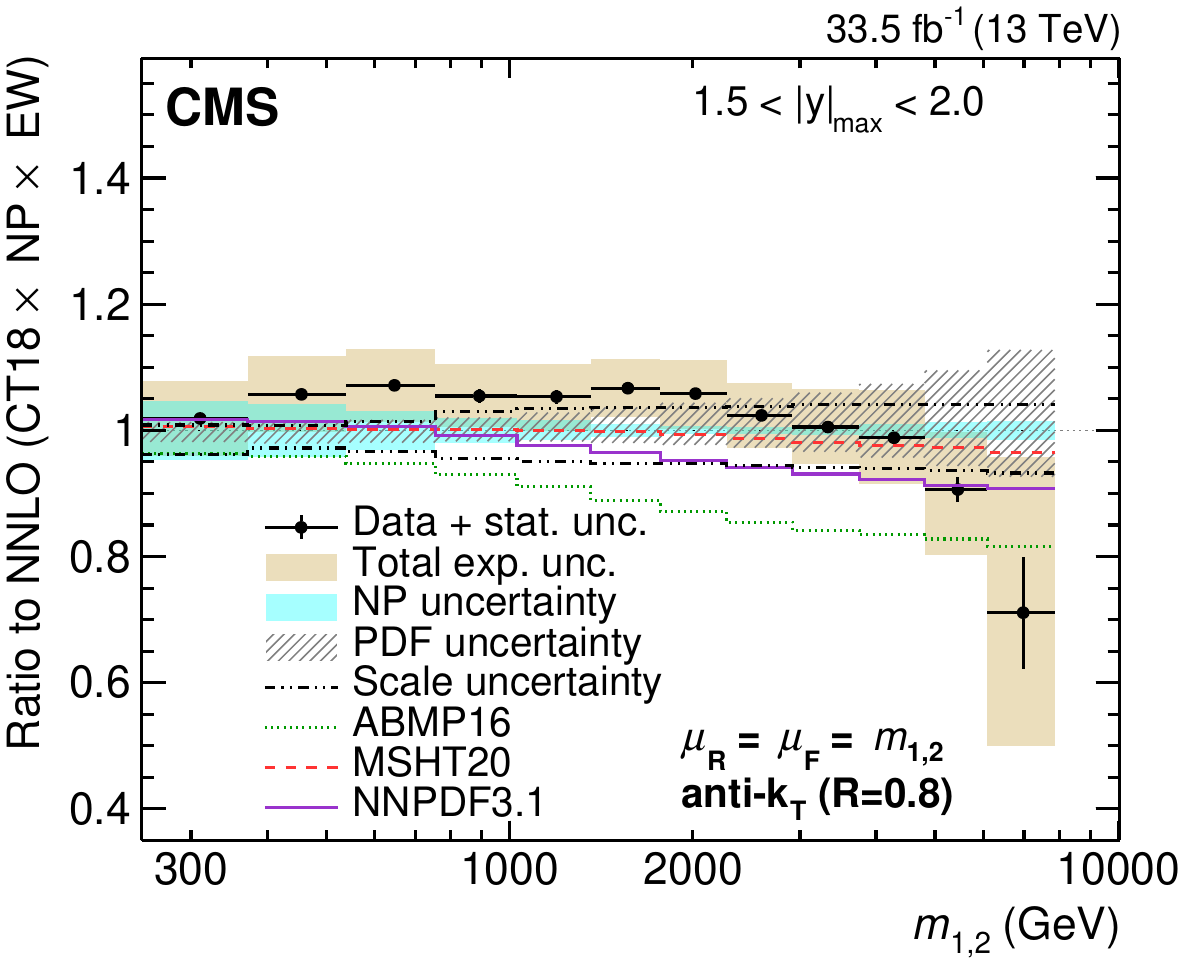}\\
    \includegraphics[width=\cmsFigWidthTwo]{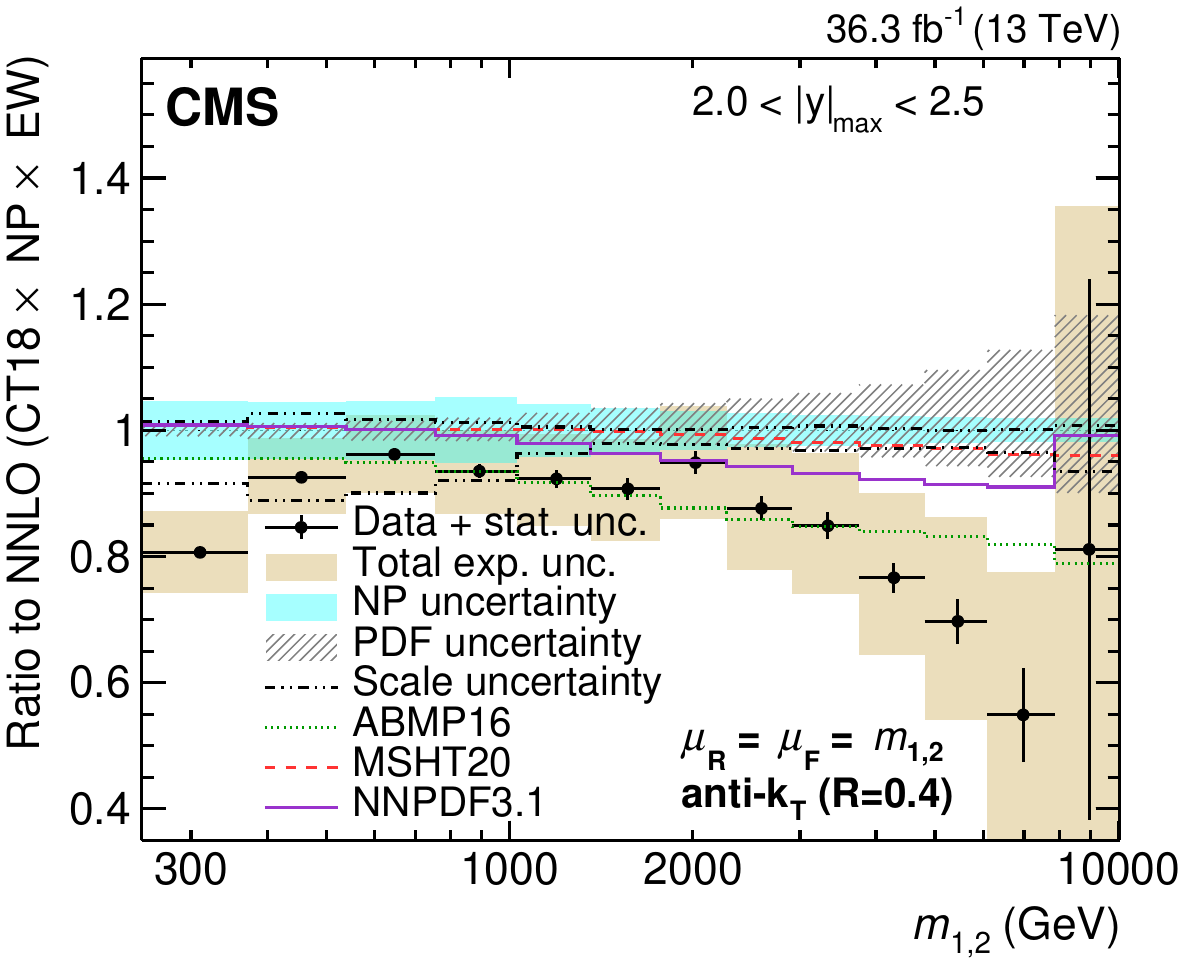}
    \includegraphics[width=\cmsFigWidthTwo]{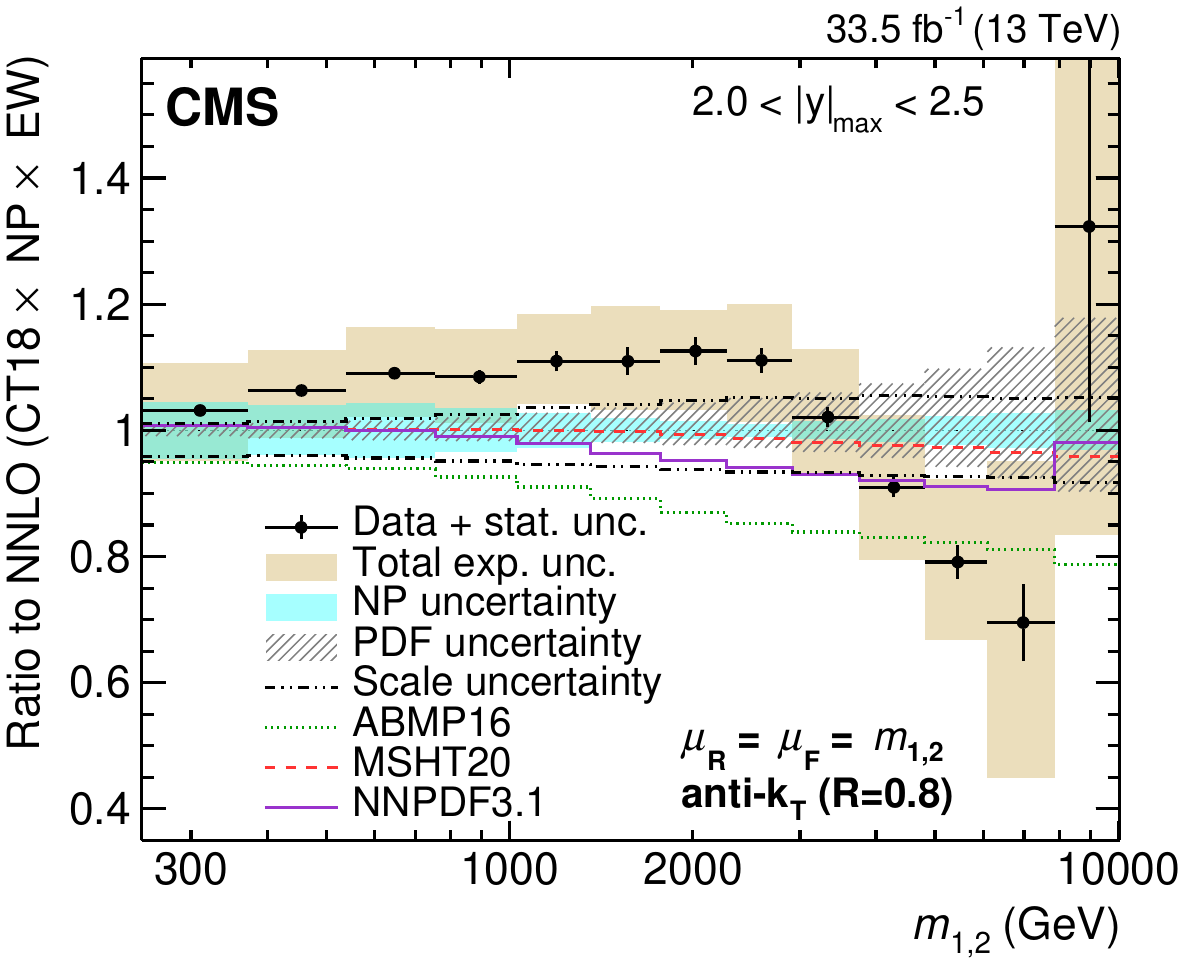}
    \caption{
    (\textit{continuation of Fig.~\ref{fig:appendix-2d-ratios-part1}})
    Comparison of the 2D dijet cross section 
    for jets with $R = 0.4$ (left) and 0.8 (right)
    as a function of $\mjj$ 
    to fixed-order theoretical calculations at NNLO, 
    shown here for two outermost $\ymax$ regions. 
    The details correspond to those of 
    Fig.~\ref{fig:2d-ratios}. 
    \label{fig:appendix-2d-ratios-part2}}
\end{figure*}

\begin{figure*}[htb]
  \centering
    \includegraphics[width=\cmsFigWidthTwo]{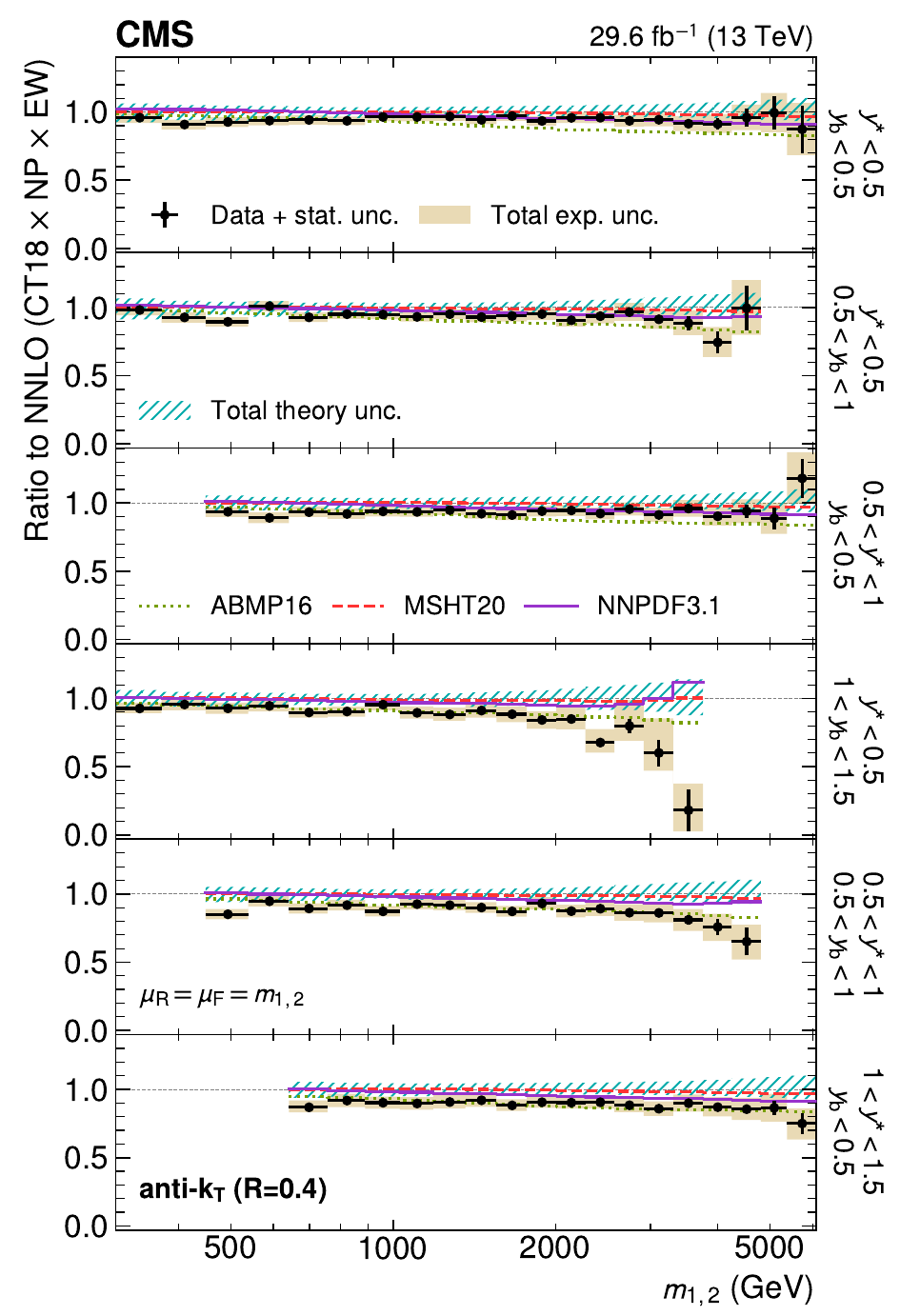}
    \includegraphics[width=\cmsFigWidthTwo]{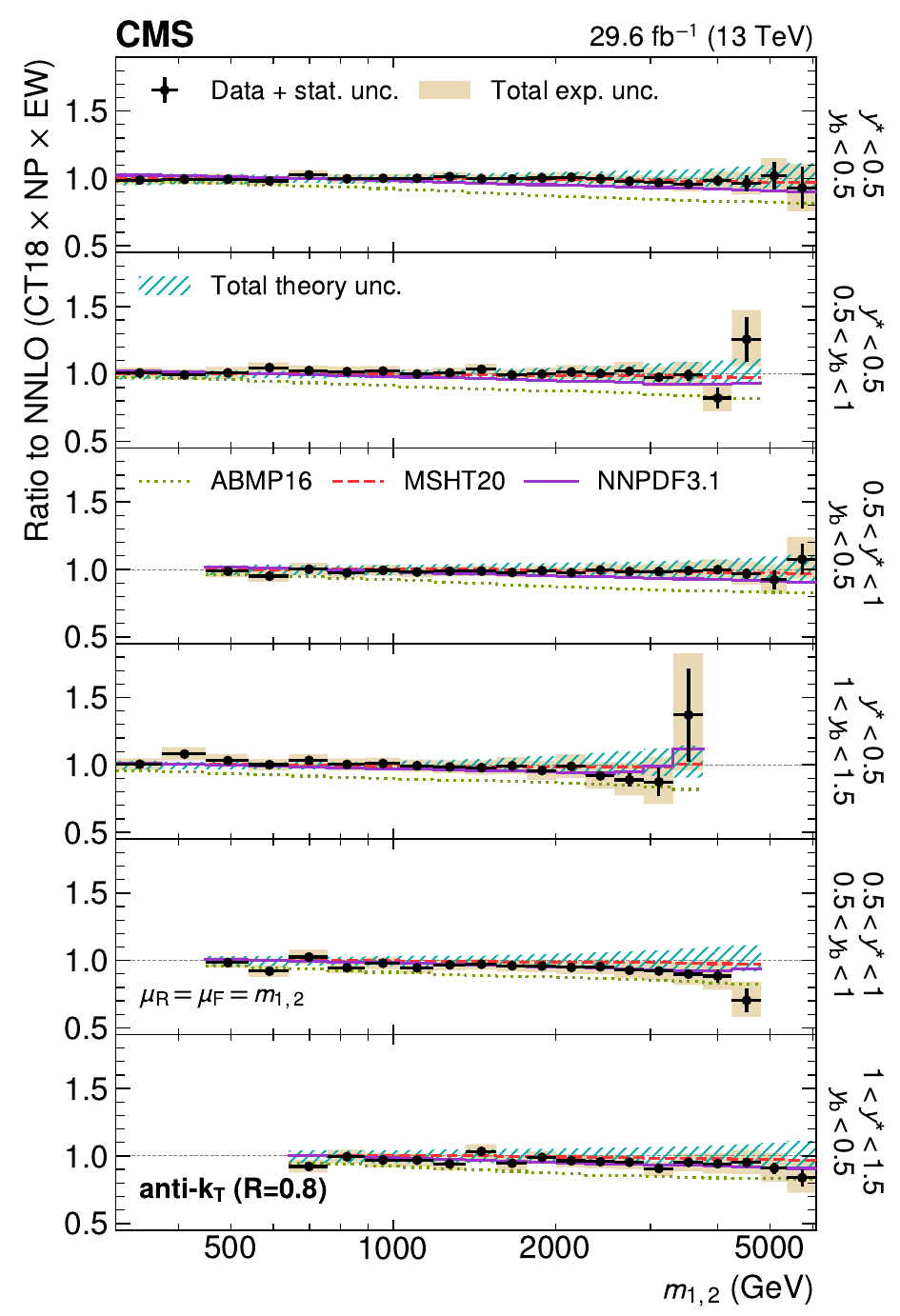}
    \caption{
    Comparison of the 3D dijet cross section 
    as a function of $\mjj$ 
    to fixed-order theoretical calculations at NNLO, 
    using jets with $R = 0.4$ (left) and 0.8 (right), 
    in six out of the total 15 $\ybys$ bins. 
    The details correspond to those of 
    Fig.~\ref{fig:3d-ratios}. 
    \label{fig:appendix-3d-ratios-mjj-part1}}
\end{figure*}

\begin{figure*}[htb]
  \centering
    \includegraphics[width=\cmsFigWidthTwo]{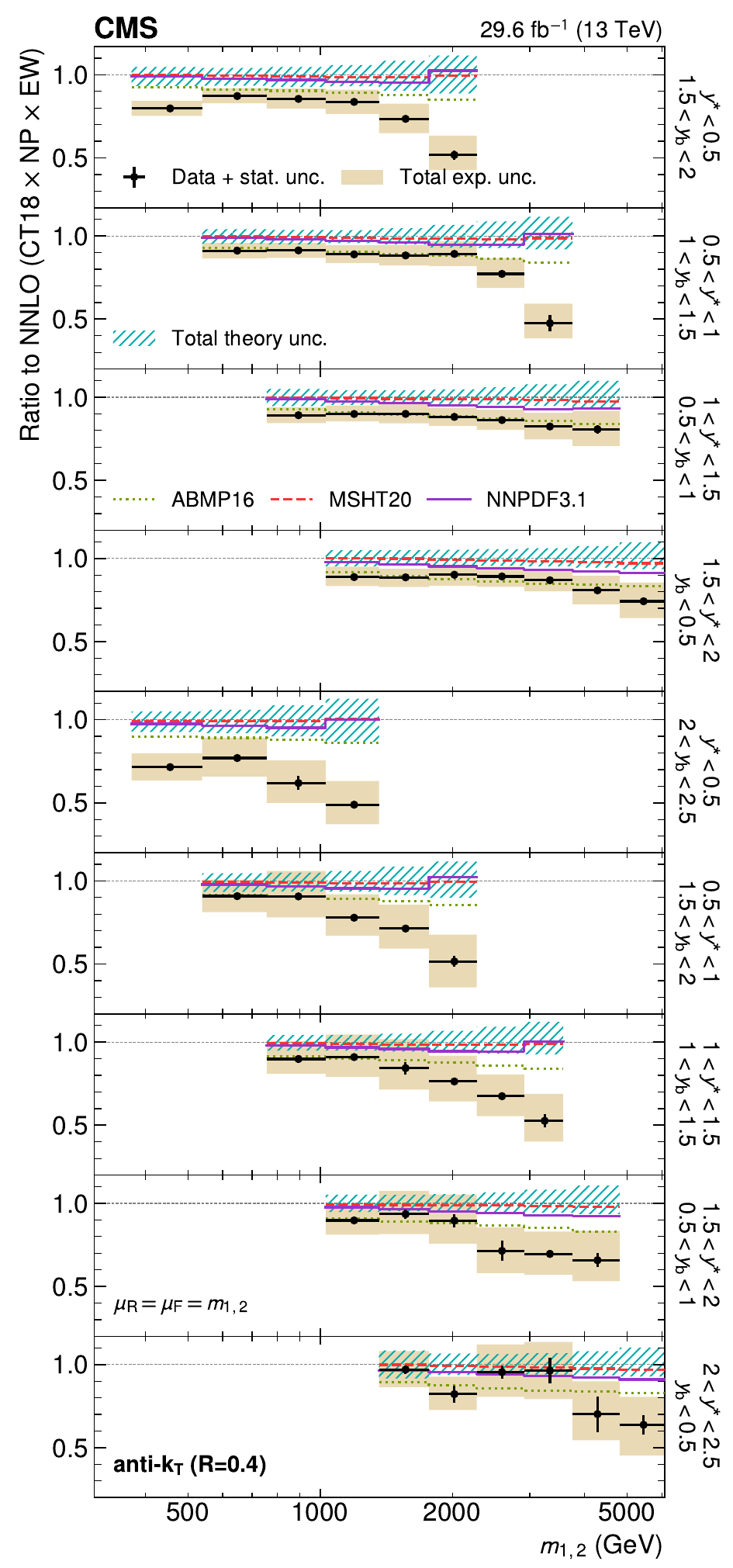}
    \includegraphics[width=\cmsFigWidthTwo]{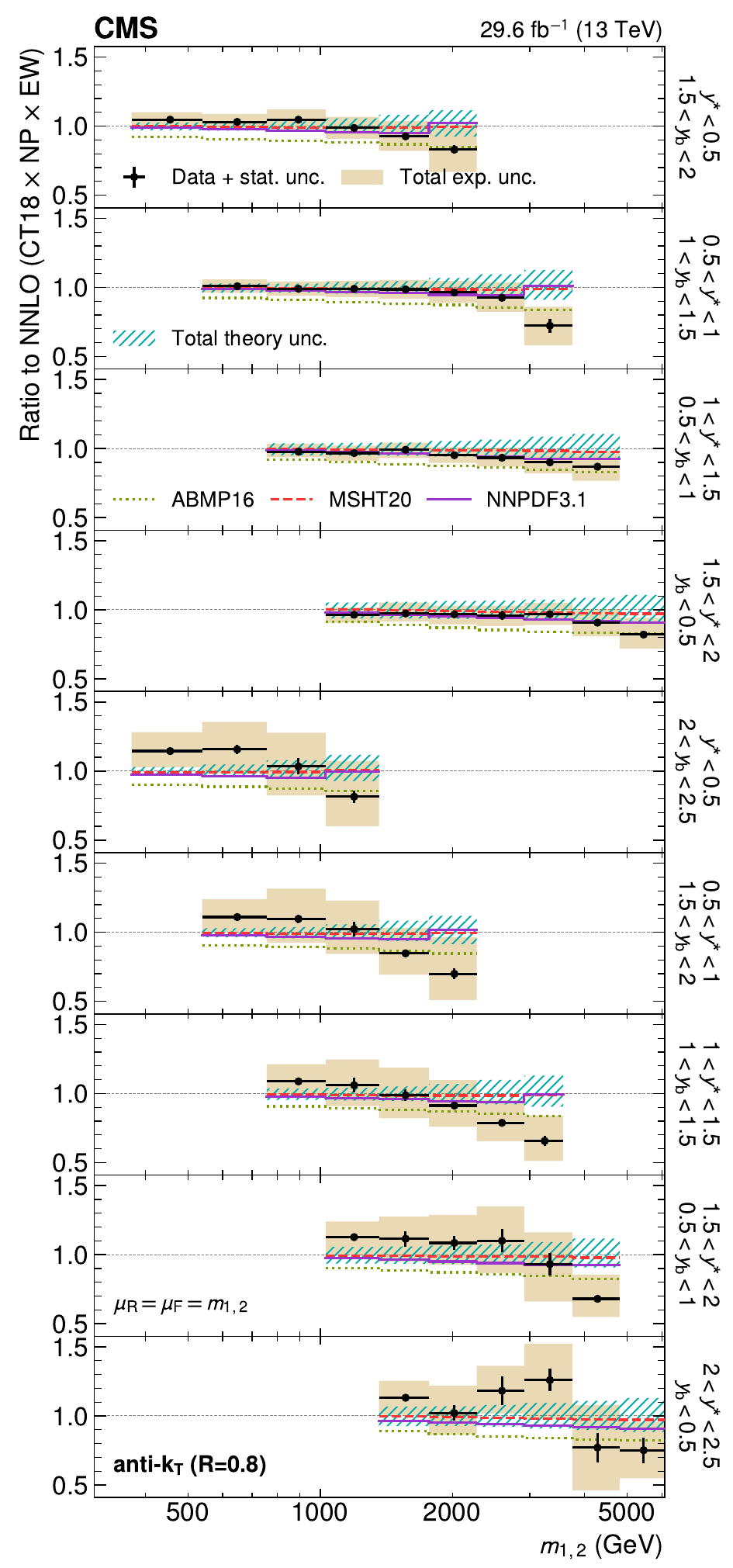}
    \caption{
    (\textit{continuation of Fig.~\ref{fig:appendix-3d-ratios-mjj-part1}})
    Comparison of the 3D dijet cross section 
    as a function of $\mjj$ 
    to fixed-order theoretical calculations at NNLO, 
    using jets with $R = 0.4$ (left) and 0.8 (right), 
    in the remaining nine out of 15 $\ybys$ bins. 
    The details correspond to those of 
    Fig.~\ref{fig:3d-ratios}. 
    \label{fig:appendix-3d-ratios-mjj-part2}}
\end{figure*}

\begin{figure*}[htb]
  \centering
    \includegraphics[width=\cmsFigWidthTwo]{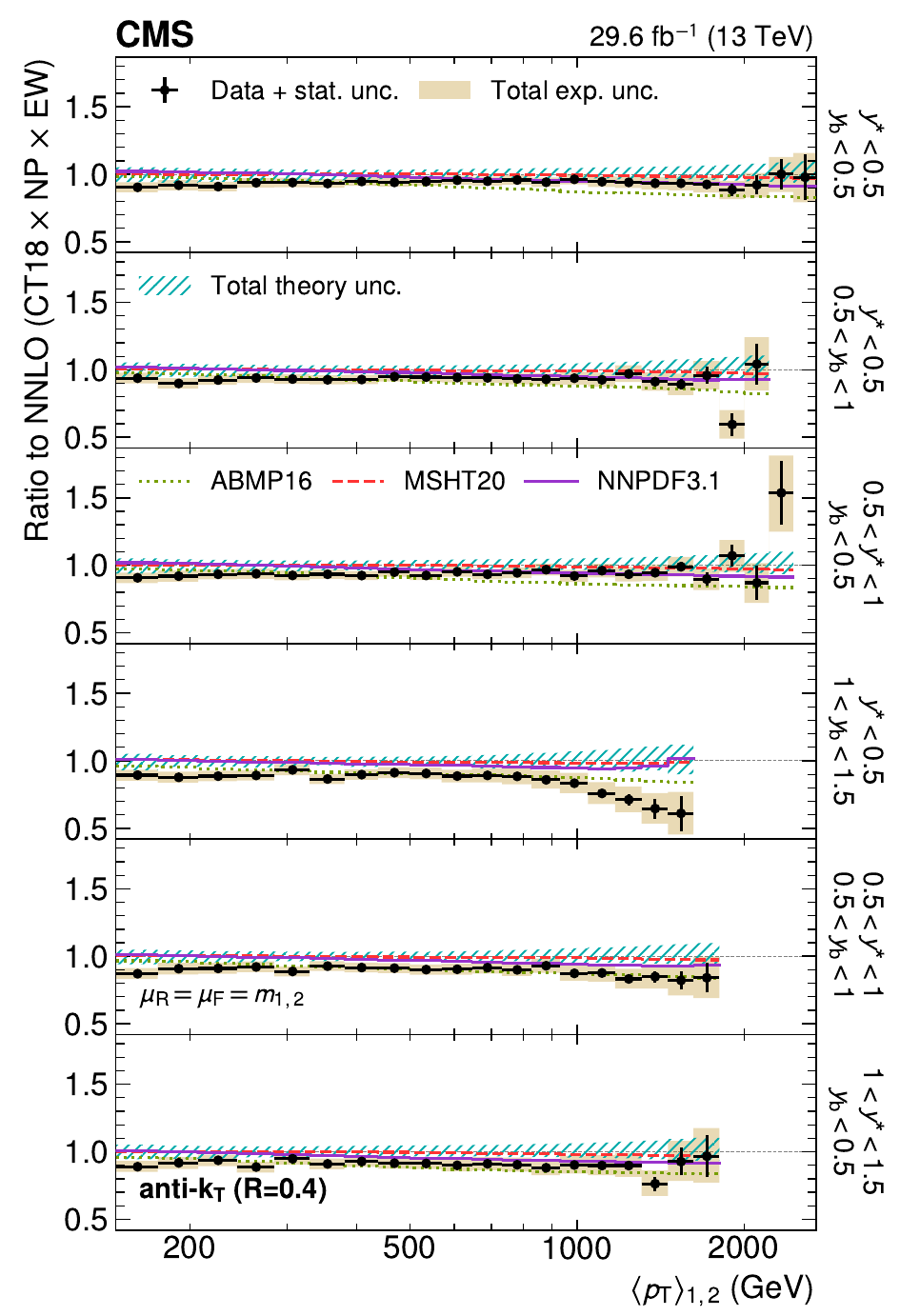}
    \includegraphics[width=\cmsFigWidthTwo]{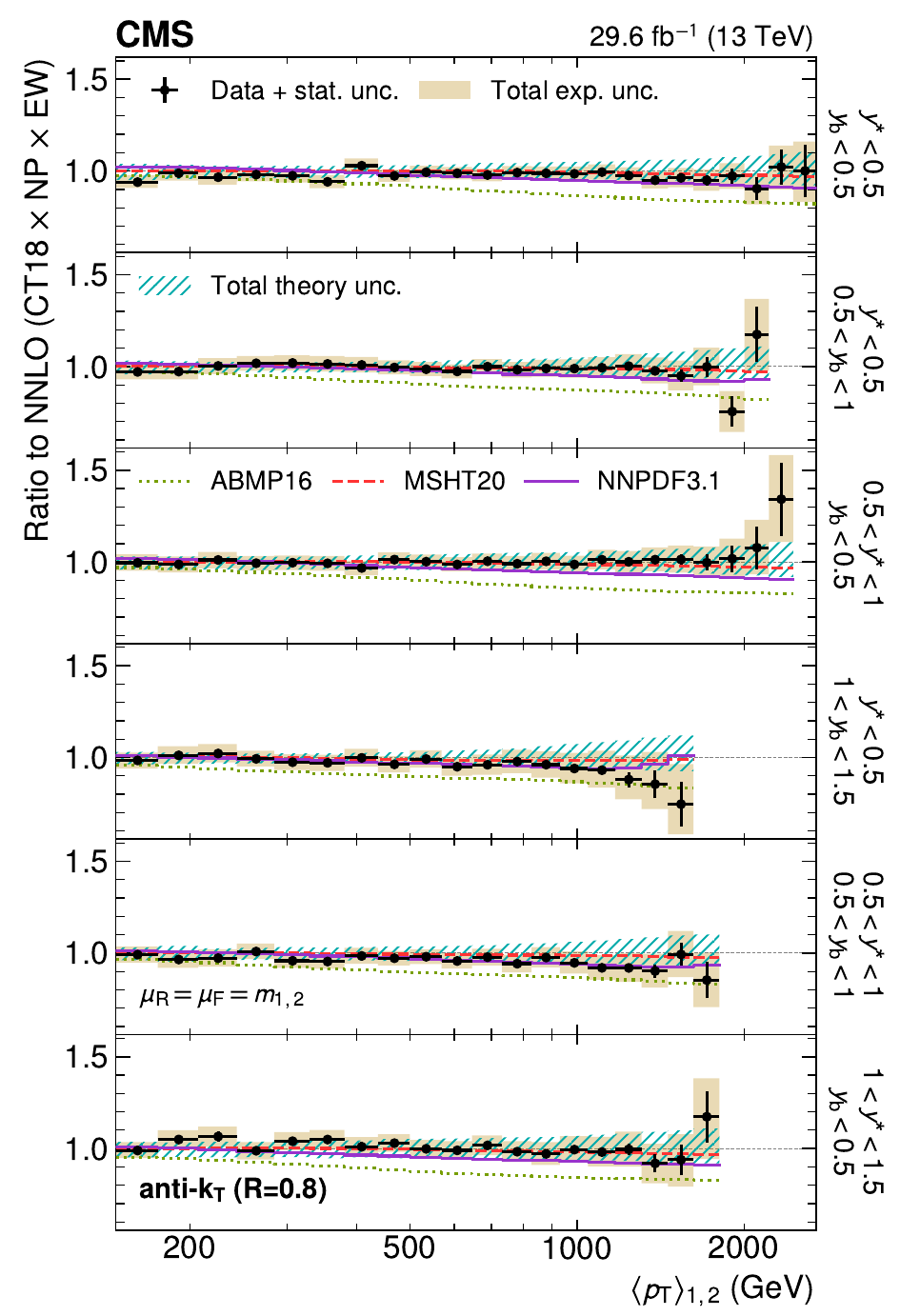}
    \caption{
    Comparison of the 3D dijet cross section 
    as a function of $\ptave$ 
    to fixed-order theoretical calculations at NNLO, 
    using jets with $R = 0.4$ (left) and 0.8 (right), 
    in six out of the total 15 $\ybys$ bins. 
    The details correspond to those of 
    Fig.~\ref{fig:3d-ratios}. 
    \label{fig:appendix-3d-ratios-ptave-part1}}
\end{figure*}

\begin{figure*}[htb]
  \centering
    \includegraphics[width=\cmsFigWidthTwo]{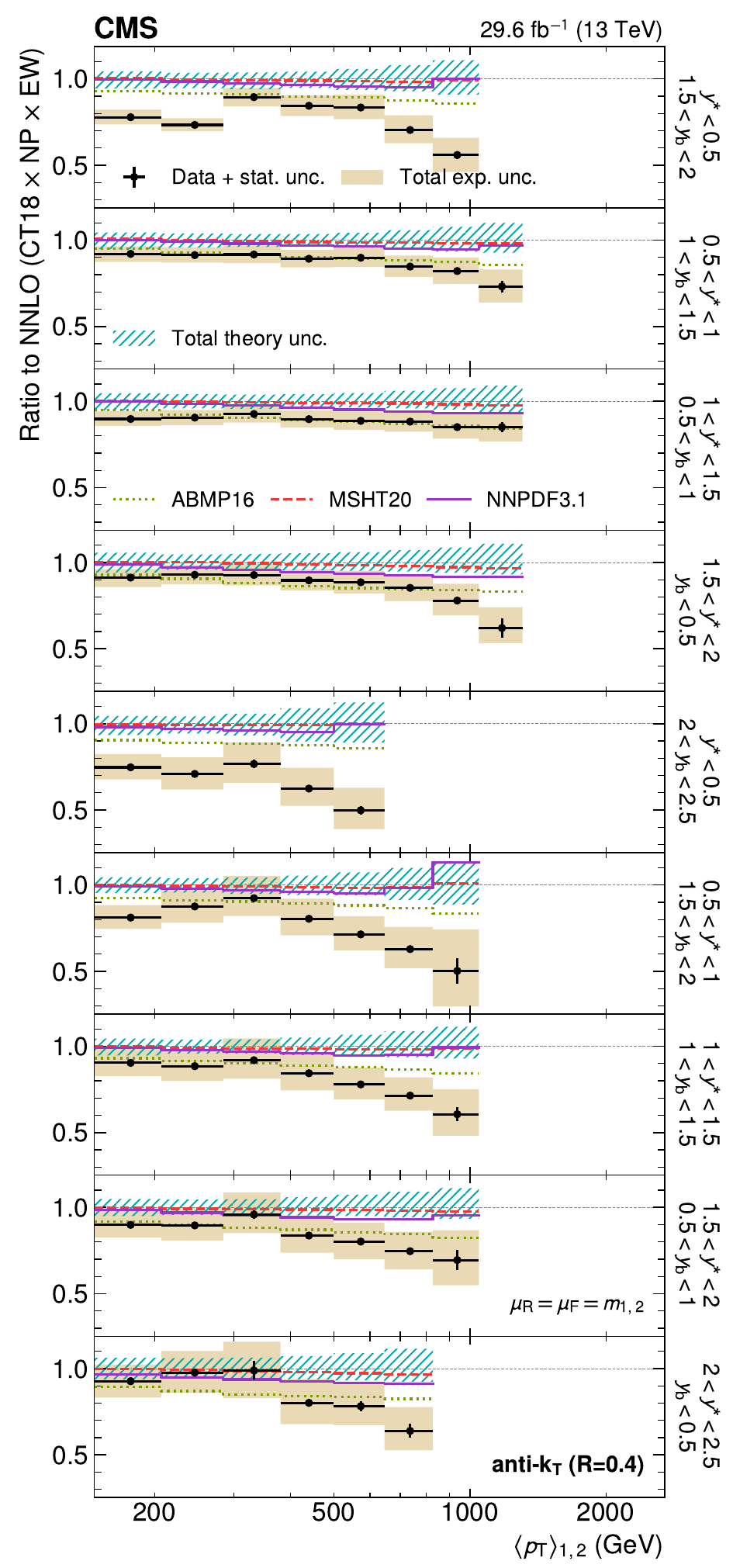}
    \includegraphics[width=\cmsFigWidthTwo]{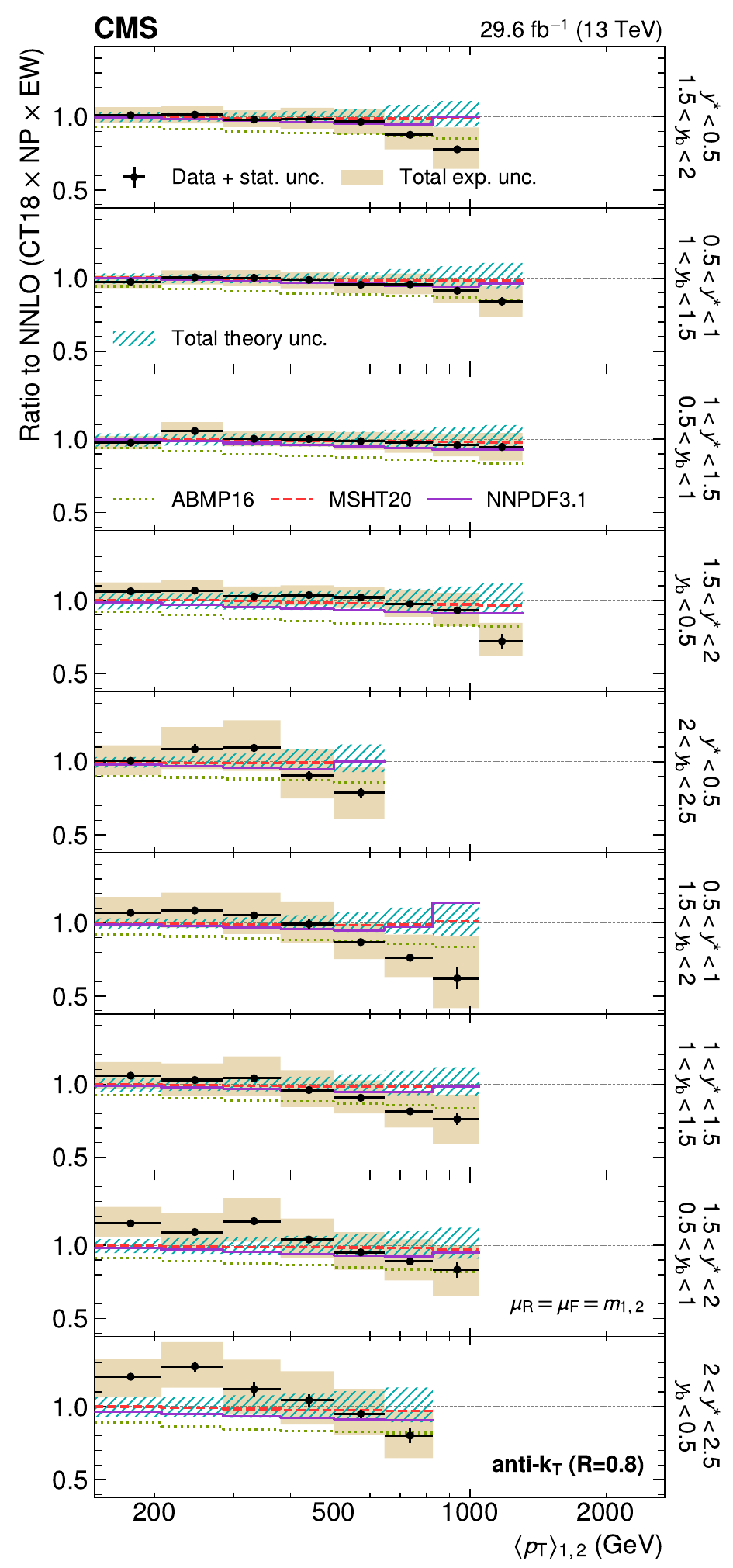}
    \caption{
    (\textit{continuation of Fig.~\ref{fig:appendix-3d-ratios-ptave-part1}})
    Comparison of the 3D dijet cross section 
    as a function of $\ptave$ 
    to fixed-order theoretical calculations at NNLO, 
    using jets with $R = 0.4$ (left) and 0.8 (right),
    in the remaining nine out of 15 $\ybys$ bins. 
    The details correspond to those of 
    Fig.~\ref{fig:3d-ratios}. 
    \label{fig:appendix-3d-ratios-ptave-part2}}
\end{figure*}
\cleardoublepage \section{The CMS Collaboration \label{app:collab}}\begin{sloppypar}\hyphenpenalty=5000\widowpenalty=500\clubpenalty=5000
\cmsinstitute{Yerevan Physics Institute, Yerevan, Armenia}
{\tolerance=6000
A.~Hayrapetyan, A.~Tumasyan\cmsAuthorMark{1}\cmsorcid{0009-0000-0684-6742}
\par}
\cmsinstitute{Institut f\"{u}r Hochenergiephysik, Vienna, Austria}
{\tolerance=6000
W.~Adam\cmsorcid{0000-0001-9099-4341}, J.W.~Andrejkovic, T.~Bergauer\cmsorcid{0000-0002-5786-0293}, S.~Chatterjee\cmsorcid{0000-0003-2660-0349}, K.~Damanakis\cmsorcid{0000-0001-5389-2872}, M.~Dragicevic\cmsorcid{0000-0003-1967-6783}, A.~Escalante~Del~Valle\cmsorcid{0000-0002-9702-6359}, P.S.~Hussain\cmsorcid{0000-0002-4825-5278}, M.~Jeitler\cmsAuthorMark{2}\cmsorcid{0000-0002-5141-9560}, N.~Krammer\cmsorcid{0000-0002-0548-0985}, L.~Lechner\cmsorcid{0000-0002-3065-1141}, D.~Liko\cmsorcid{0000-0002-3380-473X}, I.~Mikulec\cmsorcid{0000-0003-0385-2746}, J.~Schieck\cmsAuthorMark{2}\cmsorcid{0000-0002-1058-8093}, R.~Sch\"{o}fbeck\cmsorcid{0000-0002-2332-8784}, D.~Schwarz\cmsorcid{0000-0002-3821-7331}, M.~Sonawane\cmsorcid{0000-0003-0510-7010}, S.~Templ\cmsorcid{0000-0003-3137-5692}, W.~Waltenberger\cmsorcid{0000-0002-6215-7228}, C.-E.~Wulz\cmsAuthorMark{2}\cmsorcid{0000-0001-9226-5812}
% ...
\par}
\vskip\cmsinstskip
\dag:~Deceased\\
$^{1}$Also at Yerevan State University, Yerevan, Armenia\\
$^{2}$Also at TU Wien, Vienna, Austria\\
% ...
\end{sloppypar}
\end{document}